\RequirePackage{amsmath}  
\documentclass[epj,nopacs]{svjour}
\pdfoutput=1
\usepackage{graphicx}
\usepackage{amssymb}
\usepackage{xcolor}
\usepackage[utf8]{inputenc}
\usepackage{xspace}
\usepackage{hepunits}
\usepackage{multirow}
\usepackage{slashed}

\newcommand{\etal}{\textit{et al.\xspace}}
\newcommand{\ie}{\textit{i.e.}\xspace}
\newcommand{\eg}{\textit{e.g.}\xspace}
\newcommand{\apriori}{\textit{a priori}\xspace}
\newcommand{\lc}{light cone\xspace}
\newcommand{\nonpert}{nonperturbative\xspace}
\newcommand{\crossover}{crossover\xspace}
\newcommand{\EM}{energy-momentum\xspace}
\newcommand{\rhs}{RHS\xspace}
\newcommand{\lhs}{LHS\xspace}

\newcommand{\xb}{x_B}
\newcommand{\Q}{Q}
\newcommand{\im}{\mathfrak{Im}\,{}}
\newcommand{\re}{\mathfrak{Re}\,{}}

\newcommand{\refeq}[1]{eq.~(\ref{#1})}
\newcommand{\refeqs}[2]{eqs.~(\ref{#1}-\ref{#2})}
\newcommand{\reftab}[1]{table~\ref{#1}}
\newcommand{\reftabs}[2]{tables~(\ref{#1}-\ref{#2})}
\newcommand{\reffig}[1]{fig.~\ref{#1}}
\newcommand{\refcite}[1]{ref.~\cite{#1}}
\newcommand{\refcites}[1]{refs.~\cite{#1}}
\newcommand{\refsec}[1]{sec.~\ref{#1}}

\begin{document}
\title{GPD phenomenology and DVCS fitting}
\subtitle{Entering the high-precision era}
\author{Kre\v{s}imir Kumeri\v{c}ki\inst{1}
        \and Simonetta Liuti\inst{2}
        \and Herv\'e Moutarde\inst{3}
}                     
\institute{Department of Physics, Faculty of Science, University
of Zagreb, Bijeni\v{c}ka cesta 32, HR-10000 Zagreb, Croatia \and
Physics Department, University of Virginia, 382 MCormick Rd., Charlottesville, VA 22904, USA\\Laboratori Nazionali di Frascati, INFN, Frascati, Italy \and
CEA, Centre de Saclay, IRFU/Service de Physique Nucléaire, F-91191
Gif-sur-Yvette, France}
\date{Received: date / Revised version: date}
%
\abstract{
We review the phenomenological framework for accessing
Generalized Parton Distributions (GPDs) using measurements
of Deeply Virtual Compton Scattering (DVCS) from a proton target.
We describe various GPD models and fitting procedures,
emphasizing specific challenges posed both by the internal structure and properties of 
the GPD functions and by their relation to observables.
Bearing in mind forthcoming data of unprecedented accuracy,
we give a set of recommendations 
to better define the pathway 
for a precise extraction of GPDs from experiment.
} 
\maketitle
%


\section{Introduction}
\label{sec:introduction}


\subsection{The physics case}
\label{sec:gpd-motivation}

\begin{figure}[ht]
  \centering
  \includegraphics[scale=0.4]{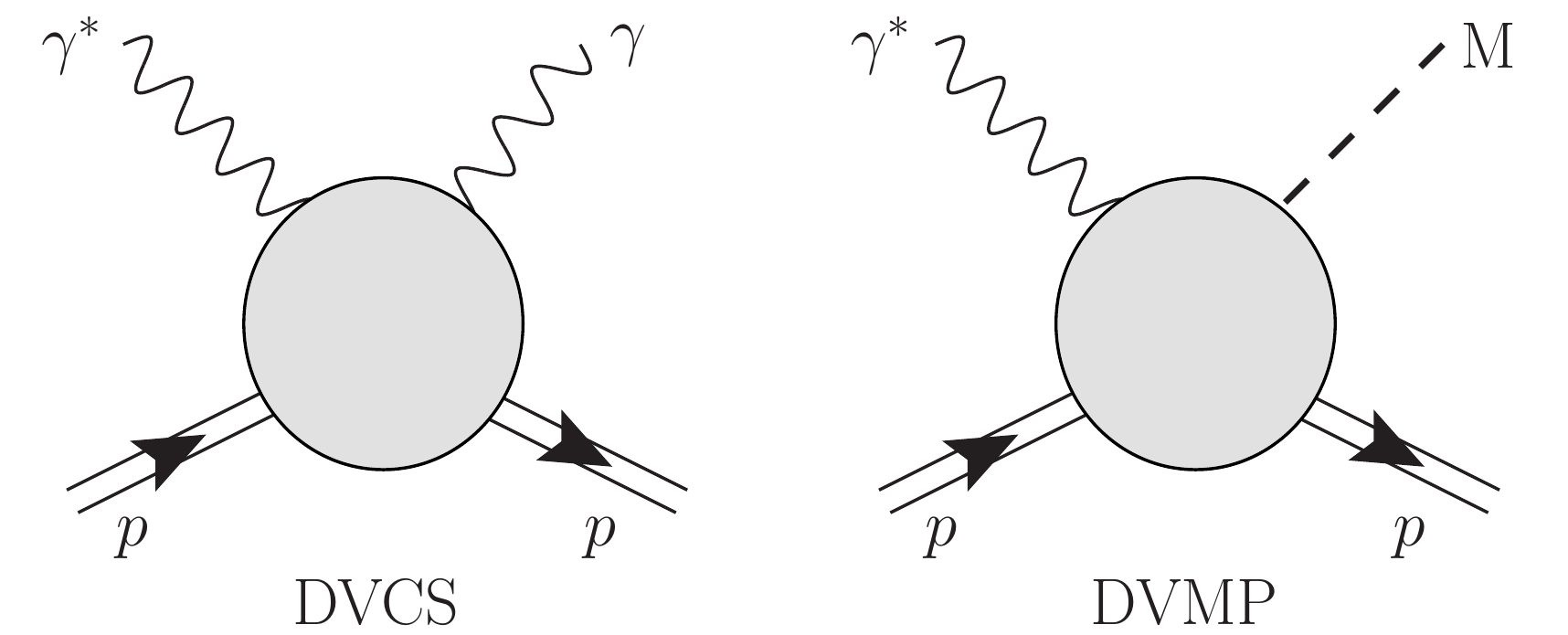}
  \caption{Two important processes facilitating access to GPDs: deeply
  virtual Compton scattering (DVCS) and deeply virtual meson production (DVMP).}
  \label{fig:dvcsVSdvmp}
\end{figure}

Generalized Parton Distributions (GPDs) were introduced in
1994 \cite{Mueller:1998fv} and rediscovered independently in
1997 \cite{Radyushkin:1996nd,Ji:1996nm}. This branch of QCD
studies grew rapidly because of their unique properties.
GPDs are related to other \nonpert objects that were studied
independently beforehand: Parton Distribution Functions
(PDFs) and Form Factors (FFs). In the infinite-momentum frame,
PDFs describe the longitudinal momentum distributions of
partons inside a hadron, and FFs are the Fourier transform
of the hadron charge distribution in the transverse plane.
GPDs naturally encompass PDFs and FFs in the case of
all hadrons, and they also extend
the notion of a Distribution Amplitude (DA) in the pion
case. This generality is remarkably complemented by one
outstanding feature: GPDs are directly related to the matrix
element of the QCD \EM tensor sandwiched between hadron
states. This is both welcome and surprising because
the \EM tensor in canonically probed through gravity. 
GPDs bring the considered energy-mo\-men\=tum matrix element 
within experimental grasp
through electromagnetic scattering. It was indeed 
realized early on that, owing to the factorization property of QCD, exclusive electroproduction of a real photon or a meson off a nucleon target at high momentum transfer
is theoretically the cleanest way to access GPDs. 
The processes of Deeply Virtual Compton Scattering (DVCS), and Deeply Virtual Meson Production (DVMP) are shown in \reffig{fig:dvcsVSdvmp}. 
The access to GPDs through DVCS and DVMP is indirect because DVCS does not depend
directly on GPDs, but on Compton Form Factors (CFFs), \ie
integrals of GPDs weighted by a specific kernel that is integrated
order by order in perturbation theory.

Nevertheless, pioneering studies
\cite{Airapetian:2001yk,Adloff:2001cn,Stepanyan:2001sm}
demonstrated the feasibility of DVCS measurements. They were
followed by numerous dedicated experiments
\cite{Chekanov:2003ya,Aktas:2005ty,Chen:2006na,Airapetian:2006zr,MunozCamacho:2006hx,Mazouz:2007aa,Aaron:2007ab,Girod:2007aa,Airapetian:2008aa,Chekanov:2008vy,Gavalian:2008aa,Aaron:2009ac,Airapetian:2009aa,Airapetian:2010aa,Airapetian:2010ab,Airapetian:2011uq,Airapetian:2012mq,Pisano:2015iqa,Jo:2015ema,Defurne:2015kxq} during a period of intense theoretical activity
which put DVCS under solid control. In particular let us
mention the full description of DVCS up to twist-3
\cite{Belitsky:2001ns,Belitsky:2008bz,Belitsky:2010jw,Belitsky:2012ch}, including the discussion of QED gauge invariance
\cite{Anikin:2000em,Radyushkin:2000ap,Kivel:2000fg,Belitsky:2000vx}
and target mass and finite momentum transfer corrections \cite{Braun:2012bg,Braun:2012hq}, the computation of higher orders in the perturbative expansion in the strong running coupling \cite{Ji:1997nk,Belitsky:1997rh,Mankiewicz:1997bk,Ji:1998xh,Belitsky:1999sg,Freund:2001hm,Freund:2001rk,Freund:2001hd,Pire:2011st,Moutarde:2013qs},
and the soft-collinear resummation of DVCS \cite{Altinoluk:2012fb,Altinoluk:2012nt}. Two closely related processes, Timelike Compton Scattering (TCS) \cite{Berger:2001xd,Boer:2015hma} and Double Deeply Virtual Compton Scattering (DDVCS) \cite{Guidal:2002kt,Belitsky:2002tf,Belitsky:2003fj} have also been discussed, and receive now considerable attention from the experimental community.

Fits to DVCS data have been successfully performed
since 2008 \cite{Kumericki:2007sa,Guidal:2008ie,Kumericki:2009uq,Moutarde:2009fg,Guidal:2009aa,Guidal:2010ig,Guidal:2010de,Goldstein:2010gu,Kumericki:2011zc,Kumericki:2011rz,GonzalezHernandez:2012jv,Kumericki:2013br,Boer:2014kya}, providing first quantitative experimental
information on CFFs. Although these fits do not give a final
word on the GPD studies, they nevertheless show that the
field is in a good shape from both theoretical and
experimental perspectives. The new era about to start will
yield data of unprecedented accuracy and with a wide kinematic
coverage. The valence region is being explored again in
Jefferson Lab (JLab) with the beginning of the experiments at
12~\GeV. The COMPASS Collaboration at CERN will soon start
DVCS data-taking. GPDs and their golden channel DVCS, are at
the heart of the physics case of a planned future 
Electron-Ion Collider (EIC).

The continuous progress in the field of GPDs
has been documented in several review articles
\cite{Ji:1998pc,Goeke:2001tz,Diehl:2003ny,Belitsky:2005qn,Boffi:2007yc,Guidal:2013rya,Muller:2014sha}. 
The present text aims at preparing the ground for fits of
forthcoming experimental data. How can GPD fitters work best
with high-precision data? The community of PDF fitters is
older and larger than its GPD analogue, and it has achieved an
impressive level of accuracy and sophistication. GPD
phenomenology is much harder, owing to the fact, in particular, that GPDs
depend on more variables and are subject to many constraints.
Considering PDF fits as an inspiring
guideline, it is nevertheless possible to see which steps
should be made to achieve a similar level of rigour over a
shorter period of time. 

Our review consists of three parts. In the remainder of 
\refsec{sec:introduction} we list 
the various constraints on GPDs, \eg, coming from discrete symmetries 
or Lorentz invariance. 
We also review several representations that fulfil such constraints, and present
selection of GPD models within each framework. The intricacy of GPD modelling is one of the
distinctive features of the field. 
In \refsec{sec:status}, we present both the theoretical and experimental
state of the art of DVCS, including various fitting strategies and
lessons obtained from fits. In the last part, \refsec{sec:preparing-future}, we 
give an outlook on future directions in the field of GPD fitting, and 
we suggest a few avenues towards improving present fitting procedures. In
particular, we stress the need for establishing code
benchmarking criteria for the various parametrizations. These would include introducing a set of uniform conventions for observable definitions, notations and data descriptions, and a thorough analysis  of
both the experimental and theoretical uncertainties.

Finally, this review is mostly dedicated to discussing GPDs and their extraction from DVCS on a nucleon
target. For other related processes, we refer the reader to the reviews in \refcite{Favart:2015epja} (DVMP),
\refcite{Dupre:2015epja} (nuclear DVCS), and to 
\refcite{Amrath:2008vx} (DVCS from a pion).


\subsection{Notations}
\label{sec:notations}

For any four-vector $a$ we define the \lc coordinates by:
\begin{equation}
\label{eq-def-lc-coordinates}
a^{\pm} = \frac{1}{\sqrt{2}} ( a^0 \pm a^3 ) \quad \textrm{ and } \quad a = ( a^+, \mathbf{a}, a^-).
\end{equation}
$(a b) = a^+ b^- + a^- b^+ - \mathbf{a} \cdot \mathbf{b}$ denotes the scalar product of two four-vectors $a$ and $b$. Indices between parentheses will mean symmetrization (and average) over indices, \eg $a^{(\mu}b^{\nu)} = (a^\mu b^\nu + a^\nu b^\mu)/2$.

We will consider hadron matrix elements of the form $\langle
P_2|O|P_1 \rangle$ for different operators $O$ sandwiched
between incoming (1) and outgoing (2) states. The total
momentum $P$ and momentum transfer $\Delta$ are:
\begin{eqnarray}
P & = & P_1 + P_2, \label{eq:def-total-momentum} \\
\Delta & = & P_2 - P_1. \label{eq:def-momentum-transfer}
\end{eqnarray}
In terms of Mandelstam variables: $t = \Delta^2$.

We will denote by $\eta$ the GPD variable known as
\emph{skewness}, see \refeq{eq:defkin} below, 
and keep the symbol $\xi$ for the kinematic
variable approximately equal to $\xb/(2-\xb)$, where $\xb$ is the usual
Bjorken scaling variable. M will stand for the proton mass, and
$e_q$ for the particle $q$ fractional electric charge in units of the
positron charge $|e|$. Furthermore, $\theta$ is the Heaviside
step function, $\gamma_\mu$ a Dirac matrix, $\sigma_{\mu\nu} =
i[\gamma_\mu, \gamma_\nu]/2$, and $g^{\mu\nu}$ is the
metric tensor. More specifically, $Q$ (resp. $Q'$) will
denote the photon virtuality in the DVCS channel (resp. TCS
channel).

We will follow the convention of Diehl \cite{Diehl:2003ny} to define GPDs in impact parameter space. The transverse plane Fourier transform $f(\mathbf{b})$ of a function $f(t)$ thus writes
\begin{equation}
\label{eq:def-transverse-plane-fourier-transform}
f(\mathbf{b}) = \int \frac{\mathrm{d}^2\mathbf{D}}{(2\pi)^2} \, e^{- i \mathbf{D} \mathbf{b}} f(t) \;,
\end{equation}
with
\begin{equation}
t = t_0 - (1-\eta^2) \mathbf{D}^2 \;,
\end{equation}
and
\begin{equation}
\label{eq:def-t-tmin}
t_0 = - \frac{4 \eta^2 M^2}{1-\eta^2} \;,
\end{equation}
being the maximal $t$ for given $\eta$.

To simplify equations, we often drop the explicit dependence on unused variables when no confusion is possible. We will simply mention LO, NLO, \ldots for "Leading Order", "Next-to-Leading Order", \ldots when referring to perturbative expansions in the strong running coupling constant.


\subsection{GPD definition and properties}
\label{sec:gpd-definition-properties}


\subsubsection{Definition}
\label{sec:gpd-definition}

\noindent GPDs are defined in the unpolarized (vector) sector as
\begin{multline}
 F^q (x, \eta, t) =
   \int \frac{d z^-}{2 \pi}\: e^{i x P^+ z^-} \\
 \times \langle P_2|\bar{q}(-z)\gamma^+ q(z)|P_1\rangle
  \Big|_{z^+=0,\, \mathbf{z}=\mathbf{0}}  \;,
   \label{eq:defquarkGPD}
\end{multline}
\begin{multline}
 F^g (x, \eta, t) =
  \frac{4}{P^+} \int \frac{d z^-}{2 \pi}\: e^{i x P^+ z^-} \\
 \times \langle P_2|G^{+\mu}_{a}(-z) G_{a \mu}^{\;\;\:+} (z)|P_1\rangle
   \Big|_{z^+=0,\, \mathbf{z}=\mathbf{0}} \;,
\label{eq:defgluonGPD}
\end{multline}
and in the polarized (axial-vector) sector as
\begin{multline}
 \tilde{F}^q (x, \eta, t) =
 \int \frac{d z^-}{2 \pi}\: e^{i x P^+ z^-} \\
 \times \langle P_2|\bar{q}(-z)\gamma^+ \gamma_5 q(z)|P_1\rangle
   \Big|_{z^+=0,\, \mathbf{z}=\mathbf{0}}  \;,
\label{eq:defquarkGPDA}
\end{multline}
\begin{multline}
 \tilde{F}^g (x, \eta, t) = \frac{4}{P^+}
 \int \frac{d z^-}{2 \pi}\: e^{i x P^+ z^-} \\
  \times \langle P_2|G^{+\mu}_{a}(-z) i \epsilon^{\perp}_{\mu\nu}
  G_{a}^{\nu +} (z)|P_1\rangle
   \Big|_{z^+=0,\, \mathbf{z}=\mathbf{0}} \;,
\label{eq:defgluonGPDA}
\end{multline}
where the \emph{skewness} $\eta$ reads
\begin{equation}
\eta = - \frac{\Delta^+}{P^+}
\label{eq:defkin}
\end{equation}
and where we suppressed polarization dependence and Wilson lines in the bilocal
operators (which serve to restore gauge invariance).

Both $F^a$ and $\tilde{F}^a$  can be decomposed as
\begin{align}
  F^a& = \frac{h^+}{P^+} H^a + \frac{e^+}{P^+} E^a  \qquad a=q, g \;, \label{eq:FtoHE}\\
  \tilde{F}^a& = \frac{\tilde{h}^+}{P^+} \tilde{H}^a + \frac{\tilde{e}^+}{P^+} \tilde{E}^a  \qquad a=q, g \;,
\label{eq:FtoHtEt}
\end{align}
where the Dirac spinor bilinears are
\begin{align}
        h^\mu& = \bar{u}(P_2) \gamma^\mu u(P_1) \,;&   e^\mu& = \frac{
i\Delta_\nu}{2 M}\bar{u}(P_2)  \sigma^{\mu\nu} u(P_1) \,, \\
\tilde{h}^\mu& = \bar{u}(P_2) \gamma^\mu \gamma_5 u(P_1) \,;& \tilde{e}^\mu& =
\frac{\Delta^{\mu}}{2 M} \bar{u}(P_2) \gamma_5 u(P_1)\,,
\label{eq:bilinears}
\end{align}
and the spinors are normalized so that $\bar{u}(p)\gamma^\mu u(p) = 2 p^\mu$.
The GPDs above are defined following 
Refs.~\cite{Diehl:2003ny,Belitsky:2005qn}. See
\reftab{tab:equivalent-GPD-kinematic-conventions} with equivalent symbols.

Four additional GPDs can be defined at twist two in the helicity flip (tensor) sector,
\begin{multline}
 F_T^{i q} (x, \eta, t) =
 \int \frac{d z^-}{2 \pi}\: e^{i x P^+ z^-} \\
 \times \langle P_2|\bar{q}(-z) i \sigma^{+i} q(z)|P_1\rangle
   \Big|_{z^+=0,\, \mathbf{z}=\mathbf{0}} \\
  = \frac{1}{P^+} \bar{u}(P_2) \left[ H_T^q i\sigma^{+i} + E_T^q  \frac{\gamma^+ \Delta^i - \gamma^i \Delta^+}{2 M} +    \right.  \\
+ \left. \tilde{H}_T^q  \frac{P^+ \Delta^i - P^i \Delta^+}{2 M^2}  + \tilde{E}_T^q  \frac{\gamma^+ P^i - \gamma^i P^+}{2 M} \right] u(P_1)\,,
\label{eq:defquarkGPDT}
\end{multline}  
where $i=1,2$.
Notice that the operator defining these GPDs is chiral-odd, {\it i.e.}, it flips quark chirality, as opposed to the chiral-even operators in \refeqs{eq:defquarkGPD}{eq:defgluonGPDA}. The chiral-odd quark GPDs cannot be measured directly in DVCS. They are accessible  through exclusive pseudoscalar meson production \cite{Ahmad:2008hp,Goloskokov:2009ia,Goloskokov:2011rd}. 
Analogously in the gluon sector
\begin{multline}
 F_T^{i j g} (x, \eta, t) =
 \frac{4}{P^+} \int \frac{d z^-}{2 \pi}\: e^{i x P^+ z^-} \\
 \times \langle P_2|\mathcal{S}G^{+i}_a(-z) G^{j+}_a(z)|P_1\rangle
   \Big|_{z^+=0,\, \mathbf{z}=\mathbf{0}} \\
  = \mathcal{S}\frac{1}{P^+} \frac{P^+ \Delta^j - \Delta^+ P^j}{2 M P^+} \bar{u}(P_2) 
  \bigg[ H_T^g i\sigma^{+i} + \\
      + E_T^g  \frac{\gamma^+ \Delta^i - \gamma^i \Delta^+}{2 M} + 
 \tilde{H}_T^g  \frac{P^+ \Delta^i - P^i \Delta^+}{2 M^2}  +  \\
 + \tilde{E}_T^g  \frac{\gamma^+ P^i - \gamma^i P^+}{2 M} \bigg] u(P_1)\,,
\label{eq:defgluonGPDT}
\end{multline}
where $i,j = 1, 2$ and the symbol $\mathcal{S}$ indicates symmetrization and trace subtraction of uncontracted indices.

For a complete classification of GPDs and of their parton correlation function substructure up to twist four see \refcites{Meissner:2008ay,Meissner:2009ww,Lorce:2013pza}.
\begin{table}
\begin{center}
\begin{tabular}{ccc}
this work & \mbox{\refcite{Belitsky:2005qn}} & \mbox{\refcite{Diehl:2003ny}} \\ \hline
$P$ & $p$  & $2 P$ \\
$\Delta$ & $-\Delta$ & $\Delta$ \\
$\eta$ & $\eta$ & $\xi$ \\ \hline
\end{tabular}
\caption{Dictionary of momentum conventions between the present text and the two top-cited reviews on GPDs.}
\label{tab:equivalent-GPD-kinematic-conventions}
\end{center}
\end{table}


\subsubsection{Forward limit}
\label{sec:forward-limit}

In the forward kinematic limit, $P_1 = P_2$, some GPDs
reduce to standard PDFs,
\begin{align}
F^{q}(x, 0, 0) = &H^{q}(x, 0, 0) = \theta(x) q(x) - \theta(-x) \bar{q}(-x)\;,
\label{eq:FqFWD} \\
F^{g}(x, 0, 0) = &H^{g}(x, 0, 0) = \theta(x) x g(x) - \theta(-x) x g(-x)\;, 
\label{eq:FgFWD} \\
\tilde{F}^{q}(x, 0, 0) = &\tilde{H}^{q}(x, 0, 0) = \theta(x) \Delta q(x) + \theta(-x) \Delta \bar{q}(-x)\;,
\label{eq:FqtildeFWD} \\
\tilde{F}^{g}(x, 0, 0) = &H^{g}(x, 0, 0) = \theta(x) x \Delta g(x) + \theta(-x) x \Delta g(-x)\;,  
\label{eq:FgtildeFWD} \\
& \tilde{H}_T^{q}(x, 0, 0) = \theta(x) \Delta_T q(x) - \theta(-x) \Delta_T \bar{q}(-x)\;.
\label{eq:FqtransvFWD}
\end{align}


\subsubsection{Discrete symmetries}
\label{sec:discrete-symmetries-properties}

Time reversal and hermiticity imply that GPDs are real and that
\begin{equation}
F (x, \eta, t) = F (x, -\eta, t) \;,
\label{eq:GPDparityeta}
\end{equation}
for all $F= F^q, F^g, \tilde{F}^q, \tilde{F}^g$. From now on, and unless explicitly specified, we will assume $\eta \geq 0$. 

\noindent The fact that the gluon is its own antiparticle implies that
\begin{align}
F^g (x, \eta, t)& = F^g (-x, \eta, t) \;, \label{eq:GPDparityx} \\
\tilde{F}^g (x, \eta, t)& = - \tilde{F}^g (-x, \eta, t)  \;.
\label{eq:tildeGPDparityx}
\end{align}
For quarks it is useful to consider the combinations,
\begin{align}
H^{q(\pm)}(x, \eta, t)& \equiv H^{q}(x, \eta, t) \mp H^{q} (-x, \eta, t) \;,\\
\tilde{H}^{q(\pm)}(x, \eta, t)& \equiv \tilde{H}^{q}(x, \eta, t) \pm
\tilde{H}^{q} (-x, \eta, t) \;,
\label{eq:singletGPD}
\end{align}
with similar relations involving $E^q$ and $\tilde{E}^q$. $H^{q(\pm)}(x)$ in forward limit reduces to $q(x)\pm \bar{q}(x)$ so $H^{q(+)}$ is called singlet
(although it has to be summed over flavors to really become singlet), and
$H^{q(-)}$ is called non-singlet or valence combination.

If one considers $C$-parity  exchanged in the $t$-channel corresponding to each GPD, then
$F^{q(+)}$, $\tilde{F}^{q(+)}$ and both gluon GPDs are $C$-even, while
$F^{q(-)}$ and $\tilde{F}^{q(-)}$ are $C$-odd.
In DVCS there is no change of $C$ going
from initial to final state, so only $C$-even GPDs contribute.


\subsubsection{Sum rules}
\label{sec:sum-rules}

\emph{Sum rules} are quite important in the GPD phenomenology.
The integrals of GPDs over $x$ are related to the quark contributions $F_1^q$ and $F_2^q$ to the elastic form factors $F_1$ and $F_2$ in the Pauli-Dirac representation,
\begin{align}
\int_{-1}^{1}dx\: H^q (x, \eta, t)& = F_{1}^{q}(t) \;, \label{eq:sumF1} \\
\int_{-1}^{1}dx\: E^q (x, \eta, t)& = F_{2}^{q}(t) \;, \label{eq:sumF2}
\end{align}
with similar relations relating $\tilde{H}$ and $\tilde{E}$
to the axial and pseudoscalar form factors $G_A$ and $G_P$.
Sum rules can be seen as a particular case of the
polynomiality property discussed in
\refsec{sec:polynomiality-property}, and the connection
of GPDs to both PDFs and FFs provides a particularly interesting physical interpretation.

Ji's sum rule \cite{Ji:1996ek} is another landmark GPD property. The Belinfante \EM tensor $T^{\mu\nu}$ \cite{Belinfante:1939em,Rosenfeld:1940em} between nucleon states can be parametrized as,
\begin{multline}
\label{eq:belinfante-em-tensor-gravitational-form-factors}
\langle P_2|T^{\mu\nu}|P_1\rangle = \bar{u}(P_2) \left[ \frac{1}{2} A(t) \gamma^{(\mu} P^{\nu)} + B(t) P^{(\mu} i \sigma^{\nu)\lambda} \frac{\Delta_\lambda}{4M} \right. \\ \left. + \frac{C(t)}{M} (\Delta^\mu \Delta^\nu - \Delta^2 g^{\mu\nu}) 
\right] u(P_1) \;,
\end{multline}
where $A$, $B$ and $C$
are called \emph{gravitational form factors}, defined for both the quark and gluon sectors. The
derivation of Ji's sum rule starts from the decomposition of
the nucleon spin into its quark and gluon contributions
\begin{equation}
\label{eq:original-ji-sum-rule-first-step}
\frac{1}{2} = \sum_q J^q + J^g \;,
\end{equation}
with both terms related to the \EM tensor
\begin{equation}
J^{q,g} = \frac{1}{2} [ A^{q,g}(0) + B^{q,g}(0) ] \label{eq:def-Jqg} \;.
\end{equation}
One can then connect the gravitational form factors with the coefficients of the correlation function defined using eqs.(\ref{eq:defquarkGPD},\ref{eq:FtoHE},\ref{eq:bilinears})
\begin{eqnarray}
&&   \int \frac{d z^-}{2 \pi}\: e^{i x P^+ z^-} 
 \langle P_2|\bar{q}(-z)\gamma^+ q(z)|P_1\rangle = \nonumber \\
&& H^q \, \bar{u}(P_2) \gamma^\mu u(P_1) + E^q \, \frac{i\Delta_\nu}{2 M}\bar{u}(P_2)  \sigma^{\mu\nu} u(P_1) 
   \label{eq:corrquarkGPD}
\end{eqnarray}
(an analogous decomposition can be made in the gluon sector).
The second Mellin moments of the GPDs $H$ and $E$ from this definition are,
\begin{eqnarray}
\int \mathrm{d}x \, x H^{q,g}(x, \eta, t) & = & A^{q,g}(t) + 4 \eta^2 C^{q,g}(t) \;, \label{eq:mellin-moment-gpd-H} \\
\int \mathrm{d}x \, x E^{q,g}(x, \eta, t) & = & B^{q,g}(t) - 4 \eta^2 C^{q,g}(t) \;, \label{eq:mellin-moment-gpd-E}
\end{eqnarray}
so that
\begin{equation}
\label{eq:original-ji-sum-rule-second-step}
2 J^q = \int_{-1}^{+1} \mathrm{d}x \, x [ H^{q} + E^{q} ](x, \eta, 0) \;.
\end{equation}
A closer look reveals that the contribution related to $H^q$ is already known from PDFs. In other words
\begin{equation}
\label{eq:original-ji-sum-rule-third-step}
2 J^q = \int_0^1 \mathrm{d}x \, x [ q(x) + \bar{q}(x) ] + \int_{-1}^{+1} \mathrm{d}x \, x E^{q}(x, \eta, 0) \;.
\end{equation}
The first term on the \rhs is the quark total momentum which can be obtained from standard measurements of PDFs in Deeply Inelastic Scattering (DIS). The second term is the new component in the sum rule: since the quark spin contribution is already known, the second term relates to both the quark orbital motion and to the nucleon's magnetic properties.
Measuring the GPD $E^q$ became one of the main motivations of the experimental GPD program, including the physics case for an EIC \cite{Kumericki:2011zc,Accardi:2012qut,Aschenauer:2013hhw}. 

More recent developments have been addressing the question of a gauge invariant decomposition of total angular momentum into its spin and orbital components,
\begin{equation}
\frac{1}{2} = L_q + S_q + L_g + S_g \;.
\label{eq:decomposition}
\end{equation}
A decomposition of the quark angular momentum in Ji's sum rule, namely,
\begin{equation}
J_q = L_q + S_q \;,
\label{eq:quark_decomposition}
\end{equation}
can be performed, where the \lhs is described by the GPDs $H^q$ and $E^q$, eq. (\ref{eq:original-ji-sum-rule-second-step}), while on the \rhs, $L_q$ is described by a specific twist-three GPD \cite{Penttinen:2000dg,Kiptily:2002nx,Hatta:2012cs,Courtoy:2013oaa}, and $2 S_q=  \Delta \Sigma_q$ is the total quark helicity. An analogous decomposition in the gluon sector is not possible in this case.
We refer to the recent reviews \cite{Leader:2013jra,Liu:2015xha} for further details on these developments.

Finally, the angular momentum sum rule was extended to a spin-1 system, \eg  the deuteron in \refcite{Taneja:2011sy}. The sum rule reads
\begin{equation}
\label{eq:original-ji-sum-rule-deuteron}
2 J^q = \int_{-1}^{+1} \mathrm{d}x \, x  H_2^{q} (x, \eta, 0) \;,
\end{equation}
where $H_2^q$ is one of the five deuteron GPDs in the vector sector \cite{Berger:2001zb}. It is interesting to notice that, analogously to the nucleon case, the angular momentum is determined by the same GPD, $H_2$ whose first Mellin moment is the magnetic form factor of the spin-1 system.


\subsubsection{Polynomiality}
\label{sec:polynomiality-property}

Related to sum-rules is the important \emph{polynomiality} property
of GPDs. Namely, using identities
\begin{multline}
\int_{-1}^{1}dx\: x^{j} F^q (x, \eta, t) = \\
\frac{1}{(P^+)^{j+1}}\langle P_2|\bar{q}
\gamma^+ (i \stackrel{\leftrightarrow}{\partial^+})^{j} q|P_1\rangle \;,
\label{eq:momquark}
\end{multline}
\begin{multline}
\frac{1}{2}\int_{-1}^{1}dx\: x^{j-1} F^g (x, \eta, t) = \\
\frac{2}{(P^+)^{j+1}}\langle P_2|
G^{+\mu}_{a} (i \stackrel{\leftrightarrow}{\partial^+})^{j-1}
G_{a \mu}^{\;\;\:+} |P_1\rangle \;,
\label{eq:momgluon}
\end{multline}
and the behavior of GPDs under discrete symmetries (see \refsec{sec:discrete-symmetries-properties}), one can show that $x^{j}$ moments of \emph{quark} GPDs are even polynomials in $\eta$ with leading powers given in \reftab{tab:quarkpolynomiality} and that $x^{j-1}$ moments of \emph{gluon} GPDs are polynomials in $\eta$ with leading powers given in \reftab{tab:gluonpolynomiality}.

\begin{table}
\begin{center}
\begin{tabular}{ccc}
GPD & even $j$ & odd $j$ \\ \hline
$H^q$, $E^q$ &  $\eta^j$ & $\eta^{j+1}$ \\
$H^q + E^q$ &  $\eta^{j}$ & $\eta^{j-1}$ \\
$\tilde{H}^q$, $\tilde{E}^q$ & $\eta^{j}$ & $\eta^{j-1}$
\end{tabular}
\caption{\label{tab:quarkpolynomiality} Leading powers of the $x^j$ Mellin moments of the twist-2 chiral-even GPDs in the quark sector.}
\end{center}
\end{table}

\begin{table}
\begin{center}
\begin{tabular}{ccc}
GPD & even $j$ & odd $j$ \\ \hline
$H^g$, $E^g$ & 0 & $\eta^{j+1}$   \\
$H^g + E^g$ & 0 & $\eta^{j-1}$  \\
$\tilde{H}^g$, $\tilde{E}^g$ &  $\eta^{j}$  & 0
\end{tabular}
\caption{\label{tab:gluonpolynomiality} Leading powers of the $x^{j-1}$ Mellin moments of the twist-2 chiral-even GPDs in the gluon sector.}
\end{center}

\end{table}

Zeros in \reftab{tab:gluonpolynomiality} are due to the $x \to - x$ (anti)symmetry
of gluon GPDs, see eqs. (\ref{eq:GPDparityx}) and (\ref{eq:tildeGPDparityx}).
Note that the zeroth ($x^{j=0}$) moment of quark GPD
leads to $\eta$-independence explicated by the
sum rules (\ref{eq:sumF1}) and (\ref{eq:sumF2}).
Also note that the first moment of $H+E$ combination is also $\eta$-independent, as explicated in Ji's sum rule (\ref{eq:original-ji-sum-rule-second-step}).


\subsubsection{Positivity}
\label{sec:positivity-property}

Positivity bounds emerge from the definition of the norm on a Hilbert space, and thus are fundamental properties of GPDs. For the sake of simplicity, we will discuss the case of spinless hadrons. In essence, positivity bounds are inequalities between GPDs and the corresponding PDFs at well-defined kinematic configurations \cite{Pire:1998nw,Radyushkin:1998es,Pobylitsa:2002iu}, \eg 
\begin{equation}
\label{eq:example-positivity-pobylitsa-first-paper}
|H^q(x, \eta, t)| \leq \sqrt{q(x^{\textrm{in}})q(x^{\textrm{out}})} \;,
\end{equation}
where the naming convention of $x^{\textrm{in}}$ and $x^{\textrm{out}}$
\begin{eqnarray}
x^{\textrm{in}} & = & \frac{x+\eta}{1+\eta} \;, \label{eq:def-xin} \\
x^{\textrm{out}} & = & \frac{x-\eta}{1-\eta} \;, \label{eq:def-xout} 
\end{eqnarray}
will become evident in \refsec{sec:overlap-representation}. This is a strong model-independent constraint. Consider for example a pion PDF computed in the Bethe-Salpeter approach with a proper implementation of the symmetry $x \leftrightarrow 1 - x$ typical of two-body problems \cite{Chang:2014lva}.  From perturbative QCD, we know that the PDF vanishes like $(1 - x)^2$ when $x$ is close to 1. From the exchange symmetry $x \leftrightarrow 1 - x$, we observe that the PDF should vanish at the same pace when $x$ is close to 0. In particular, we conclude from \refeq{eq:example-positivity-pobylitsa-first-paper} that a GPD computed consistently in that framework should vanish on the \crossover line $x = \eta$. This is completely consistent with the result of \refcite{Ji:2006cr} where all possible $q\bar{q}g$ states were consistently introduced to obtain a pion GPD model with a nonzero value on the \crossover line.

The derivation of an inequality such as \refeq{eq:example-positivity-pobylitsa-first-paper} proceeds from the Cauchy-Schwarz inequality. The matrix element defining a GPD can be identified as an inner product of two different states. Its absolute value is smaller than the product of the norms of these two states, and each of these two terms is recognized as the matrix element defining a PDF. This is  basically the underlying reasoning of \refcites{Pobylitsa:2001nt,Pobylitsa:2002gw} and refs. therein. From this derivation, the positivity bounds are restricted to the DGLAP regions $|\eta| \leq |x| \leq 1$.

This argument can however be made more general: as a guideline, we may remember the proof of Cauchy-Schwarz inequality. Consider a real inner product $(.|.)$, two vectors $a$, $b$ in a real Hilbert space, and a real $\lambda$. From the positivity of $\|a + \lambda b\|^2 = \|a\| ^2+ 2 \lambda (a|b) + \lambda^2 \|b\|^2$ for all $\lambda$, we derive $|(a|b)| \leq \|a\| \|b\|$. The positivity of the norm of the Hilbert space of quark-hadron states is at the heart of the argument. In \refcite{Pobylitsa:2002iu} Pobylitsa derived inequalities from the positivity of the norm of arbitrary superpositions of states $\sum \int \frac{\mathrm{d}P^+ \mathrm{d}^2\mathbf{P} \mathrm{d}\lambda}{2P^+} \, g_\sigma(\lambda, P) q(\lambda n) |H(P)\rangle$, where the sum runs over various hadron states $H$ of momentum $P$ and spin $\sigma$, with the (good component of the) quark field taken at point $\lambda n$ and weighted by arbitrary functions $g_\sigma$. This procedure yields infinitely many inequalities, all translating in various forms the positive definiteness of the norm. From the model-building point of view, this fact makes positivity bounds an even more severe constraint. All of these inequalities admit the following generic form in the impact parameter representation \cite{Pobylitsa:2002iu}
\begin{multline}
\label{eq:pobylitsa-positivity-generic-form}
\int_{-1}^{+1} \mathrm{d}\eta \int_{|\eta|}^{+1} \frac{\mathrm{d}x}{1-x} p^*\left(x^{\textrm{out}}, \frac{\mathbf{b}}{1-x}\right) p\left(x^{\textrm{in}}, \frac{\mathbf{b}}{1-x}\right) \\ F(x, \eta, \mathbf{b}) \geq 0\;,
\end{multline}
for arbitrary function $p$, and where $F(x, \eta, \mathbf{b})$ represents the matrix element defining GPDs in impact parameter space. At last, representing explicitly a GPD as an inner product \cite{Pobylitsa:2002vi}
\begin{multline}
\label{eq:positivity-constraint-gpd-form}
(1-\eta^2) H(x, \eta, \mathbf{b}) = \\
\sum_k Q_k^*\left(x^{\textrm{out}}, \frac{\mathbf{b}}{1-x}\right) Q_k\left(x^{\textrm{in}}, \frac{\mathbf{b}}{1-x}\right) \;,
\end{multline}
where the sum can range over a discrete or continuous collection of functions, guarantees the fulfillment of \refeq{eq:pobylitsa-positivity-generic-form} (after the change of variables which maps $(\eta, x)$ such that $1 \geq x \geq |\eta|$ to $(x^{\textrm{in}}, x^{\textrm{out}})$ in $[-1, +1]^2)$.

For completeness, we mention that the stability of positivity bounds under LO evolution is established in \refcite{Pobylitsa:2002iu}. The sign of the norm of (unphysical) quark-hadron states is questioned in \refcite{Pobylitsa:2002ru} where an alternative proof of the positivity bounds is given.


\subsection{GPD parametrizations}
\label{sec:gpd-parameterization}

In the years following  the introduction of GPDs and the definition of their fundamental physical properties, several frameworks have been defined that provide parametric forms to be used for their extraction from experimental data. This theoretical progress has been proceeding simultaneously to the development of various experimental programs to measure GPDs (see \refsec{sec:experimental-data}). In what follows we give a list of the frameworks or representations that have been used for data interpretation, followed by a description of specific models within each framework. 


\subsubsection{Overlap}
\label{sec:overlap-representation}

This representation bears its name from the description of a GPD as an overlap of light front wave functions. This representation was derived by Diehl \etal \ \cite{Diehl:2001xx}. Here again we will only discuss the quark sector, the gluon sector following \textit{mutatis mutandis}. We will mostly use the notations of \refcite{Diehl:2001xx}, but will restrict ourselves to spinless hadrons for brevity.

A Fock state made of $N$ partons is generically denoted $|N, \beta; k_1, \ldots, k_N\rangle$ where $\beta$ encode the information about the partons: their type, their helicity and their color. A hadron state $H$ with momentum $P$ is made of an arbitrary number of partons, weighted by corresponding light front wave functions $\psi_{N,\beta}$
\begin{equation}
\label{eq:def-light-front-wave-function}
|H; P\rangle = \sum_{N, \beta} \int [\mathrm{d}x]_N [\mathrm{d}^2\mathbf{k}]_N \psi_{N,\beta} |N, \beta; k_1, \ldots, k_N\rangle \;,
\end{equation}
where the symbols $[\mathrm{d}x]_N$ and $[\mathrm{d}^2\mathbf{k}]_N$ are compact notations for
\begin{eqnarray}
\ [\mathrm{d}x]_N & = & \prod_{i=1}^N \mathrm{d}x_i \delta\left(1 - \sum_{i=1}^N x_i\right) \;, \label{eq:not-integration-measure-momentum-fraction} \\
\ [\mathrm{d}^2\mathbf{k}]_N & = & \frac{1}{(16\pi^3)^{N-1}} \prod_{i=1}^N \mathrm{d}^2\mathbf{k}_i \delta\left(\mathbf{P} - \sum_{i=1}^N \mathbf{k}_i\right) \;. \label{eq:not-integration-measure-transverse-momentum}
\end{eqnarray}
The light front wave function normalization is derived from the hadron state covariant normalization, \ie including contributions from all parton states
\begin{equation}
\label{eq:def-wave-function-normalization}
\sum_{N, \beta} \int [\mathrm{d}x]_N [\mathrm{d}^2\mathbf{k}]_N |\psi_{N,\beta}|^2 = 1 \;.
\end{equation}
The next step consists in expanding the good component of the quark field in terms of operators creating Fock states with given plus and transverse momenta, helicity and color. The active parton $j$ is emitted from the hadron, and later absorbed by it, while the other partons $i \neq j$ are spectators. The wave functions depend on momentum components relative to the considered hadron momentum. This kinematic matching is made in frames where the incoming or outgoing hadron have zero transverse momentum, hence the terminology "in" and "out" for kinematic variables relevant to the DGLAP region $\eta \leq x \leq 1$
\begin{eqnarray}
(x^{\textrm{in}}_i, \mathbf{k}^{\textrm{in}}_i) 
& = & 
\left(\frac{x_i}{1+\eta}, \mathbf{k}_i - \frac{x_i}{1+\eta} \mathbf{P}_1\right) \;, \label{eq:in-kinematic-spectator} \\
(x^{\textrm{in}}_j, \mathbf{k}^{\textrm{in}}_j) 
& = & 
\left(\frac{x_j+\eta}{1+\eta}, \mathbf{k}_j + \frac{1-x_j}{1+\eta} \mathbf{P}_1\right) \;. \label{eq:in-kinematic-active} 
\end{eqnarray}
The "out" variables are simply obtained by changing $\eta$ to $-\eta$ and $\mathbf{P}_1$ to $\mathbf{P}_2$.

In this region the overlap representation of the GPD $H$ writes
\begin{multline}
\label{eq:gpd-overlap-positive-dglap}
H^q(x, \eta, t) = \sum_{N, \beta} \sqrt{1-\eta^2}^{2-N} \sum_{j=q} \int [\mathrm{d}x]_N [\mathrm{d}^2\mathbf{k}]_N \\ \delta(x-x_j) \psi_{N, \beta}^*(x^{\textrm{out}}_i, \mathbf{k}^{\textrm{out}}_i) \psi_{N, \beta}(x^{\textrm{in}}_i, \mathbf{k}^{\textrm{in}}_i) \;.
\end{multline}
The overlap representation has a similar structure in the other DGLAP region $-1 \leq x \leq -\eta$. Considering \refeq{eq:positivity-constraint-gpd-form}, this is enough to ensure that every model built from the overlap representation will fulfil positivity bounds. Furthermore, the overlap representation can even be used as a first principle statement to establish a general form for the positivity bounds, \eg as in \refcite{Diehl:2003ny}. 

However, the result in the ERBL region involves the overlap of wave functions with $N-1$ and $N+1$ constituents. The polynomiality property relates in this case wave functions with different partonic contents. This poses a constraint on the building of GPD models from the overlap representation. Recent progress in this direction will be discussed in \refsec{sec:double-distribution-representation}.

\subsubsection{Covariant Scattering Matrix Approach}
\label{sec:covariant-scattering-matrix}

The backdrop for this approach is a gauge invariant extension at leading twist of the covariant parton model \refcite{Landshoff:1970ff,Brodsky:1973hm}. The structure functions/PDFs for DIS processes as well as the CFFs/GPDs in DVCS result directly from the analytic behavior of the quark/gluon-proton scattering amplitude. 
The parton-proton amplitude defined in this framework is a holomorphic function of the parton's four-momentum component, $k^-$. 

An important aspect of the covariant scattering matrix approach is in its covariant regularization which can be carried out in different ways depending on the specific model. An even more interesting feature is that it provides a natural framework where Regge behavior of the structure functions can be naturally connected to their Bjorken scaling property \cite{Brodsky:1973hm}. The models we consider here correspond to the lowest order in perturbation theory. 
At this order the scattering amplitude includes two vertices with a proton, a parton undergoing the hard scattering, and a spectator system. The latter corresponds to a scalar or an axial vector spectator/recoiling system, namely a diquark for the valence quark distribution, or a tetraquark or higher diquark excited states for the sea quarks. It is instead an octet three quarks system, for the gluon distribution. 

The propagator structure of the covariant amplitude is given by, 
\begin{eqnarray}
i T(s,t,u, k^2, k'^2) & = &   \frac{i\Gamma(k)}{(k^2-m_q^2) + i \epsilon} \, \frac{i\Gamma(k')}{(p-k)^2-M_X^2 + i \epsilon} \nonumber \\ & \times & \frac{i}{(k'^2-m_q^2)+ i \epsilon},
\label{eq:covariantamplitude}
\end{eqnarray}
where $s,t,u$ are the Mandelstam invariants, $k^2$ and $k'^2$ are the initial and final quark four-momentum squared, $m_q$ is quark mass, $(p-k)^2$ is the spectator system four-momentum squared, $M_X$ is its mass, and $\Gamma(k)$ is a vertex function. As we will see in the model section $\Gamma(k)$ can be taken as a pointlike coupling, in which case a regularization \textit{\`{a} la} Pauli Villars applies, or as falling of with $k^2$, thus providing a covariant ultraviolet cutoff. 

The analytic behavior of $T(s,t,u, k^2, k'^2)$ is such that the spectator is placed on-shell while the struck parton is off-shell in the DGLAP region ($\eta \leq x \leq 1$ and $-1 \leq x \leq -\eta$). Vice versa in the ERBL region it is the struck parton to be placed on-shell.   

Taking into account the spin structure of the particles involved results in a more complicated structure in the numerator of eq. (\ref{eq:covariantamplitude}), without, however, changing its analytic behavior. In this case, the quark/gluon-proton scattering amplitude depends directly on the initial (final) parton helicity, $\lambda(\lambda')$ and the initial (final) proton helicity, $\Lambda (\Lambda')$, namely,
\[ T_{\Lambda'\lambda', \Lambda \lambda}(s,t,u, k^2, k'^2). \]

The condition of polynomiality in this approach is satisfied automatically, due to the covariance of the amplitude. In practical models which rely on approximations, this property has to nevertheless be tested. 

The advantage of the covariant scattering matrix approach is in that it allows one to describe Regge behavior of the GPDs at low $x$. This is accomplished by allowing for a spectral distribution for the spectator mass characterized by a peak at low mass values $\approx 1 $ GeV, and a behavior $\propto (M_X^2)^\alpha$, at $M_X>> 1$ GeV (where $\alpha$ is the Regge intercept parameter). As we show in \refsec{sec:spectator-models}, the $t$ dependence can also be described in this scenario.


\subsubsection{Double Distributions}
\label{sec:double-distribution-representation}

Double Distributions (DDs) were introduced first by M\"uller \etal \ under the name \emph{spectral functions} \cite{Mueller:1998fv} and later rediscovered by Radyushkin \cite{Radyushkin:1998es,Radyushkin:1998bz}. They offer the attractive feature of naturally solving the polynomiality constraint exposed in \refsec{sec:polynomiality-property}. We will explain below why it is so by considering the quark sector, but the extension to the gluon sector is straightforward.

The quark DDs $F^q$ and $G^q$ of a spinless hadron are defined by the following matrix element \cite{Polyakov:1999gs}:
\begin{multline}
\langle P_2|\bar{q}(-z) \slashed{z} q(z)|P_1\rangle \Big|_{z^2=0} = \\
(P z) \int_{\Omega} \mathrm{d}\beta\mathrm{d}\alpha \, e^{- i \beta (P z) + i \alpha (\Delta z)} F^q( \beta, \alpha, t ) \\
\, - (\Delta z)\int_{\Omega} \mathrm{d}\beta\mathrm{d}\alpha \, e^{- i \beta (P z) + i \alpha (\Delta z)} G^q( \beta, \alpha, t ).
\label{eq-def-DD-F-G-spinless-target}
\end{multline}
They are related to the GPD $H^q$ through:
\begin{equation}
\label{eq:relation-gpd-H-dds-spinless-hadron}
H^q(x, \eta) = \int_{\Omega} \mathrm{d}\beta\mathrm{d}\alpha \, \big( F(\beta, \alpha) + \eta G(\beta, \alpha) \big) \delta(x - \beta - \alpha \eta).
\end{equation}
The physical domain of GPDs $|x|, |\eta| \leq 1$ (with $x = \beta + \alpha \eta$) restricts the support of the DDs to the rhombus $\Omega = \{ (\beta, \alpha) \in \mathbb{R}^2, |\beta| + |\alpha| \leq 1 \}$. As emphasized by Teryaev \cite{Teryaev:2001qm} and Tiburzi \cite{Tiburzi:2004qr}, there are infinitely many parameterizations for DDs yielding the \emph{same} GPDs. Consider for example an arbitrary function $\sigma^q$ vanishing\footnote{We refer to \refcite{Tiburzi:2004qr} for a detailed discussion of boundary conditions on $\sigma^q$.} on the boundary of the rhombus $\Omega$. The transformation:
\begin{eqnarray}
F^q(\beta,\alpha) & \rightarrow & F^q(\beta,\alpha) + \frac{\partial \sigma^q}{\partial \alpha}(\beta,\alpha),  \label{eq-def-gauge-transform-F} \\
G^q(\beta, \alpha) & \rightarrow & G^q(\beta, \alpha) - \frac{\partial \sigma^q}{\partial \beta }(\beta,\alpha),  \label{eq-def-gauge-transform-G}
\end{eqnarray}
leaves the GPD $H^q$ in \refeq{eq:relation-gpd-H-dds-spinless-hadron} unchanged. In particular, there is one particular transformation \cite{Belitsky:2000vk} allowing the description of the two DDs $F^q$ and $G^q$ in terms of one single function $f^q$:
\begin{eqnarray}
F^q(\beta, \alpha) & = & \beta f^q(\beta, \alpha), \label{eq:def-F-1CDD} \\
G^q(\beta, \alpha) & = & \alpha f^q(\beta, \alpha). \label{eq:def-G-1CDD}
\end{eqnarray}
This choice is referred to as One-Component Double Distribution (1CDD) in \refcite{Belitsky:2005qn} and was recently used for model building and theoretical considerations \cite{Radyushkin:2011dh,Radyushkin:2012gba,Radyushkin:2013hca,Radyushkin:2013bba,Mezrag:2013mya}. The relation (\ref{eq:relation-gpd-H-dds-spinless-hadron}) between the GPD $H^q$ and the 1CDD $f^q$ now is:
\begin{equation}
\label{eq:relation-gpd-ocdd}
H^q(x, \eta) = x \int_{\Omega} \mathrm{d}\beta\mathrm{d}\alpha \, f^q(\beta, \alpha) \delta(x - \beta - \alpha \eta).
\end{equation}
Introducing the variables $s \in [-1, +1]$ and $\phi \in [0, 2\pi]$ such that $\eta = \tan \phi$ and $s = x \cos \phi$, we obtain the canonical form of the Radon transform:
\begin{equation}
\label{eq:relation-gpd-ocdd-radon}
H^q(x, \eta) = \frac{x}{\sqrt{1+\eta^2}} \int_{\Omega} \mathrm{d}\beta\mathrm{d}\alpha \, f^q(\beta, \alpha) \delta(s - \beta \cos \phi - \alpha \sin \phi).
\end{equation}
The inversion of this integral transform has been first discussed by Teryaev \cite{Teryaev:2001qm} and requires the prior knowledge of the GPD both inside and outside the physical region $|x|, |\eta| \leq 1$. It can be shown \cite{helgason:1999radonbook} that any smooth function satisfying a polynomiality condition is the Radon transform of another smooth function. In that sense DDs should not only be seen as a way to model GPDs consistently with respect to polynomiality. On the contrary, polynomiality exactly means that a GPD is the Radon transform of a DD: DDs naturally solve the polynomiality condition, and this statement is \emph{model-independent}.

On the contrary, the positivity constraint on GPDs is not manifest in the DD representation. Pobylitsa investigated the possibility to fulfil both positivity and polynomiality \cite{Pobylitsa:2002ru,Pobylitsa:2002vw}. In particular, Pobylitsa's solution in \refcite{Pobylitsa:2002ru} relies on a modified DD representation in the Polyakov-Weiss gauge, \ie \ the specific choice of DDs $F^q_{PW}$ and $G^q_{PW}$ for which $G^q_{PW}(\beta, \alpha) = \delta(\beta) D^q(\alpha)$, where $D^q$ is a function supported in $[-1, +1]$ called $D$-term. The relation \refeq{eq:relation-gpd-ocdd} between $H$ and the DDs is changed to:
\begin{multline}
\label{eq:relation-gpd-dd-pobylitsa}
H^q(x, \eta) = (1-x) \int_{\Omega} \mathrm{d}\beta\mathrm{d}\alpha \, F^q_P(\beta, \alpha) \delta(x - \beta - \alpha \eta) \\ + D^q\left(\frac{x}{\eta}\right) \;.
\end{multline}
A relation between the DDs $F^q_P(\beta, \alpha)$ and $\delta(\beta) D^q(\alpha)$ in \refeq{eq:relation-gpd-dd-pobylitsa} on the one hand, and the DDs $F^q(\beta, \alpha)$ and $G^q(\beta, \alpha)$ in \refeqs{eq:def-F-1CDD}{eq:def-G-1CDD} on the other hand, has been given (up to some assumptions on the behavior of the 1CDD at the boundary of the rhombus) in \refcite{Muller:2014sha}. To the best of our knowledge, this particular representation has never been used as a starting point for model-building.

An alternative line of research has been pursued in \refcites{Hwang:2007tb,Muller:2014tqa}. Its aim is the identification of DDs from the description of a GPD as an overlap of \lc wave function. If this program is successful, both polynomiality and positivity constraints are \textit{a priori} satisfied. In between a DD model has been constructed. This promising program allows so far the construction of GPDs and DDs starting from a model-dependent form of the \lc wave function, where the DD can be read off by inspection. However this form is too restrictive yet to be used with \eg the mathematically consistent 2-body \lc wave functions $\psi(x, \mathbf{k})$ encountered in Bethe-Salpeter modeling, which may possess the exchange symmetry $(x, \mathbf{k}) \leftrightarrow (1-x, -\mathbf{k})$ (for example in the case of a pion \lc wave function). This should nevertheless not undermine the merit of the approach, which opens a new path to flexible GPD modeling satisfying polynomiality and positivity.


\subsubsection{Conformal moments}
\label{sec:conformal-moments-representation}

Another representation of GPDs is in terms of conformal moments,
which are defined by convolution
of momentum-fraction GPDs with Gegenbauer polynomials
\begin{align}
F_{j}^q(\eta, t)& = \frac{1}{k_{j}^{3/2}} \int_{-1}^{1}dx\, \eta^j \,C_{j}^{3/2}(x/\eta) \, F^{q}(x, \eta, t) \;,
\label{eq:defFjq} \\
F_{j}^g(\eta, t)& = \frac{1}{2 k_{j-1}^{5/2}} \int_{-1}^{1}dx\, \eta^{j-1} \, C_{j-1}^{5/2}(x/\eta) \, F^{g}(x, \eta, t) \,.
\label{eq:defFjg} \\
F_{j}^\Sigma(\eta, t)& = \frac{1}{2 k_{j}^{3/2}} \int_{-1}^{1}dx\, \eta^j \, C_{j}^{3/2}(x/\eta) \, F^{\Sigma}(x, \eta, t) \;,
\label{eq:defFjSigma}
\end{align}
for integer $j$, and  where the normalization coefficients are
given in terms of Euler gamma functions:
\begin{equation}
k_{j}^{3/2} = \frac{3}{j}k_{j-1}^{5/2} = \frac{2^j \Gamma(j+3/2)}{
\Gamma(3/2)\Gamma(1+j)}\;.
\label{eq:GegenCoeff}
\end{equation}
$F^q$ and $F^g$ have been introduced in \refeqs{eq:defquarkGPD}{eq:defgluonGPD}, and their relation to the usual $H$ and $E$ GPDs is in \refeq{eq:FtoHE}. Here
\begin{align}
\label{eq:defFxSigma}
F^{\Sigma}(x)& = \sum_{q=u,d,s} F^{q}(x) - F^{q}(-x) \;, \\
F_{j}^\Sigma& = \sum_{q=u,d,s} F_{j}^q \;,
\end{align}
and the normalization coefficients above are chosen so
that for odd $j$ the forward limit is
\begin{align}
F_{j}^{q}& \to q_j + \bar{q}_j \;, \nonumber \\
F_{j}^{g}& \to g_j \;, \\
F_{j}^{\Sigma}& \to \Sigma_j = \sum_{q} q_j + \bar{q}_j \;,
\label{eq:FjFWD}
\end{align}
where on the \rhs, there are usual Mellin moments $\int_{0}^{1} dx x^j $ of PDFs,
and $\Sigma(x) = \sum_{q} q(x) + \bar{q}(x)$.
Since for integer $j$ the conformal moments above are just linear combinations
of Mellin moments, they are polynomials in $\eta$, where the order of the polynomial
can be read of from \reftabs{tab:quarkpolynomiality}{tab:gluonpolynomiality}.
Conformal moments are equal to matrix elements of \emph{local} conformal
operators
\begin{align}
O^{q}_j& = \frac{1}{k_{j}^{3/2}} (i \partial^+)^j \, \bar{q}\,
\gamma^+ \, C_{j}^{3/2}
 \bigg(\frac{\stackrel{\leftrightarrow}{D^+}}{\partial^+}\bigg)
\, q \;,  \label{eq:defOq} \\
O^{g}_j& = 2 \frac{1}{k_{j-1}^{5/2}} (i \partial^+)^{j-1} \, G^{+\mu}_{a} \,
C_{j-1}^{5/2} \bigg(\frac{\stackrel{\leftrightarrow}{D^+}}{\partial^+}\bigg) \,
G_{a \mu}^{\;\;\:+} \;,
\label{eq:defOg}
\end{align}
where $\stackrel{\leftrightarrow}{D}_\mu \equiv \stackrel{\rightarrow}{D}_\mu -
\stackrel{\leftarrow}{D}_\mu$
and $\partial_\mu \equiv \stackrel{\rightarrow}{\partial}_\mu +
\stackrel{\leftarrow}{\partial}_\mu$.
In particular,
\begin{multline}
\frac{1}{(P^+)^{j+1}} \langle P_2|O_{j}^a|P_1\rangle = F_{j}^a \\
= \frac{h^+}{P^+} H_{j}^a + \frac{e^+}{P^+} E_{j}^a  \qquad a=q, g, \Sigma \;.
\label{eq:defConME}
\end{multline}

In terms of conformal moments, momentum-fraction GPDs are given by formal
series expansion, \eg for quark GPDs
\begin{equation}
F(x,\eta,t) = \sum_{j=0}^{\infty} (-1)^j p_{j}(x,\eta)
F_{j}(\eta, t) \;,
\label{eq:cPWE}
\end{equation}
where $p_{j}(x, \eta)$ are Gegenbauer polynomials with absorbed
Gegenbauer weight function and normalization constant
\begin{equation}
p_{j}(x, \eta) = \eta^{-j-1}
\frac{2^j \Gamma(5/2+j)}{\Gamma(3/2) \Gamma(3+j)}
\left(1-\frac{x^2}{\eta^2}\right)
C_{j}^{3/2}\left(-\frac{x}{\eta}\right) \;.
\label{eq:defpj}
\end{equation}
Series \refeq{eq:cPWE} is formally divergent, so one needs to
specify the prescription for resumming it, where also the full
GPD support region $-1 \le x \le 1$ should be restored (if the series in \refeq{eq:cPWE} were converging, the resulting GPD would have support in $[-\eta, +\eta]$.).
Various resummation prescriptions are put forth in
\refcites{Belitsky:1997pc,Shuvaev:1999fm,Noritzsch:2000pr,Mueller:2005ed}.

Although conformal symmetry is broken in QCD at loop level, some
residual effects of this symmetry make conformal moment representation
of GPDs convenient for phenomenology.
Foremost, at LO there is no renormalization mixing of conformal moments of different
conformal spin $j+2$ so their evolution is given by diagonal operator.
(Mixing between gluon and singlet quark GPDs is of course still present.)
At NLO, operators from \refeqs{eq:defOq}{eq:defOg} start to mix, which
leads to non-diagonal evolution of conformal GPD moments.
Still, even this can be countered by special choice of renormalization
scheme (called \emph{conformal scheme}, $\overline{CS}$
\cite{Mueller:1993hg,Melic:2002ij}) so that
non-diagonal evolution can be pushed to NNLO level.
This non-mixing has been utilized to write efficient
computer code for GPD evolution \cite{Kumericki:2007sa}.
Also, conformal moments, being given by matrix elements of
local operators \refeqs{eq:defOq}{eq:defOg}, are computable on
the lattice.

Working within conformal moment representation one can perform
separation of variables using SO(3) partial wave expansion in
the $t$-channel \cite{Polyakov:1998ze,Polyakov:2002wz},
where the center-of-mass scattering angle corresponds at leading order to the
inverse GPD skewness variable
\begin{equation}
    \theta_{\rm cm} = \frac{1}{\eta} \;.
\end{equation}
This expansion can then be implemented working with so-called
quintessence functions
whose Mellin moments give conformal GPD moments, leading to
``dual" GPD representation \cite{Polyakov:2002wz}.
Another implementation uses
Mellin-Barnes integral resummation of series \refeq{eq:cPWE}
\cite{Mueller:2005ed},
\begin{equation}
  F(x, \eta, t) = \frac{i}{2} \int_{c-i\infty}^{c+i\infty}
 dj \frac{1}{\sin \pi j} p_j(x, \eta) F_{j}(\eta, t)\;,
\end{equation}
leading to Mellin-Barnes SO(3) partial wave GPD representation.
Prescriptions on how to analytically extend $p_j(x, \eta)$ from \refeq{eq:defpj}
to complex $j$ can be found in \cite{Mueller:2005ed}.
The mathematical connection between these two GPD representations and
their relation to the double distribution representation described in \refsec{sec:double-distribution-representation} has been recently
elucidated in \refcite{Muller:2014wxa}.


\subsection{A selection of models}
\label{sec:gpd-models}

Here we briefly describe some contemporary models which are specific versions of the frameworks discussed in Section \ref{sec:gpd-parameterization}. 
In the present context, an ideal theory to experiment comparison favors building relatively simple models that allow one to numerically estimate both the GPD behavior in the various kinematic variables, and the size of the 
observables for different processes in various kinematic regimes.
Our review is therefore not aimed at representing a comprehensive list of the many GPD models that have been worked out by various groups.
We have selected models according to the following criteria:
\begin{enumerate}
\item they satisfy the   physical constraints listed in the previous sections, either entirely, or within well defined approximations; 
\item they can provide useful guidance for disentangling physical situations where the theory might show interesting aspects (see, for instance, the discussion of dispersion relations in section \ref{sec:DR}); 
\item they feature various tunable parameters that make them apt for a direct phenomenological application through data comparison. 
\end{enumerate}


\subsubsection{Double Distribution models}
\label{sec:double-distribution-models}

From 1999 on, GPD models have been built on the basis of the Radyushkin Double Distribution Ansatz (RDDA) \cite{Musatov:1999xp}. DDs in the Polyakov-Weiss gauge, mentioned in \refsec{sec:double-distribution-representation}, have been used continuously, apart from some recent attempts \cite{Radyushkin:2011dh,Radyushkin:2012gba,Radyushkin:2013hca,Radyushkin:2013bba,Mezrag:2013mya,Szczepaniak:2007af}. The general idea is exposed below with the example of the GPD $H$ in the quark sector
\begin{multline}
\label{eq:relation-gpd-dd-polyakov-weiss}
H^q(x, \eta) = \int_{\Omega} \mathrm{d}\beta\mathrm{d}\alpha \, F^q_{PW}(\beta, \alpha) \delta(x - \beta - \alpha \eta) \\ + D^q\left(\frac{x}{\eta}\right) \;.
\end{multline}
The RDDA relates the DD $F^q_{PW}(\beta, \alpha, t)$ to the $t$-dependent PDF $q(x, t)$ through:
\begin{equation}
F^q_{PW}( \beta, \alpha, t ) = \pi_N( \beta, \alpha )\, q( \beta, t ),
\label{eq-rdda}
\end{equation}
where \emph{profile functions}  $\pi_N$ reads
\begin{equation}
\label{eq-def-profile-function}
\pi_N( \beta, \alpha ) = \frac{\Gamma(3/2+N)}{\sqrt{\pi}\Gamma(1+N)} \frac{[ ( 1 - |\beta| )^2 - \alpha^2 ]^N}{( 1 - |\beta| )^{2 N + 1}} \;,
\end{equation}
and is normalized like
\begin{equation}
    \int_{-1+|\beta|}^{1-|\beta|} d\alpha \;\pi_{N}(\beta,\alpha) = 1 \;.
    \label{eq:profile-normalization}
\end{equation}
In practice, the $t$-dependent PDF is modeled either with a Regge-type behavior $q(x, t) \propto q(x) x^{- \alpha' t}$, or a factorized (uncorrelated) form $q(x, t) \propto q(x) F^q_1(t)$. In the former case, $\alpha'$ is chosen to approximately describe the quark contribution to the hadron form factor $F^q_1(t)$, while this ingredient is directly incorporated in the latter case. This is the basis of the popular Goloskokov Kroll (GK) \cite{Goloskokov:2009ia,Goloskokov:2005sd,Goloskokov:2007nt} and Vanderhaeghen Guichon Guidal (VGG) \cite{Goeke:2001tz,Vanderhaeghen:1998uc,Vanderhaeghen:1999xj,Guidal:2004nd} models, which are described in great details in \refcite{Guidal:2013rya}.

We will illustrate the explicit building of a RDDA model with the example of the GPD $H$ in the quark sector in the GK and in the VGG model. It will demonstrate that the RDDA is an efficient way to generate realistic\footnote{We mean realistic in the phenomenological sense, \ie the model predictions have the correct order of magnitude, and can be used (at least) as a reliable first estimate. However, from \refsec{sec:double-distribution-representation}, it is clear that such a model generally cannot be expected to fulfill all theoretical constraints.} GPD models "on the fly", implementing at least the properties of polynomiality (\refsec{sec:polynomiality-property}), discrete symmetries (\refsec{sec:discrete-symmetries-properties}), and forward limit (\refsec{sec:forward-limit}).

In the GK model, the exponent $N$ of the profile function $\pi_N$ (\ref{eq-def-profile-function}) is taken as 1 for valence quarks and 2 for sea quarks. This exponent is set to 1 in the VGG model. However this difference is not expected to be quantitatively important, as we can infer \eg from the evaluations of \refcite{Mezrag:2013mya}. The $t$-dependence is expressed (at $\eta = 0$) as
\begin{equation}
H_i(x, 0, t)=q_i(x) x^{-\alpha^\prime t} \, e^{ b_{i} t } \quad \textrm{ with } i = \textrm{val or sea} \;.
\label{eq:gk-t-dependence}
\end{equation}
The VGG $t$-dependence of $H_{\textrm{val}}(x, 0, t)$ is different because there is an $x$-dependent term in the exponential which allows the recovering of the large-$t$ behavior of the form factor $F_1$
\begin{equation}
H_{\textrm{val}}(x, 0, t) = q_{\textrm{val}}(x) x^{-\alpha^\prime (1-x) t}.
\label{eq:vgg-t-dependence}
\end{equation}

Data sensitive mostly to the GPD $H^q$ are available over a large $Q^2$ range. 
Therefore its dependence  on the factorization scale $\mu$ cannot be neglected, and is tentatively accounted for through the $\mu$-dependence of the PDF $q(x, \mu)$ used in the RDDA approach \refeq{eq-rdda}. 

Note that positivity bounds are checked \textit{a posteriori} but cannot be guaranteed \apriori. 

The $D$-term is not fixed by QCD first principles. It is tied to the question of the $J=0$ fixed pole contribution which has been discussed recently in great details in \refcite{Muller:2014wxa,Muller:2015vha}. A flavor-singlet $D$-term $D$ can be defined by considering all active quark flavors
\begin{equation}
D( \alpha, t ) = \sum_{q} D^q( \alpha, t ),
\label{eq:def-flavor-singlet-D-term}
\end{equation}
projected on the basis of Gegenbauer polynomials $C^{3/2}_n$:
\begin{equation}
D( \alpha, t ) = ( 1 - \alpha^2 ) \sum_{n = 0, n \textrm{ odd}}^\infty d_n( t, \mu^2 ) C^{3/2}_n( \alpha ).
\label{eq:def-D-term-projection-Gegenbauer-polynomials}
\end{equation}
The Chiral Quark Soliton Model ($\chi$QSM) yields estimates (see \refcite{Goeke:2001tz} and references therein) of the first three non-vanishing terms of this expansion at a very low scale $\mu_0 \simeq 600~\MeV$ and zero momentum transfer
\begin{eqnarray}
d_1( t = 0~\GeV^2, \mu_0^2 ) & \simeq & - 4.0, \label{eq:ChQSM-d1-low-scale} \\
d_3( t = 0~\GeV^2, \mu_0^2 ) & \simeq & - 1.2, \label{eq:ChQSM-d3-low-scale} \\
d_5( t = 0~\GeV^2, \mu_0^2 ) & \simeq & - 0.4. \label{eq:ChQSM-d5-low-scale} 
\end{eqnarray}
However note that, at the low scale $\mu_0$, Schweitzer \etal \ \cite{Schweitzer:2002nm} report a value $d_1^{u+d} \simeq -9.46$  while Wakamatsu predicts $d_1^{u+d} \simeq - ( 4.9~-~6.2)$ for the $\chi$QSM and $d_1^{u+d} \simeq -0.716$ for the MIT Bag model \cite{Wakamatsu:2007uc}. The $D$-term is set to 0 in the GK model

There were only few attempts to model DDs not following the RDDA, which somehow gave the feeling that DD modeling was reducible to RDDA modeling. On the contrary, few studies \cite{Tiburzi:2002tq,Tiburzi:2002kr,Dorokhov:2011ew,Mezrag:2014tva,Mezrag:2014jka} directly computed DDs to implement polynomiality by construction. Since they were restricted to pion DDs and GPDs, they were not constrained by DVCS data. Generally, such studies proceed by evaluating triangle diagrams yielding matrix elements of quark twist-2 operators. It has been shown in \refcite{Mezrag:2014jka} that such a procedure implements the soft pion theorem (identifying the GPD at $t=0~\GeV^2$ and $\eta = 1$ with the pion DA) as soon as the pion-quark-antiquark vertices obey Bethe-Salpeter equations with a proper implementation of chiral symmetry. It is one of the few examples where GPDs computed from triangle diagrams fulfill \apriori the soft pion theorem.


\subsubsection{Models in conformal moments space}
\label{sec:MB-SO3-PWE}

Several GPD models constructed in conformal moments space
have been used for studying GPD properties,
properties of QCD perturbation expansion of GPD evolution operators
and Compton form factors, and for global fitting.
As described in \refsec{sec:conformal-moments-representation},
they are based on Mellin-Barnes integral GPD representation
\cite{Mueller:2005ed} expanded in
$t$-channel SO(3) partial waves \cite{Kumericki:2007sa}:
\begin{equation}
H_{j}(\eta, t) = \sum_{J = J_{\rm min}}^{j+1}
H_{j}^J(t) \eta^{j+1-J} \hat{d}^J(\eta) \;,
\label{eq:HPWE}
\end{equation}
where summation is over $t$-channel angular momentum $J$,
and $\hat{d}^J(\eta)$ are crossed version of appropriate
Wigner rotation matrices normalized as $\hat{d}(0)=1$.
For example, for the $t$-channel helicity conserved
``electric'' GPD moment combination $H_j + (t/4M_{p}^2)E_j$,
we have
\begin{equation}
\hat{d}^J(\eta) = \frac{\Gamma(1/2)\Gamma(J+1)}{2^J \Gamma(J+1/2)}
\eta^J C_{J}^{1/2}\left(\frac{1}{\eta}\right)\;.
\label{eq:dWigner}
\end{equation}
The leading partial wave amplitude $H_{j}^{j+1}(t)$ is the Mellin moment
of the zero-skewness GPD so in the forward limit it is equal
to the Mellin moment of the corresponding PDF
\begin{equation}
H_{j}^{j+1}(0) = q_j \equiv  \int_{0}^{1} dx\; x^j q(x) \;.
\label{eq:def-qj}
\end{equation}
If we take a standard simple ansatz for PDFs
\begin{equation}
  q(x) = \frac{N}{B(2-\alpha, \beta+1)} x^{-\alpha}(1-x)^\beta\;
\label{eq:PDFansatz}
\end{equation}
where the Euler beta function $B$ is factored out so that parameter 
$N=N_{q},N_{G}$
corresponds to average longitudinal momentum fraction
$\langle x \rangle$ for given flavor of quarks or gluons, 
satisfying the sum rule
($\Sigma$ is singlet flavor combination, cf.  \refeq{eq:FjFWD})
\begin{equation}
N_{\Sigma} + N_{G} = 1 \;,
\label{eq:N-sum-rule}
\end{equation}
then the Mellin moment $q_j$ \refeq{eq:def-qj} is
\begin{equation}
q_j = N \frac{B(1-\alpha+j, \beta+1)}{B(2-\alpha,\beta+1)}\;.
\label{eq:qj}
\end{equation}
\begin{figure}[ht]
  \centering
  \includegraphics[scale=0.4]{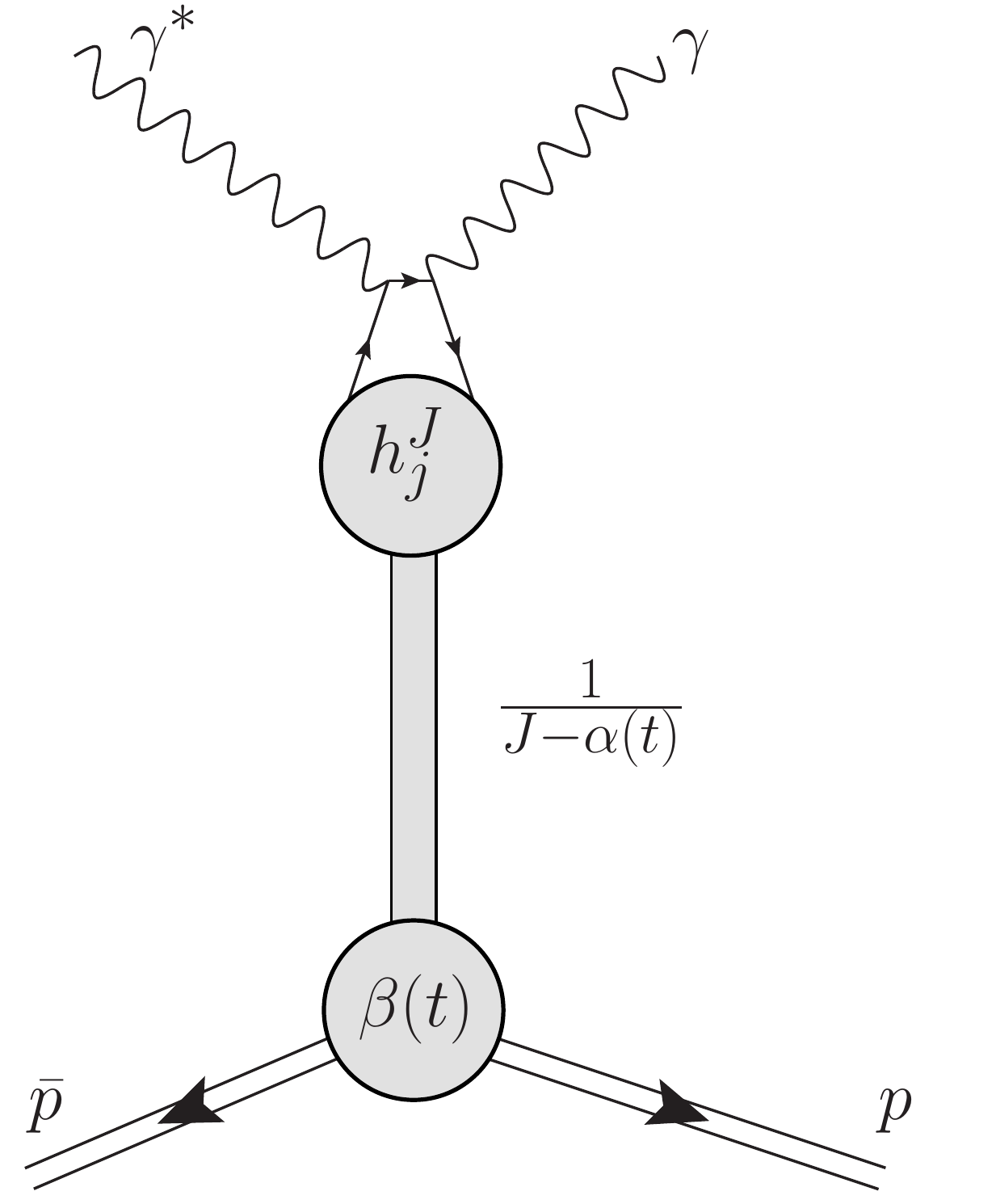}
  \caption{Modelling partial waves of conformal moments using crossed
    process $\gamma^{*}\gamma\to p\bar{p}$.}
  \label{fig:pwe}
\end{figure}
Concerning the dependence on $t$, the partial wave amplitudes $H_{j}^J(t)$
are modelled by relying on a Regge-inspired picture of $t$-channel exchanges
of mesonic states of total angular momentum $J$ which are
\begin{itemize}
\item coupled with strength parameter $h_{j}^{J}$ to quark-antiquark
state (formed at short distance by colliding photons),
\item propagating as appropriate Reggeon
\begin{equation}
\frac{1}{m^{2}(J)-t} \propto \frac{1}{J-\alpha(t)}
\end{equation}
with trajectory
\begin{equation}
   \alpha(t) = \alpha + \alpha' t
\label{eq:Regge-trajectory}
\end{equation}
\item and which is coupled to nucleon--anti-nucleon pair with
strength described by a $p$-pole impact form factor
\begin{equation}
  \beta(t) = \frac{1}{\left(1 - \frac{t}{M^2}\right)^p}\;,
  \label{eq:ppole}
\end{equation}
parameterized by cut-off mass $M$ (not to be confused with
proton mass), giving the total Ansatz
\begin{equation}
H_{j}^J(t) = \frac{h_{j}^{J}}{J-\alpha(t)}
\frac{1}{\left(1 - \frac{t}{M^2}\right)^p}\;.
\label{eq:PWAansatz}
\end{equation}
\end{itemize}
This is illustrated on \reffig{fig:pwe}.
Using again the simple PDF ansatz of \refeqs{eq:PDFansatz}{eq:qj},
and restoring full Regge trajectory $\alpha \to \alpha(t)$
as in \refeq{eq:Regge-trajectory}, one gets
for the leading partial wave amplitude
\begin{equation}
H_{j}^{j+1}(t) \equiv q_{j}(t) =
q_j \frac{1+j-\alpha}{1+j-\alpha - \alpha' t} \beta(t) \;.
\label{eq:qjt}
\end{equation}

In some studies, the residual dependence on $t$ has been described by
an exponential ansatz $\beta(t) = \exp(B t)$, often used in
Regge phenomenology, instead by a
multipole impact form factor \refeq{eq:ppole}.
Such an exponential an\-satz brings no advantage in fits
to present data and is more difficult to advocate from
field-theoretic perspective, so a multipole ansatz is
favored.
Note that the $t$ cut-off mass parameter $M$ could also
in principle depend on angular momentum, $M = M(J)$,
but the additional parametrization describing this also
brings no advantage in fits so this is presently usually
ignored.

Modelling of all GPDs relevant for present phenomenology
within the framework of full Mellin-Barnes SO(3)
partial wave decomposition has not yet been undertaken.
Models presently on the market truncate the SO(3) series
\refeq{eq:HPWE} to one or few leading terms, \ie terms
with highest $J=j+1, j-1, \ldots$, corresponding to
smallest powers of $\eta^{j+1-J} = \eta^0, \eta^2, \ldots$.
Furthermore, these models were originally devised for
description of small-$\xb$ collider DVCS data so
further expansion around $\eta=0$ was made to obtain
simplified model of form
\cite{Kumericki:2007sa,Kumericki:2009uq}
\begin{equation}
H_j(\eta, t) = q_{j}(t) + s_{2}\, \eta^2 q_{j}(t)
                        + s_{4}\, \eta^4 q_{j}(t) + \cdots \;,
\label{eq:lowx-KM-model}
\end{equation}
where $t$ and $j$ dependence of subleading partial waves
is for simplicity taken to be
equal to that of leading one, \ie, given by \refeq{eq:qjt}.
Strength of subleading partial waves $s_{2,4,\ldots}$ are
free parameters of the model. They essentially control
the skewness property of the GPD, i.e. the ratio of the
GPD on the so-called \crossover line $\eta=x$ and
GPD at $\eta=0$. Since the normalization parameter $N$ and
the Regge intercept $\alpha$ are fixed by DIS data, the $t$-dependence of $q_j$, and thus
that of the GPD, is controlled by $\alpha'$ and the cut-off mass
$M$. In the singlet sector, relevant for small-$\xb$ region,
the values $\alpha'_{\textrm{sea}\approx\Sigma} = \alpha'_{G} = 0.15\,
\GeV^{-2}$ are fixed.
Such values are favored by fits,
but fits are not very sensitive to them, so leaving them
as free parameters proved inefficient.
Cut-off masses $M$ are in present fits strongly correlated
with values of multipole power $p$ in \refeq{eq:ppole},
so it makes sense to also fix values for $p$.
Counting rules would give $p_{\rm sea} = 4$ and $p_G = 3$,
but to facilitate direct comparison of the cut-off mass $M$ with
the characteristic proton size coming from the dipole parametrization
of Sachs form factors, one can take $p_{\rm sea}=p_G=2$.
Fits are also not sensitive to the gluon cut-off mass $M_G$, which can
be fixed at $M_{G}^2 = 0.7\,\GeV^2$, value suggested
by the analysis of $J/\Psi$ production collider data \cite{Aktas:2005xu}.
This leaves the final set of free model parameters:
\begin{equation}
\{M_{\rm sea}, s_{2}^{\rm sea}, s_{2}^G, s_{4}^{\rm sea},
s_{4}^G \}\,,
\label{eq:KM-free-params}
\end{equation}
for models of sea quark and gluon GPDs with three partial waves,
(sometimes called nnl-PW) used in newest published fits
\cite{Kumericki:2013br}.
In \refcite{Kumericki:2009uq} two partial waves (nl-PW) are used.
The third partial wave does not bring much more flexibility
to fits but the resulting gluon GPDs turn out to be more realistic
considering their partonic interpretation (fits without
third wave tend to have negative gluon GPD at low $Q^2$).
The inclusion of second partial wave is, however, essential. Namely,
the skewness ratio of the minimal, leading partial wave model (l-PW) is
fixed at a too large a value to describe simultaneously
DIS and DVCS data.
The described models (with two or more partial waves) can
successfully reproduce all available low-$\xb$
DVCS data, see \refsec{sec:fits-global-lowx}.
They are also used as a sea parton component
of hybrid models \cite{Kumericki:2009uq}, where the valence component
is described by the simpler modelling of just the GPD on
the \crossover line $\eta=x$, and using dispersion relation
technique, see \refsec{sec:DR}, to recover the remaining
needed part.
As described in \refsec{sec:fits-global-world}, such hybrid models
are able to describe essentially all presently available
DVCS data.


\subsubsection{Spectator models}
\label{sec:spectator-models}
GPD spectator models are specific applications of the covariant scattering matrix approach (\refsec{sec:covariant-scattering-matrix}) stemming from a more phenomenological view of the problem, or from a bottom-up perspective. 

In these models the parton-proton amplitude is described by a holomorphic function which exhibits four poles in the $k^-$ complex plane. GPDs in the DGLAP region, $|x|\geq \eta$, are determined by the $u$-channel simple pole from the spectator propagator; conversely, in the ERBL region, $|x|\leq \eta$, they are obtained setting either quark on-shell, {\it i.e.} from the two poles in the proton-quark vertex functions. 

In \refcite{Brodsky:2008qu} a model for the GPD $H$ was given for a scalar diquark spectator, where the vertex functions reproduce the perturbative asymptotic behavior of the Dirac form factor, $F_1(t)$. This model is by construction ``hard", and it does not reproduce the electromagnetic form factor behavior at small momentum transfer. However, a set of parameters is provided which give the correct normalization, $F_1(0)=1$, while providing functional forms for the GPD behavior at different values of $(x,\eta)$. 

A more general version of the perturbative diquark model was given in \refcites{Goldstein:2010gu,GonzalezHernandez:2012jv,Ahmad:2006gn,Ahmad:2007vw,Goldstein:2013gra} which includes a Regge-inspired spectral analysis for the parton-proton scattering amplitude. This resulted in a parametric form for the DGLAP region  based on a ``Regge improved" diquark model, or ``reggeized diquark model", that also provides a framework for reproducing the low $t$ behavior of the proton form factors. The ERBL region  is treated in this model so that the properties of crossing symmetry, continuity at the crossover points with the DGLAP region ($x=\pm \eta$) and approximate polynomiality are satisfied. 
Regge behavior is obtained by letting the spectator system's
invariant mass, $M_X$, vary according to a spectral distribution, at variance with most models where the recoiling system’s mass is kept fixed. The variable mass spectator systems exhibit different structure as one goes from low to high mass values: at low mass values one has a simple scalar or axial vector spectator, whereas at large mass values one has more complex correlations. The occurrence of the latter, also known as reggeization \refcite{GonzalezHernandez:2012jv}, is regulated by a spectral distribution, $\rho(M_X^2)$ which upon insertion in the correlation function yields for small $x$ the desired $x^{-\alpha}$ behavior.
The resulting parametrization was summarized in the following expression for the quark sector, 
\begin{equation}
F_q(x,\eta,t)  = {\cal N}_q G_{M_X,m}^{M_\Lambda}(x,\eta,t) \,  
R^{\alpha,\alpha^\prime}_{p_q}(x,\eta,t) 
\label{fit_form}
\end{equation}
where $q=u,d$, $F_q \equiv H_q, E_q, \tilde{H}_q, \tilde{E}_q$; the functions $G_{M_X,m}^{M_\Lambda}$, and  
$R^{\alpha,\alpha^\prime}_{p_q}$, are the quark-diquark and Regge contributions, respectively. They depend on mass parameters for: the struck quark, $m$, the (diquark) spectator, $M_X$, the diquark form factor cut-off parameter, $M_\Lambda$. The Regge trajectory parameters are: $\alpha$ (intercept), $\alpha'$ (slope), and $p_q$ (x-dependent modulation).

The parametrization can be extended to the valence quark, sea quark, and gluon contributions. For valence quarks the spectator is given by scalar and axial-vector diquarks, which, through the $SU(4)$ symmetry, allow one to perform a flavor analysis by distinguishing
between isoscalar ($ud$) and isovector ($uu$) spectators. For scattering from sea quarks, the spectator is a tetra-quark state, namely a $uudq(\bar{q})$ state with $q=u,d,s,c$. For gluons it is a three quark system in a color octet state. The possibility of distinguishing among different flavors in this model reflects the underlying color symmetry which can be seen as an indirect manifestation of chiral symmetry breaking. 

\begin{figure}[ht]
  \centering
  \includegraphics[scale=0.45]{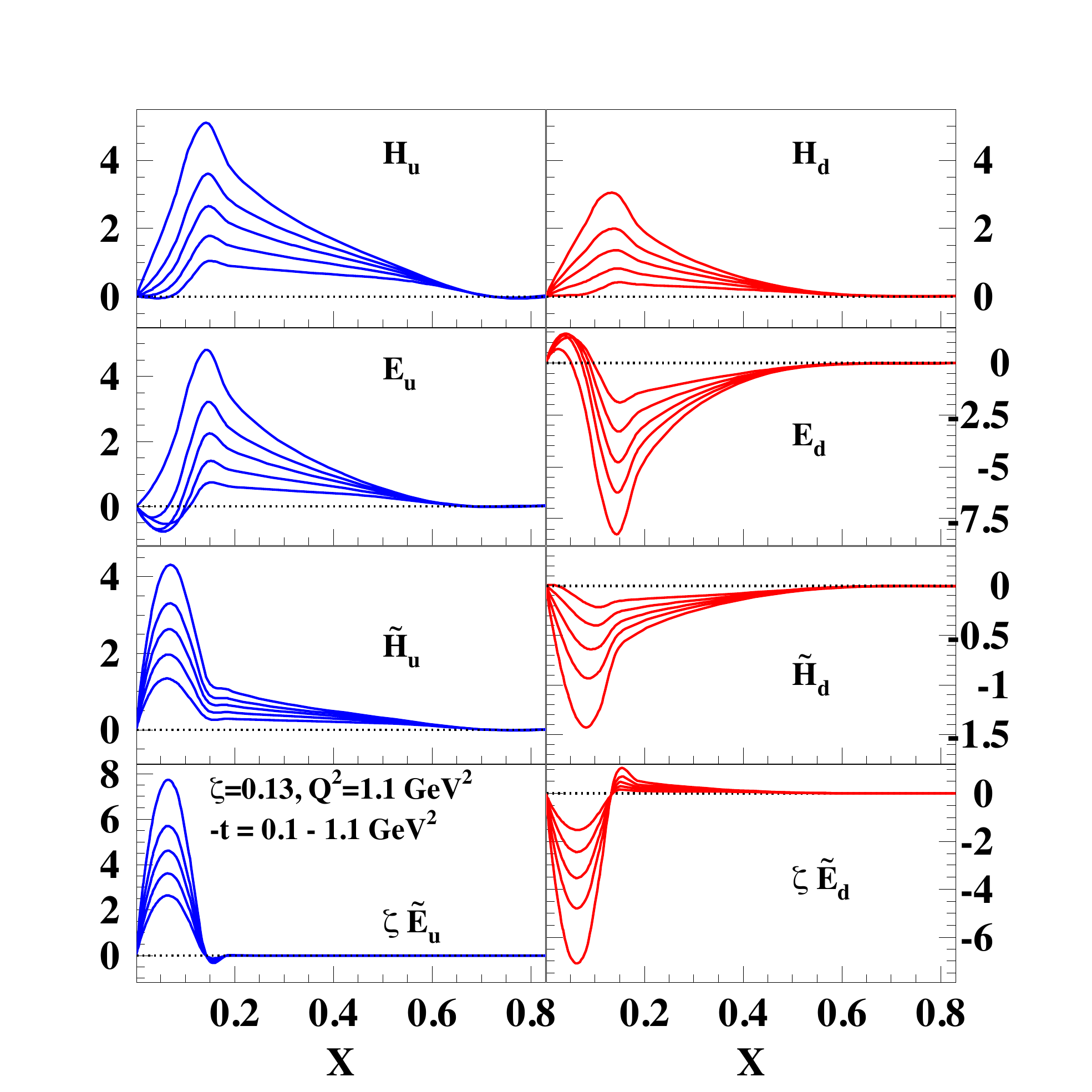}
  \caption{GPDs evaluated using the spectator model described in the text plotted vs. $x$ at $\xb=\zeta=2 \eta/(1+\eta) = 0.13$, $Q^2=  2$ GeV$^2$. The range in $-t$ is: $0.1  \leq -t \leq 1.1 \, {\rm GeV}^2$. Curves with the largest absolute values correspond to the lowest $t$ (adapted from \refcite{Goldstein:2013gra}).}
  \label{fig:spectator}
\end{figure}

The set of chiral even GPDs, $H_q, E_q,\tilde{H}_q, \tilde{E}_q$ from the model is shown in \reffig{fig:spectator}. The quark-diquark components are given by,
\begin{eqnarray}
G_{M_X,m}^{M_\Lambda,(H)} & = &  \displaystyle\mathcal{N} \int \frac{d^2k_\perp}{1-x_1} 
\frac{%
\left[  \left(m+M x_1 \right)  \left(m + M x_2 \right) + k_\perp\cdot \tilde{k}_\perp\right]}{(k^2-M_\Lambda^2)^2(k^{\prime \, 2}-M_\Lambda^2)^2} \;, \nonumber \\
\label{GPDH}
\end{eqnarray} 
\begin{eqnarray}
G_{M_X,m}^{M_\Lambda, (E)} & = & \displaystyle \mathcal{N}  2M \frac{1-\eta}{1+\eta}  \int  \frac{d^2k_\perp}{1-x_1} 
\nonumber \\
& \times & \frac{\left[\left(m + M x_1 \right) \frac{k_\perp \cdot \Delta}{\Delta^2} - \left(m+M x_2 \right) \displaystyle   \frac{\tilde{k}_\perp \cdot \Delta}{\Delta^2} \right] }{(k^2-M_\Lambda^2)^2(k^{\prime \, 2}-M_\Lambda^2)^2} \nonumber \\
 \label{GPDE}
\end{eqnarray}
\begin{eqnarray}
G_{M_X,m}^{M_\Lambda,(\widetilde{H})} & = &  \displaystyle\mathcal{N}  \int \frac{d^2k_\perp}{1-x_1}
\frac{%
\left[  \left(m+M x_1 \right)  \left(m + M x_2 \right) -  k_\perp\cdot \tilde{ k}_\perp\right]}{(k^2-M_\Lambda^2)^2(k^{\prime \, 2}-M_\Lambda^2)^2} 
\label{GPDHTILDE} \nonumber \\
\end{eqnarray} 
\begin{eqnarray}
G_{M_X,m}^{M_\Lambda,(\widetilde{E})} & = & - \displaystyle\mathcal{N} \displaystyle \, \frac{4M}{\eta} (1-\eta)  \int \frac{d^2k_\perp}{1-x_1}
\nonumber \\
& \times &  \frac{\left[  \left(m+ M x_1 \right) \displaystyle\frac{\tilde{k} \cdot \Delta}{\Delta^2}  + \left(m + M x_2 \right) \frac{k_\perp \cdot \Delta}{\Delta^2} \right] }{(k^2-M_\Lambda^2)^2(k^{\prime \, 2}-M_\Lambda^2)^2} \nonumber \\
 \label{GPDETILDE}
\end{eqnarray}
where $x_{1(2)} = (x\pm\eta)/(1 \pm \eta)$, $\tilde{k}_\perp = k_\perp - (1-x)/(1-\eta) \Delta$, $k^2$ and $k'^2$, the incoming and outgoing quark virtualities, depend on $M_X$, $M$, and $k_T^2$; 
$\mathcal{N}$, the normalization constant, is in \GeV$^4$. Although the parametrization is written for the ``asymmetric'' choice of kinematics with the initial proton momentum along the $z$-axis, this can easily be connected to the ``symmetric" choice, adopted in our review, where the average of the initial and final proton momenta are along $z$ \refcite{Diehl:2003ny}.
The Regge term is given by
\begin{equation}
R^{\alpha,\alpha^\prime}_{p_q}=  x^{-[\alpha + \alpha^\prime(x) t  ]},
\label{regge}
\end{equation}
where  
\begin{equation}
\alpha^\prime(x) =  \alpha^\prime  (1-x)^{p_q}. 
\end{equation}

A quantitative fit using experimental information from DIS, the nucleon electroweak form factors, and a selection of available DVCS data from Jefferson Lab \cite{Girod:2007aa} was developed using the valence component of the parametrization in \refcite{Goldstein:2010gu}: the model is consistent with theoretical constraints imposed numerically, and the experimental data are let to guide the shape of the parametrization as closely as possible.


\section{Theoretical and experimental status}
\label{sec:status}


\subsection{Theory of DVCS}
\label{sec:theory-toolbox}

\begin{figure}[t]
\centering
\includegraphics[scale=0.33]{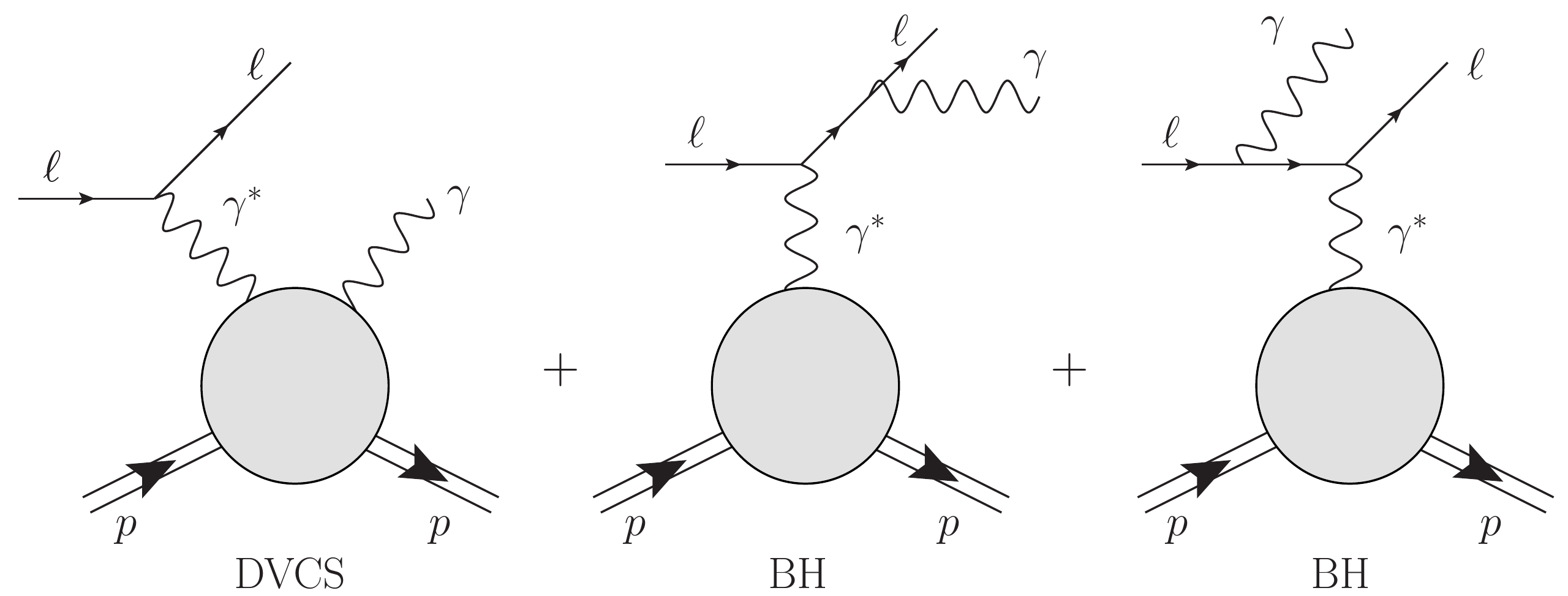}
\caption{Leptoproduction of a real photon as a coherent superposition
of DVCS and Bethe-Heitler amplitudes.}
\label{fig:leptoproduction}
\end{figure}

\begin{figure}  
\centerline{\includegraphics[scale=0.4]{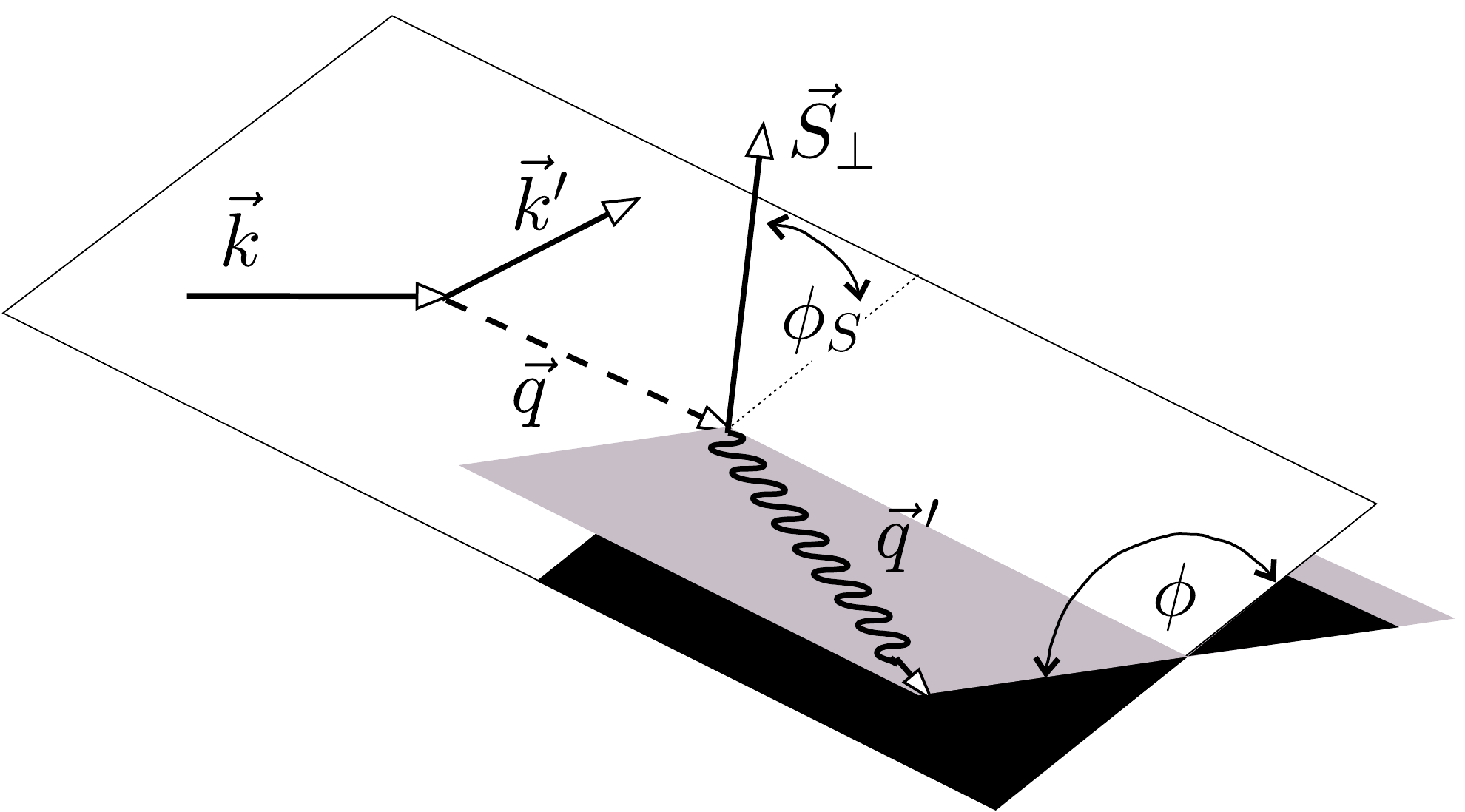}}
\caption{Definition of momenta and angles relevant for leptoproduction
of a real photon in the Trento convention \protect\cite{Bacchetta:2004jz}.
(Fig. taken from \cite{Airapetian:2011uq}.)}
\label{fig:trento}       
\end{figure}

Measurements of DVCS are mostly realized via the process of
leptoproduction of
a real photon, where also an interference with the Bethe-Heitler radiation occurs, as displayed in \reffig{fig:leptoproduction}.
The general cross-section is differential
in  $\xb$, the negative squared momentum of virtual photon
$\Q^2 = - q^2$, the squared momentum transfer
$t$, and two azimuthal angles measured relatively to the lepton
scattering plane: the angle $\phi$ to the photon-target scattering plane and
the angle $\phi_S$ to the transversal component of the target polarization
vector, as displayed on \reffig{fig:trento}.
The cross section is given by
\begin{equation}
\frac{d^5\sigma}{d\xb d\Q^2 d|t| d\phi d\phi_S } =
\frac{\alpha^3\, \xb}{16\pi^2 \Q^4 \sqrt{1+\epsilon^2}}
\big|\mathcal{T}\big|^2 \;,
\label{eq:xs5fold}
\end{equation}
where $\alpha$ is the electromagnetic fine structure constant,
$\epsilon = 2\xb M/\Q$, $M$ is the mass of the target,
and $\mathcal{T}$ is
coherent superposition of DVCS and Bethe-Heitler amplitudes
\begin{equation}
|\mathcal{T}|^2 = |\mathcal{T}_{\rm BH} + \mathcal{T}_{\rm DVCS}|^2
=|\mathcal{T}_{\rm BH}|^2 + |\mathcal{T}_{\rm DVCS}|^2 + \mathcal{I}\;.
\label{eq:coherent}
\end{equation}
The DVCS amplitude $\mathcal{T}_{\rm DVCS}$
can be decomposed either
in helicity amplitudes or, equivalently, in
complex valued Compton Form Factors (CFFs)
which are to be measured in experiments. The latter are usually denoted as
\begin{displaymath}
\label{twelveCFFs}
\mathcal{H},\mathcal{E},\tilde{\mathcal{H}},\tilde{\mathcal{E}},
\mathcal{H}_{\rm T},\mathcal{E}_{\rm T},
\tilde{\mathcal{H}}_{\rm T},\tilde{\mathcal{E}}_{\rm T}\,.
\end{displaymath}
Owing to the validity of QCD factorization theorems \cite{Collins:1998be}
the CFFs can be written at leading order in perturbative QCD, as the following convolution ($\mathcal{F}=\mathcal{H}, \mathcal{E}, ...$),
\begin{eqnarray}
\mathcal{F}(\eta,t) & = & \sum_q e_q^2 \int_{-1}^{1} dx \nonumber \\
& \times & \left[ \frac{1}{\eta-x - i\epsilon} -  \frac{1}{\eta+x - i\epsilon} \right] F^q(x,\eta,t) \,, \label{eq:convF} \\
\tilde{\mathcal{F}}(\eta,t) & = & \sum_q e_q^2 \int_{-1}^{1} dx \nonumber \\
& \times & \left[ \frac{1}{\eta-x - i\epsilon} +  \frac{1}{\eta+x - i\epsilon} \right] \tilde{F}^q(x,\eta,t) \,, \label{eq:convFt} 
\end{eqnarray}
where the choice of symbols is motivated by their relation to GPDs. As a consequence of helicity conservation at the photon vertex, the transversity GPDs appear only at NLO in DVCS. They can be measured, however, in processes that directly allow for helicity flip, for instance in $\pi^0$ and $\eta$ production \cite{Ahmad:2008hp}. In particular, $\pi^0$ electroproduction  constitutes the main background process for DVCS.
In the experimentally accessible range of $Q^2$, $1/Q$ power corrections can be relatively large; for these the formalism is extended to include a set of higher twist GPDs and CFFs denoted $\mathcal{H}_{3}, \mathcal{E}_{3},...$. The transversity gluon GPDs also appear at this order \cite{Belitsky:2001ns}. It should be noted that the literature does not provide a uniform naming scheme for the twist three GPDs: three different notations appear in \refcites{Belitsky:2001ns,Meissner:2009ww,Kiptily:2002nx}, respectively. A conversion table among schemes for the vector sector was given in \refcite{Courtoy:2013oaa}.  Finally, a recent analysis of twist four corrections including kinematic power corrections $\mathcal{O}(t/Q^2)$  and $\mathcal{O}(M^2/Q^2)$ in terms of double distributions and Mellin-Barnes integrals has been recently made available in \refcite{Braun:2014sta}. 

The \nonpert part of the Bethe-Heitler amplitude
$\mathcal{T}_{\rm BH}$ is, on the other
hand, given in terms of the (in the relevant kinematical
region) well-known elastic form factors $F_1(t)$ and $F_2(t)$.
This then, through the interference term $\mathcal{I}$, gives
an experimental access to both the real and imaginary parts of the CFFs.

The various CFFs/GPDs can be disentangled by measuring several independent observables in exclusive lepton–proton scattering experiments where both the lepton beam and the target 
can be polarized. The general framework is given in terms of helicity amplitudes for the $\gamma^* p \rightarrow \gamma p$ process, as described \eg in \refcite{Diehl:2005pc}. 
These in turn factor into a hard scattering amplitude for the process, $\gamma^* q \rightarrow \gamma q$, $g_{\lambda \lambda'}^{\Lambda_{\gamma^*} \Lambda_\gamma}$, which depends on the initial and final photon and quark helicities, and a quark-proton helicity amplitude $A_{\Lambda'\lambda',\Lambda\lambda}$, which contains the GPDs. In DVCS, in particular, only the chiral even GPDs can be tested.

The helicity structure of the GPDs is described systematically in \refcite{Diehl:2003ny}. 
At twist two the relevant amplitudes are, 
\begin{eqnarray}
A_{++,++} & = & \sqrt{1-\eta^2} \left[\frac{H+\tilde{H}}{2} - \frac{\eta^2}{1-\eta^2} \frac{E+\tilde{E}}{2}  \right] \\
A_{-+,-+} & = & \sqrt{1-\eta^2} \left[\frac{H-\tilde{H}}{2} - \frac{\eta^2}{1-\eta^2} \frac{E-\tilde{E}}{2}  \right] \\
A_{++,-+} & = & - e^{-i \phi} \frac{\sqrt{t_0-t}}{2M} \frac{E - \eta \tilde{E}}{2} \\
A_{-+,++} & = & e^{i \phi} \frac{\sqrt{t_0-t}}{2M} \frac{E + \eta \tilde{E}}{2} 
\end{eqnarray}
The remaining helicity configurations are obtained by parity relations: $A_{-\Lambda' - \lambda' ,-\Lambda - \lambda} = (-1)^{\Lambda' - \lambda' - \Lambda + \lambda} \, A^*_{\Lambda'  \lambda' ,\Lambda \lambda}$.
The phase factor contains the angle $\phi$ between the lepton and hadron planes (see \reffig{fig:trento}). Using products of the helicity amplitudes to form the various contributions to the cross sections 
in \refeq{eq:coherent}, one obtains an expression that depends on various modulations of the type $\sin n\phi$, and $\cos n \phi$.
The final expressions provided in 
 \refcite{Belitsky:2001ns} read,
\begin{multline}
\label{eq:TBH2}
|{\cal T}_{\rm BH}|^2 = \frac{1}{\xb^2 t (1+\epsilon^2)^2 \,
\mathcal{P}_1(\phi)\mathcal{P}_2(\phi)}
 \\
  \times\bigg\{{c}_0^{\rm BH}  +\sum_{n=1}^2 {c}_n^{\rm BH} \cos(n \phi) +
    {s}_1^{\rm BH} \sin\phi\bigg\}\,
\end{multline}
\begin{multline}
\label{eq:TI}
{\cal I} = \frac{- e_\ell}{\xb t y \,
\mathcal{P}_1(\phi)\mathcal{P}_2(\phi)}
 \\
\times  \bigg\{ {c}_0^{\cal I}  + \sum_{n=1}^3 \left[ {c}_n^{\cal I} \cos(n \phi)+
    {s}_n^{\cal I} \sin(n \phi) \right]\bigg\}\,,
\end{multline}
\begin{multline}
\label{eq:TDVCS2}
|{\cal T}_{\rm DVCS}|^2 = \frac{1}{\Q^2}
 \\
 \times \bigg\{ {c}_0^{\rm DVCS}   + \sum_{n=1}^2\left[
{c}_n^{\rm DVCS} \cos(n \phi) +
{s}_n^{\rm DVCS}\right]\bigg\}\,,
\end{multline}
where $y$ is the lepton energy loss in the target frame,
$e_\ell$ in \refeq{eq:TI} is the lepton beam charge
in units of positron charge, and
$1/(\mathcal{P}_1(\phi) \mathcal{P}_2(\phi))$ originate
from the lepton propagators in Bethe-Heitler amplitude
(see \refcite{Belitsky:2001ns} for expressions).
The CFFs enter
quadratically the harmonic coefficients $c_n^{\rm DVCS}$ and $s_n^{\rm DVCS}$
of $|{\cal T}_{\rm DVCS}|^2$, and linearly those of ${\cal I}$, while they don't enter
the (often dominant) Bethe-Heitler squared part.
Detailed expressions for the coefficients $c_n$ and $s_n$ are given
in \refcites{Belitsky:2001ns,Belitsky:2008bz,Belitsky:2010jw,Belitsky:2012ch}, labeled BMK. Note that the cross section in \refeq{eq:xs5fold} undergoes a similar decomposition into its BH, DVCS and interference terms whether the target is unpolarized, longitudinally polarized, or transversely polarized, the coefficients $c_n$ and $s_n$ being given by different expressions with sensitivities to different GPDs in each case.   

\begin{figure} 
\centerline{\includegraphics[scale=0.36]{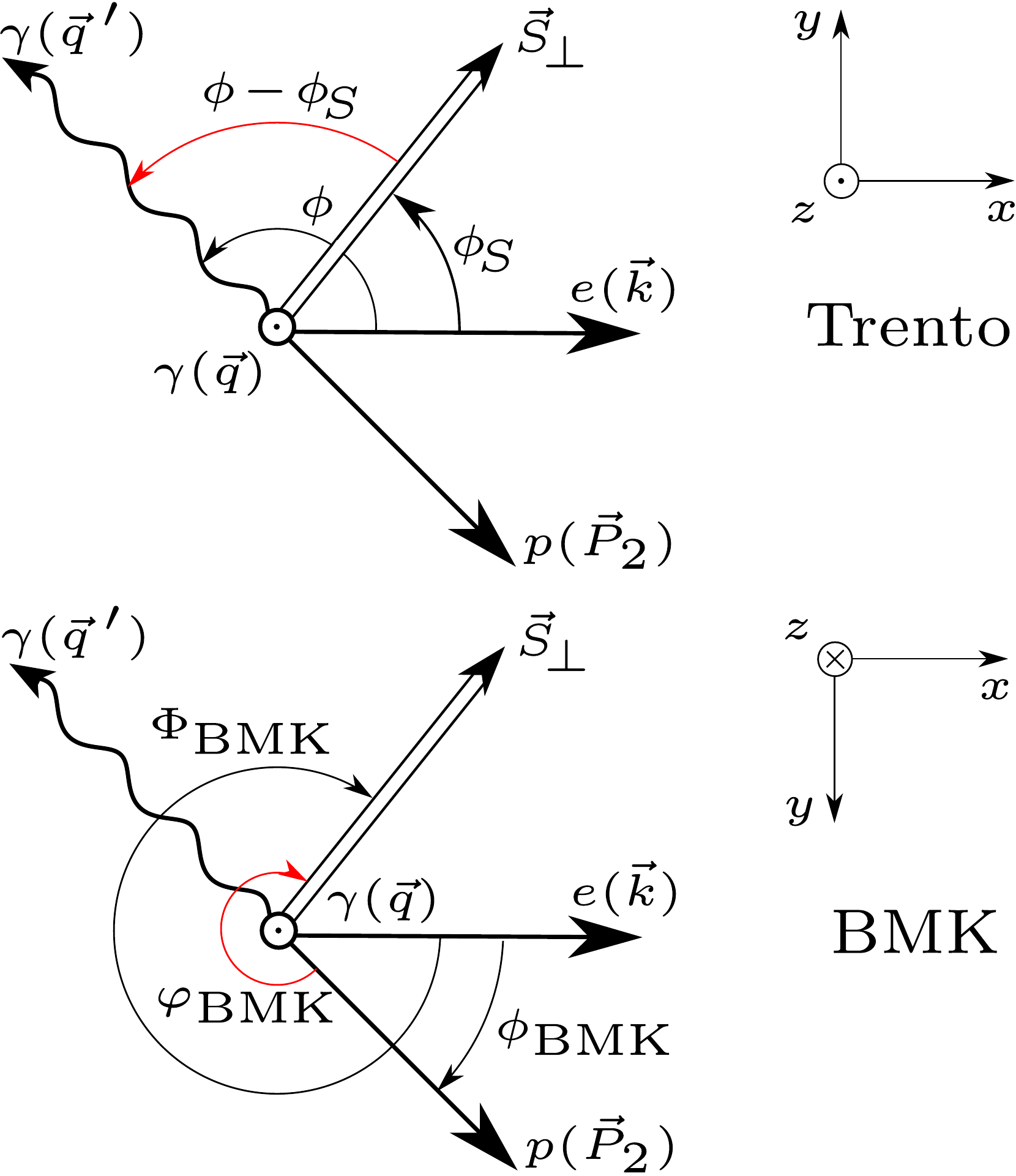}}
\caption{Comparison of the same leptoproduction event in
the Trento \protect\cite{Bacchetta:2004jz} and BMK
\protect\cite{Belitsky:2001ns} frames. The virtual photon
momentum is coming out of the paper along the
positive (negative) $z$-axis in Trento (BMK) frame. The emphasis is
on various azimuthal angles, so everything is projected
onto $x-y$ plane and the outgoing lepton is omitted. See
\protect\reffig{fig:trento} for the 3D view of the same
event.}
\label{fig:TrentoVsBMK}       
\end{figure}

One should keep in mind that BMK use 
a different coordinate system, which is related to the
``Trento'' coordinate system of \reffig{fig:trento} by
\begin{align}
\phi_{\rm BMK}& = \pi - \phi \\
\varphi_{\rm BMK}& \equiv \Phi_{\rm BMK} - \phi_{\rm BMK} = \phi - \phi_S - \pi \;,
\label{eq:BMK2Trento}
\end{align}
see \reffig{fig:TrentoVsBMK}.
One of the biggest challenges in DVCS analyses is a precise determination of the  $\phi$ dependence of the various asymmetries and/or cross section terms from which to extract the CFFs. 
As we explain in the following sections, these are particularly hard to disentangle in asymmetry measurements, since they contain ``competing'' $\phi$-dependent terms both in the numerator and denominator of their expressions. 
Among the harmonic coefficients in \refeqs{eq:TI}{eq:TDVCS2} the ones which are expected to be dominant because they contain leading twist GPDs are
\begin{displaymath}
 c_{0,1}^{\cal I},\; s_{1}^{\cal I},\;
\text{and}\; c_{0}^{\rm DVCS}\;.
\end{displaymath}
Each coefficient has a different form depending on the target polarization and is dominated, in turn, by a specific GPD.
The coefficient $s_{2}^{\cal I}$ is particularly interesting, even if non leading, because it has been singled out  as a direct probe of the twist-three GPD which measures orbital angular momentum \cite{Courtoy:2013oaa}.
Finally, for BH, 
$c_{0,1}^{\rm BH}$
are dominant, the rest of the coefficients being suppressed by kinematics.


\subsection{DVCS observables}
\label{sec:dvcs-observables}
To access particular CFFs via leptoproduction measurements,
one uses some kind of harmonic analysis and various
choices of beam and target polarizations and charges, if
available.

Using the notations of \refcite{Kroll:2012sm}, the cross section for the leptoproduction of a real photon by a lepton $l$ (with charge $e_l$ in units of the positron charge, and helicity $h_l/2$) off an unpolarized  target can be written as
\begin{multline}
d\sigma^{h_l,e_l}(\phi) = d\sigma_{\rm UU}(\phi)\left[1 + h_l A_{\rm LU,DVCS}(\phi)
                      \right. \\ \left. + e_lh_l A_{\rm LU,I}(\phi) + e_l A_{\rm C}(\phi)\right] \;,
\label{eq:airapetian-asymmetries}
\end{multline}
where only the $\phi$-dependence of the observables is explicit. In facilities where 
longitudinally polarized, positively and negatively charged beams are available, 
the asymmetries $A_{\rm LU,DVCS}$, $A_{\rm LU,I}$ and $A_{\rm C}$ can be isolated. This is the case for a large part of HERMES data, see \refsec{sec:experimental-data}. It is quite conventional to use the first subscript to refer to the beam and second to the target polarization ($U$ for unpolarized, $L$ for
longitudinal, etc.). For example, the beam charge asymmetry 
is singled out from the combination
\begin{equation}
A_{\rm C}(\phi) = \frac1{4d\sigma_{\rm UU}(\phi)}\,\left[   
             ( d\sigma^{\stackrel{+}{\rightarrow}} + d\sigma^{\stackrel{+}{\leftarrow}} )
           - ( d\sigma^{\stackrel{-}{\rightarrow}} 
           + d\sigma^{\stackrel{-}{\leftarrow}}) \right]\,.
\label{eq:A_C}
\end{equation}

Analogous combinations yield the two beam spin asymmetries $A_{LU,I}$ and 
$A_{LU,DVCS}$:
\begin{eqnarray}
A_{\rm LU,I}(\phi) & = &\frac{(d\sigma^{\stackrel{+}{\rightarrow}} - d\sigma^{\stackrel{+}{\leftarrow}}) - (d\sigma^{\stackrel{-}{\rightarrow}} - d\sigma^{\stackrel{-}{\leftarrow}})}{4d\sigma_{\rm UU}(\phi)} \;, \; \\
A_{\rm LU,DVCS}(\phi) & = &\frac{(d\sigma^{\stackrel{+}{\rightarrow}} - d\sigma^{\stackrel{+}{\leftarrow}}) + (d\sigma^{\stackrel{-}{\rightarrow}} - d\sigma^{\stackrel{-}{\leftarrow}})}{4d\sigma_{\rm UU}(\phi)} \;. \;
\end{eqnarray}

If an experiment has access
to only one value of $e_l$, such as in Jefferson Lab, the asymmetries defined in
\refeq{eq:airapetian-asymmetries} cannot be isolated. One can only measure
the beam spin asymmetry $A_{\rm LU}^{e_l}$, which depends on the combined charge-spin cross section as
\begin{equation}
A_{\rm LU}^{e_l}(\phi) = \frac{d\sigma^{\stackrel{e_l}{\rightarrow}} 
     - d\sigma^{\stackrel{e_l}{\leftarrow}}} {d\sigma^{\stackrel{e_l}{\rightarrow}}
      + d\sigma^{\stackrel{e_l}{\leftarrow}}} \;.
\end{equation}
In this equation we use the usual notation of labelling the combined charge-spin cross section with the sign of the beam charge $e_l$ and an arrow $\rightarrow$ ($\leftarrow$) for the helicity plus (minus).
$A_{\rm LU}^{e_l}$ can be written as a function of the asymmetries defined in \refeq{eq:airapetian-asymmetries}
\begin{equation}
A_{\rm LU}^{e_l}(\phi) = \frac{e_l A_{\rm LU,I}(\phi)+A_{\rm LU,DVCS}(\phi)}{1+e_lA_{\rm C}(\phi)} \;.
\label{eq:alu-alui-aludvcs}
\end{equation}

The target longitudinal spin asymmetry reads
\begin{equation}
A_{\rm UL}^{e_l}(\phi)  = \frac{ [ d\sigma^{\stackrel{e_l}{\leftarrow\Rightarrow}} 
     + d\sigma^{\stackrel{e_l}{\rightarrow\Rightarrow}} ] - [ d\sigma^{\stackrel{e_l}{\leftarrow\Leftarrow}} 
     + d\sigma^{\stackrel{e_l}{\rightarrow\Leftarrow}} ] } { [ d\sigma^{\stackrel{e_l}{\leftarrow\Rightarrow}} 
     + d\sigma^{\stackrel{e_l}{\rightarrow\Rightarrow}} ] + [ d\sigma^{\stackrel{e_l}{\leftarrow\Leftarrow}} 
     + d\sigma^{\stackrel{e_l}{\rightarrow\Leftarrow}} ] } \;,
\end{equation}
where the double arrows $\Leftarrow$ ($\Rightarrow$) indicates the target polarization
state parallel (anti-parallel) to the beam momentum. The double longitudinal target spin asymmetry
is defined similarly
\begin{equation}
A_{\rm LL}^{e_l}(\phi) = \frac{ [ d\sigma^{\stackrel{e_l}{\rightarrow\Rightarrow}} 
     + d\sigma^{\stackrel{e_l}{\leftarrow\Leftarrow}} ] - [ d\sigma^{\stackrel{e_l}{\leftarrow\Rightarrow}} 
     + d\sigma^{\stackrel{e_l}{\rightarrow\Leftarrow}} ] } { [ d\sigma^{\stackrel{e_l}{\rightarrow\Rightarrow}} 
     + d\sigma^{\stackrel{e_l}{\leftarrow\Leftarrow}} ] + [ d\sigma^{\stackrel{e_l}{\leftarrow\Rightarrow}} 
     + d\sigma^{\stackrel{e_l}{\rightarrow\Leftarrow}} ] } \, ,
\end{equation}

The HERMES collaboration also had access to a transversely polarized target with both electrons and positrons. They therefore
were able to measure two types of observables
\begin{multline}
A_{\rm UT,I}(\phi,\phi_S) = \\
 \frac{ d\sigma^+(\phi_S)- d\sigma^+(\phi_S+\pi)+d\sigma^-(\phi_S)
 - d\sigma^-(\phi_S+\pi) } 
{ d\sigma^+(\phi_S)+ d\sigma^+(\phi_S+\pi)+d\sigma^-(\phi_S)
+ d\sigma^-(\phi_S+\pi) } \,,
\end{multline}
\begin{multline}
A_{\rm UT,DVCS}(\phi,\phi_S)  = \\
\frac{ d\sigma^+(\phi_S)- d\sigma^+(\phi_S+\pi)
-d\sigma^-(\phi_S)+ d\sigma^-(\phi_S+\pi) } 
{ d\sigma^+(\phi_S)+ d\sigma^+(\phi_S+\pi)
+d\sigma^-(\phi_S)+ d\sigma^-(\phi_S+\pi) } \,, 
\label{eq:A_UTDVCS}
\end{multline}
where dependence on $\phi$ is suppressed on \rhs. 

For experiments which cannot deliver
cross-sections, but asymmetries, one can often use the
dominance of the Bethe-Heitler term in the denominator to
still obtain more
or less direct linear dependence on CFFs. For example, the first
sine harmonic of the beam spin asymmetry, as measured \eg in Jefferson Lab, is defined as
\begin{equation}
A_{LU}^{-, \sin\phi} \equiv
\frac{1}{\pi} \int_{-\pi}^{\pi} d\phi\: \sin\phi\:
A^-_{LU}(\phi) \;.
\label{eq:ALUm1}
\end{equation}
This harmonic is then approximately proportional to
linear combination of CFFs:
\begin{equation}
A_{LU}^{-, \sin\phi} \propto
\im\bigg(F_{1}\mathcal{H} - \frac{t}{4 M^2} F_{2}
\mathcal{E} +\frac{\xb}{2} (F_1 + F_2)
\tilde{\mathcal{H}}\bigg) \;,
\label{eq:CIunp}
\end{equation}
and will be dominated by $\im\mathcal{H}$. Similarly,
beam \emph{charge} asymmetry $A_C$ gives access to
$\re\mathcal{H}$ etc.

If one measures cross-sections, one can also
perform normal Fourier analysis, or it may be
favorable to work with specially weighted Fourier
integral measure \cite{Belitsky:2001ns}
\begin{equation}
d\phi \;\to\; dw \equiv \frac{2\pi \mathcal{P}_1(\phi) \mathcal{P}_2(\phi)}
{\int_{-\pi}^{\pi} d\phi\, \mathcal{P}_1(\phi) \mathcal{P}_2(\phi)}
d\phi \;,
\label{eq:w}
\end{equation}
thus cancelling strongly oscillating factors
$1/(\mathcal{P}_1(\phi) \mathcal{P}_2(\phi))$
in Bethe-Heitler and interference terms,
\refeqs{eq:TBH2}{eq:TI}.
Series of such \emph{weighted} harmonic terms, \eg
\begin{equation}
\sigma^{\sin n\phi,w} \equiv
\frac{1}{\pi} \int_{-\pi}^{\pi} dw\:
\sin n\phi\:
\sigma(\phi) \;,
\label{eq:weighted}
\end{equation}
converges then faster with increasing $n$ than normal
Fourier series.


\subsection{Evaluation of Compton Form Factors}
\label{sec:cff-evaluation}

\begin{figure}[t]
\centering
\includegraphics[scale=0.42]{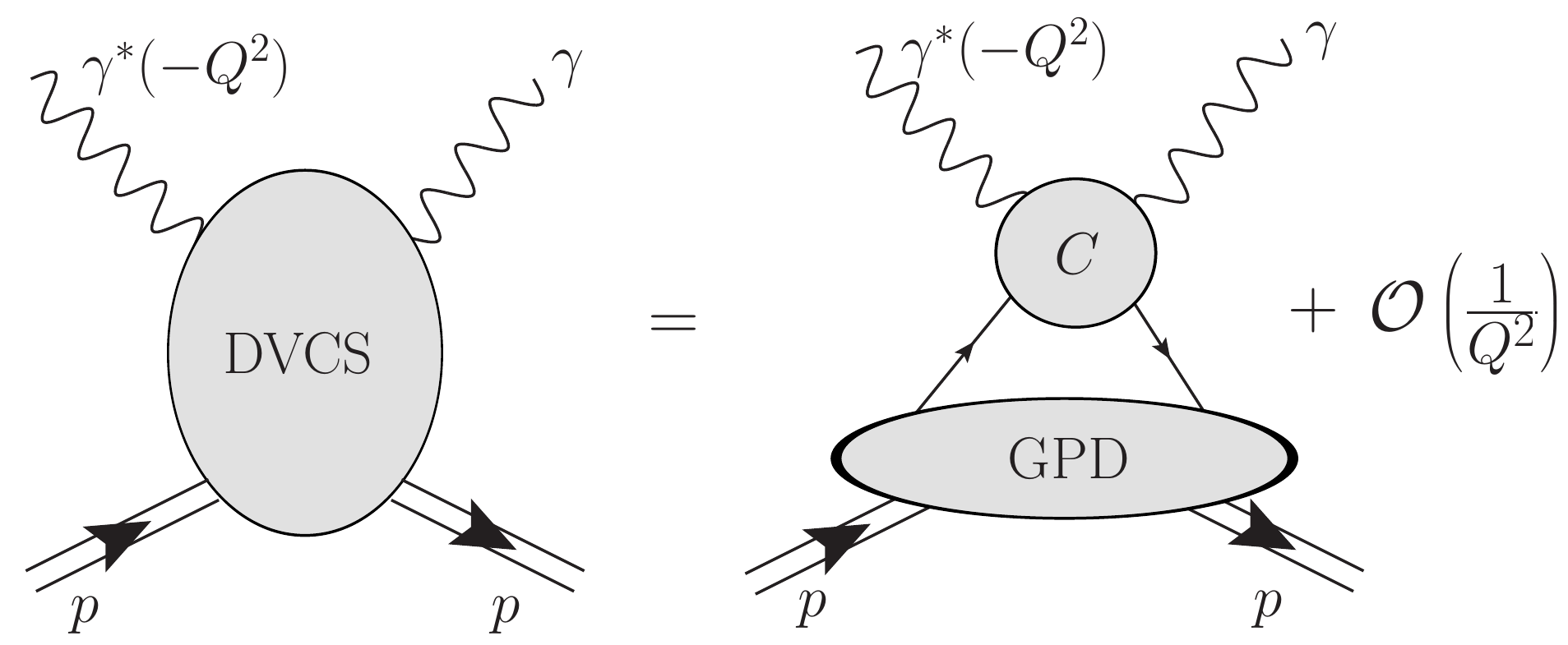}
\caption{Factorization of DVCS amplitude into convolution of
perturbative hard scattering on parton and non-perturbative
GPD function.}
\label{fig:factorization}
\end{figure}
For the twist-two related first four CFFs ($\mathcal{F}
\in \{\mathcal{H}, \mathcal{E}, \tilde{\mathcal{H}}, \tilde{\mathcal{E}}\}$)
we have a factorization theorem \cite{Collins:1998be},
see \reffig{fig:factorization},
expressing them to
leading order in $1/\Q^2$ as convolution
of the perturbatively calculable hard-scattering coefficient and
the non-perturbative GPD function, \emph{e.g.}, for flavor
singlet ($S$) contribution,
\begin{multline}
\label{eq:SIfactorization}
\mathcal{F}_{\rm S}(\xi,t,\Q^2) =
\int_{-1}^{1} \frac{dx}{\xi} \\
\times \mbox{\boldmath $C$}(x/\xi,
\Q^2/\mu^2,\alpha_s(\mu)) \mbox{\boldmath $F$}(x,\eta=\xi,
t,\mu^2)\,,
\end{multline}
where $\mu$ is the factorization scale, usually set equal
to photon virtuality $\mu^2 = \Q^2$, and to the LO
scaling variable $\xi$ is
\begin{equation}
\xi = \frac{\xb}{2-\xb}\;.
\label{eq:def-xi}
\end{equation}
Here we organize the singlet quark and gluon GPDs in a column vector
\begin{equation}
\mbox{\boldmath $F$} =
\begin{pmatrix}
F_{\Sigma} \\[1ex]  F_{\rm G}
\end{pmatrix}\,, \quad
F \in \{{H},{E},\widetilde {H},\widetilde {E}\}\;,
\label{eq:defFvec}
\end{equation}
and the hard scattering coefficient functions in a row vector,
\begin{equation}
\mbox{\boldmath $C$}=(C_{\Sigma}, \; \frac{1}{\xi} {C_{\rm G}}) \;,
\label{eq:defCvec}
\end{equation}
whose QCD perturbation series starts as
\begin{equation}
\label{eq:CvecLO} \frac{1}{\xi} \mbox{\boldmath $C$}
(x/\xi,\Q^2/\mu^2,\alpha_s(\mu))
=\left(\frac{1}{\xi-x-i\epsilon},0\right) + {\cal O}(\alpha_s)\,.
\end{equation}
Note that the crossed-diagram contribution to \refeq{eq:CvecLO}
is absorbed into the symmetrized quark singlet distribution
\begin{multline}
F_{\Sigma}(x,\eta,t,\mu^2) = \\
\sum_{q=u,d,\cdots} \Big[F_{q}(x,\eta,t,\mu^2)\mp
F_{q}(-x,\eta,t,\mu^2)\Big]\,,
\end{multline}
where the upper sign in the bracket is valid for
$F \in \{H, E\}$, while lower is for $F \in \{\widetilde H,
\widetilde E\}$. Formulas for the non-singlet sector are
analogous, and the total CFF is
\begin{equation}
 \mathcal{F} = e_{\rm NS}^2 \mathcal{F}_{\rm NS}
             + e_{\rm S}^2 \mathcal{F}_{\rm S} \;,
        \quad
\mathcal{F}_{\rm S} = \mathcal{F}_{\Sigma} +
  \mathcal{F}_{\rm G}\;,
  \label{eq:NSplusS}
\end{equation}
with charge factors $e^{2}_{\rm NS}$ and
$e^{2}_{S}=e^{2}_{\Sigma}$ determined using the
decomposition of sum over $N_f$ active light quark flavors
\begin{equation}
  \sum_{q = u, d, s, \dots} e_{q}^2 \mathcal{F}_q =
   e_{\rm NS}^2 \mathcal{F}_{NS}
   + e_{\Sigma}^2 \mathcal{F}_{\Sigma} \;,
  \label{eq:defQs}
\end{equation}
so that in LO the familiar ``handbag'' approximation
formulas \refeqs{eq:convF}{eq:convFt}
are recovered.

\begin{figure}[ht]
  \centering
  \includegraphics[scale=0.7]{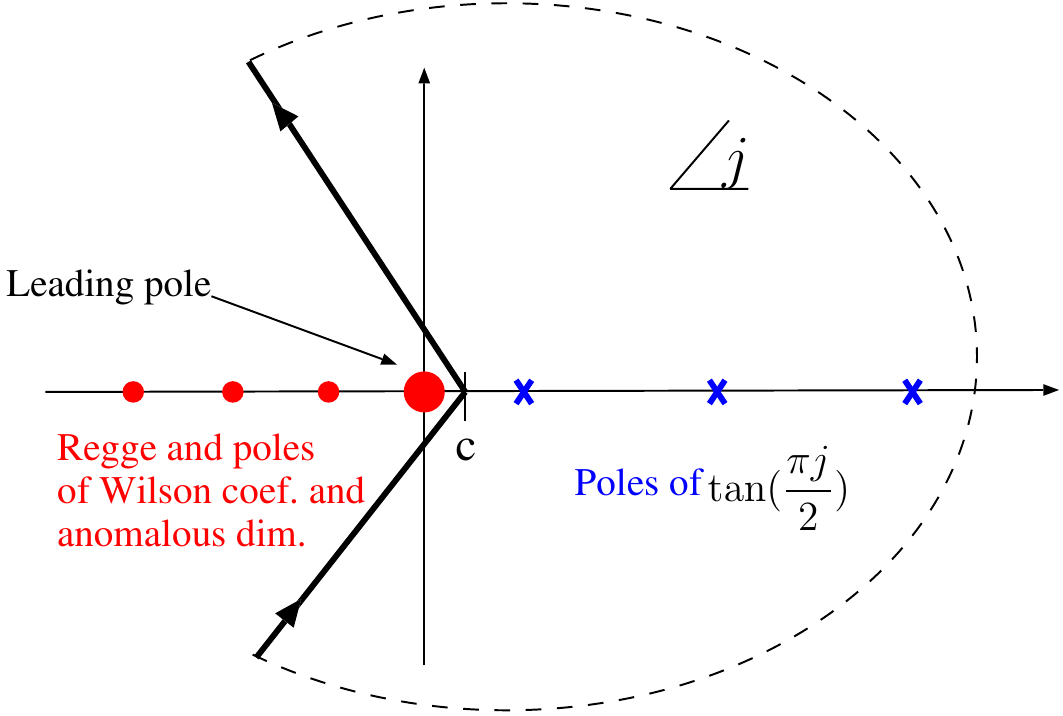}
  \caption{Contour for Mellin-Barnes integration. Slant
  to the left can improve numerical convergence.}
  \label{fig:MB}
\end{figure}

If one is working in the conformal moment GPD representation, see
\refsec{sec:conformal-moments-representation},
factorization formula for CFFs (\ref{eq:SIfactorization}) can
be transformed in the space of conformal moments using
transforms (\ref{eq:defFjg}--\ref{eq:defFjSigma}) for GPDs
and transforms
\begin{multline}
\label{eq:ConMomC}
 \mbox{\boldmath $C$}_{j}
(\Q^2/\mu^2,\alpha_s(\mu)) = \frac{
2^{j+1}\Gamma(j+5/2)}{\Gamma(3/2) \Gamma(j+4)} \\
\times \frac{1}{2} \int_{-1}^1\! dx\;\mbox{\boldmath $C$}
(x,\Q^2/\mu^2,\alpha_s(\mu)) \\
\times \left(
\begin{array}{cc}
 (j+3)\big[1-x^2\big] C_j^{3/2}(x) & 0 \\
0 & 3\big[1-x^2\big]^2 C_{j-1}^{5/2}(x)
\end{array}
\right)\,,
\end{multline}
for coefficient
functions. Then \refeq{eq:SIfactorization} is transformed into
a divergent infinite sum
\begin{multline}
\label{def:conmomsum}
 \mathcal{F}_{S}(\xi,t,\Q^2) =
 2 \sum_{j=0}^\infty \xi^{-j-1} \\
 \times \mbox{\boldmath $C$}_{j}(
 \Q^2/\mu^2,\alpha_s(\mu))\; \mbox{\boldmath
$F$}_{j}(\xi,t,\mu^2) \;.
\end{multline}
We can resum it using the Mellin-Barnes integration
along the complex-$j$ plane contour shown on \reffig{fig:MB} and (with
the help of dispersion relations connecting real
and imaginary part of CFF) we get
\begin{multline}
\label{eq:MBcalF} \mathcal{F}_{S}(\xi,t,\Q^2)
 = \frac{1}{2 i}\int_{c-i \infty}^{c+ i \infty}
dj\; \xi^{-j-1} \left[i + \tan
\left(\frac{\pi j}{2}\right)\right] \\[2ex]
\times\mbox{\boldmath $C$}_{j}(\Q^2/\mu^2,\alpha_s(\mu))
\mbox{\boldmath $F$}_{j}(\xi, t,\mu^2)\,.
\end{multline}
In the conformal moments approach, evolution of GPDs from some
fixed input scale $\mu_0$ to the scale of interest $\mu$ is
given by
\begin{equation}
  \mbox{\boldmath $F$}_{j}(\eta, t, \mu) = \sum_k
  \mbox{\boldmath ${\cal E}$}_{jk} (\mu, \mu_0; \eta)
  \mbox{\boldmath $F$}_{k}(\eta, t, \mu_0) \;,
  \label{eq:moment-space-evolution}
\end{equation}
where evolution operator
$\mbox{\boldmath$\mathcal{E}$}_{jk}$ mixes gluon and
singlet quark components but is diagonal 
($\propto \delta_{jk}$)
at LO, and
in the special $\overline{CS}$ scheme also at NLO.
Explicit form of this operator (including diagonal
NNLO part) for non-singlet and singlet case can be found in
\refcites{Mueller:2005nz,Kumericki:2006xx,Kumericki:2007sa}.


\subsection{Dispersion relation technique}
\label{sec:DR}

Using analytic properties of the DVCS amplitude via
dispersion relations provides a convenient modelling tool
\cite{Frankfurt:1997ha,Teryaev:2005uj,Kumericki:2007sa,%
Anikin:2007yh,Diehl:2007jb,Polyakov:2007rv}.
Note that since GPDs are real
functions, the handbag formula \refeq{eq:convF} leads
to a simple LO relation between the GPD calculated at
the cross-over line $\eta=x$ and the imaginary part
of the CFF, \eg, for $\mathcal{H}$,
\begin{multline}
\im{\cal H}(\xi,t,\Q^2)
\stackrel{\rm LO}{=} \pi \sum_{q=u,d,s,\dots} e_{q}^2
\\ \times \Big[H_{q}(\xi,\xi,t,{\cal Q}^2)-
H_{q}(-\xi,\xi,t,{\cal Q}^2)\Big]\;.
\label{eq:DR-imag}
\end{multline}
On the other hand, the dispersion relation connects this
to $\re{\cal H}$,
\begin{multline}
\re{\cal H}(\xi,t,\Q^2) =
\frac{1}{\pi}\,\text{P.\! V.}\!\int_{0}^{1}\!
d\xi^\prime \\
\times\left(\frac{1}{\xi-\xi^\prime} -
\frac{1}{\xi+\xi^\prime} \right) \im{\cal
H}(\xi^\prime,t,\Q^2)
+ {\cal C}_{\cal H}(t,\Q^2) \;,
\label{eq:DR-real}
\end{multline}
and at the most one subtraction constant
\begin{gather}
\mathcal{C}_\mathcal{H}=-\mathcal{C}_\mathcal{E}\,;
\qquad
\mathcal{C}_{\tilde{\mathcal{H}}}=
\mathcal{C}_{\tilde{\mathcal{E}}}=0\,.
\end{gather}
Instead of $H(x,\eta,t)$ one can model the simpler
functions $H(x,x,t)$ and $C(t)$, in a LO and leading-twist approximation, ignoring the effects of GPD
evolution,
which are all acceptable
approximations when trying to describe presently
available data in fixed-target kinematics
(for a critique of the use of dispersion
relations beyond LO, see
\cite{Goldstein:2009ks}.) 
The dispersion relation technique has
been utilized in \cite{Kumericki:2009uq} for
modelling the valence part of GPDs in hybrid models,
see \refsec{sec:fits-global-world}.

\subsection{Existing experimental data}
\label{sec:experimental-data}
\begin{figure*}  
\centerline{\includegraphics[scale=0.75]{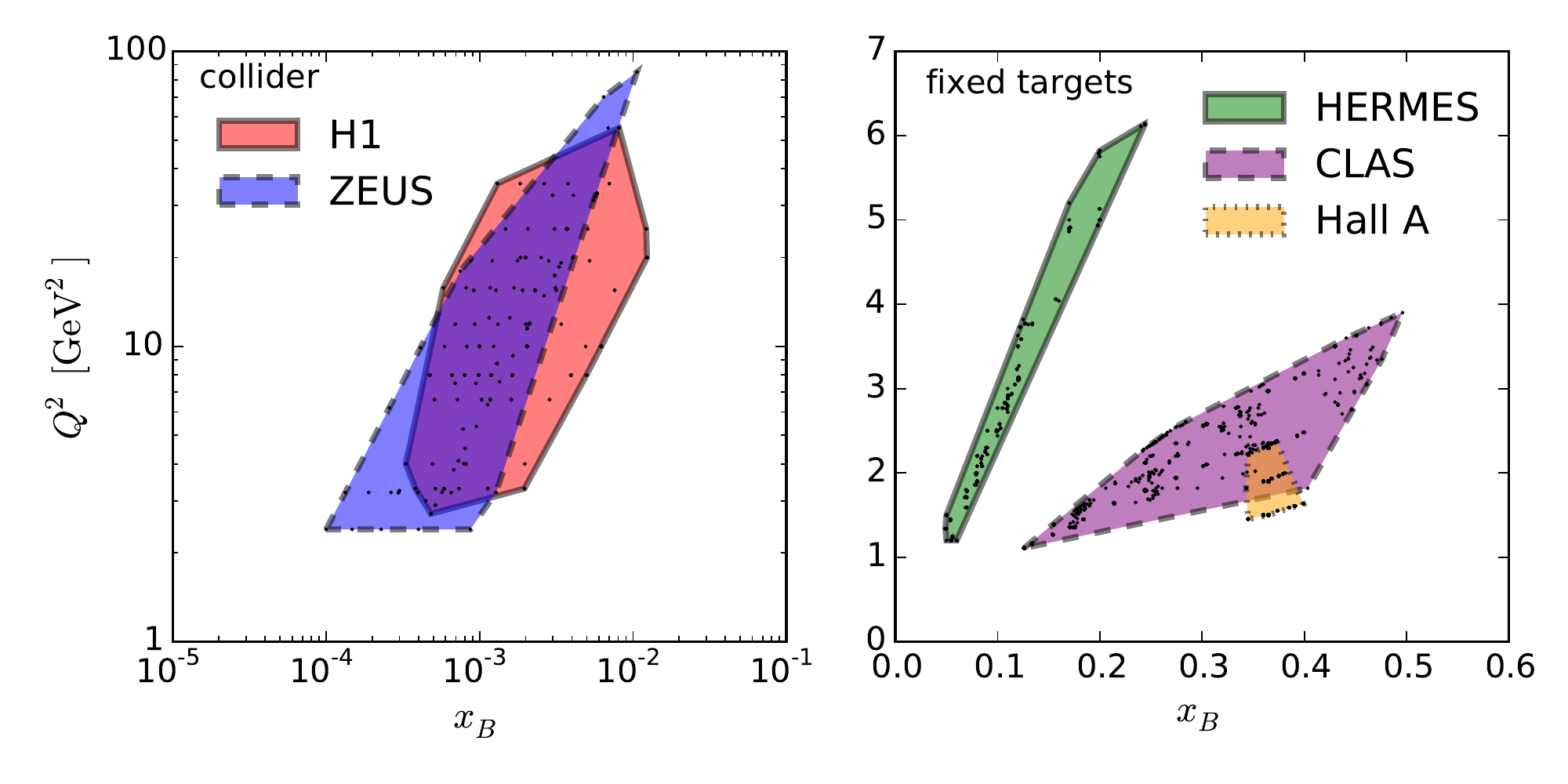}}
\caption{World DVCS data kinematic coverage.}
\label{fig:coverage}       
\end{figure*}

We will not go into a detailed review of
DVCS experiments, but will just display a simple overview of available
data on proton targets in the form of tables \ref{tab:exp-collider} and \ref{tab:exp-fixed}. 
The current kinematic coverage is summarized in fig.~\ref{fig:coverage}.

\begin{table*}  
\renewcommand{\arraystretch}{1.2}
\caption{Overview of DVCS on proton experiments at HERA collider.
Observable $\sigma$ is the cross section for the leptoproduction of real
photon $\ell p \to \ell \gamma p$, whereas $\sigma_{\rm
DVCS}$ is the cross section for the sub-process $\gamma^{*}p\to\gamma p$.
Last two columns give total number of published data points
corresponding to each observable, and number of those points
which are statistically independent.
}
\label{tab:exp-collider}       
\begin{center}
\begin{tabular}{cllccccccc}
\hline\noalign{\smallskip}
\multirow{2}{*}{Collab.} & \multirow{2}{*}{Year} &
\multirow{2}{*}{Ref.} & \multirow{2}{*}{Observables} &
\multicolumn{3}{c}{Kinematics} &\multirow{2}{*}{}&
\multicolumn{2}{c}{No. of points}  \\
\cline{5-7}\cline{9-10}\noalign{\smallskip}
& &  & & $\Q^2$ [$\GeV^2$] & $W$ [$\GeV$] & $|t|$ [$\GeV^2$] &&
total & indep. \\
\noalign{\smallskip}\hline\noalign{\smallskip}
H1   & 2001 & \cite{Adloff:2001cn}   &
\parbox{10em}{
$d\sigma/d\Q^2$, $d\sigma/dW$, \\
$\sigma_{\rm DVCS}(\Q^2)$, $\sigma_{\rm DVCS}(W)$
}&
2--20 & 30--120 & $<$1 &&
\parbox{6em}{\centering 4+4\\4+4} & 4 \\
ZEUS & 2003 & \cite{Chekanov:2003ya} &
\parbox{10.5em}{
$\sigma_{\rm DVCS}(\Q^2)$, $\sigma_{\rm DVCS}(W)$,\rule{0pt}{4ex}\\
$\sigma_{\rm DVCS}(\Q^2, W)$
}&
5--100 & 40--140&  && \parbox{6em}{\centering 10+13\rule{0pt}{4ex}\\12} &
 13 \\
H1   & 2005 & \cite{Aktas:2005ty}    &
\parbox{10.5em}{
$\sigma_{\rm DVCS}(\Q^2)$, $\sigma_{\rm DVCS}(W)$,\rule{0pt}{4ex}\\
$d\sigma_{\rm DVCS}/dt$
}&
2--80 & 30--140 & $<$1 &&
\parbox{6em}{\centering 9+14\rule{0pt}{4ex}\\12}& 9 \\
H1   & 2007 & \cite{Aaron:2007ab}    &
\parbox{10.5em}{
$\sigma_{\rm DVCS}(\Q^2)$, $\sigma_{\rm DVCS}(W)$,\rule{0pt}{4ex}\\
$\sigma_{\rm DVCS}(\Q^2,W)$ \\ $d\sigma_{\rm DVCS}/dt$
}&
6.5--80 & 30--140 & $<$1 &&
\parbox{6em}{\centering 4+5\rule{0pt}{4ex}\\15\\48} &
\parbox{6em}{\centering \rule{0pt}{4ex}\\15\\24} \\
ZEUS & 2008 & \cite{Chekanov:2008vy} &
\parbox{10.5em}{
$\sigma_{\rm DVCS}(\Q^2)$, $\sigma_{\rm DVCS}(W)$,\rule{0pt}{4ex}\\
$\sigma_{\rm DVCS}(\Q^2,W)$ \\ $d\sigma_{\rm DVCS}/dt$
}&
1.5--100 & 40--170 & 0.08--0.53 &&
\parbox{6em}{\centering 6+6\rule{0pt}{4ex}\\8\\4} & 8  \\
H1   & 2009 & \cite{Aaron:2009ac}    &
\parbox{10.5em}{
$\sigma_{\rm DVCS}(\Q^2)$, $\sigma_{\rm DVCS}(W)$,\rule{0pt}{4ex}\\
$\sigma_{\rm DVCS}(\Q^2,W)$ \\ $d\sigma_{\rm DVCS}/dt$,
$A_{\rm C}(\phi)$
}&
6.5--80 & 30--140 & $<$1 &&
\parbox{6em}{\centering 4+5\rule{0pt}{4ex}\\15\\24+6} &
\parbox{6em}{\centering \rule{0pt}{4ex}\\15\\6} \\
 \noalign{\smallskip}\hline
\textbf{}\end{tabular}
\end{center}
\end{table*}

\begin{table*}  
\renewcommand{\arraystretch}{1.4}
\caption{Overview of DVCS experiments with fixed proton
target. In the last two columns, numbers in italic font
denote measurements overall consistent with zero within one standard deviation.}
\label{tab:exp-fixed}       
\begin{center}
\begin{tabular}{cllccccccc}
\hline\noalign{\smallskip}
\multirow{2}{*}{Collab.} & \multirow{2}{*}{Year} &
\multirow{2}{*}{Ref.} & \multirow{2}{*}{Observables} &
\multicolumn{3}{c}{Kinematics} & \multirow{2}{*}{} &
\multicolumn{2}{c}{No. of points}  \\
\cline{5-7}\cline{9-10}\noalign{\smallskip}
& &  & & $\xb$ & $\Q^2$ [$\GeV^2$] & $|t|$ [$\GeV^2$] &&
total & indep.  \\
\noalign{\smallskip}\hline\noalign{\smallskip}
HERMES & 2001 & \cite{Airapetian:2001yk} & $A_{\rm LU}^{\sin\phi}$ &
0.11 & 2.6 & 0.27 &&1&1\\
CLAS & 2001 & \cite{Stepanyan:2001sm} & $A_{\rm LU}^{\sin\phi}$ &
0.19 & 1.25 & 0.19 && 1&1  \\
CLAS & 2006 & \cite{Chen:2006na}  &  $A_{\rm UL}^{\sin\phi}$&
0.2--0.4 & 1.82 & 0.15--0.44 && 6&3  \\
HERMES & 2006 & \cite{Airapetian:2006zr} &$A_{\rm C}^{\cos\phi}$ &
0.08--0.12 & 2.0--3.7 & 0.03--0.42 && 4& 4 \\
Hall A & 2006 & \cite{MunozCamacho:2006hx} & $\sigma(\phi)$, $\Delta\sigma(\phi)$ &
 0.36 & 1.5--2.3 & 0.17--0.33 && 4$\times$24+12$\times$24 &
 4$\times$24+12$\times$24\\
CLAS & 2007 & \cite{Girod:2007aa} & $A_{\rm LU}(\phi)$ &
0.11--0.58 & 1.0--4.8 & 0.09--1.8 && 62$\times$12 &
62$\times$12 \\
HERMES & 2008 & \cite{Airapetian:2008aa} &
\parbox{10em}{$A_{\rm C}^{\cos (0,1)\phi}$,
$A_{\rm UT,DVCS}^{\sin(\phi-\phi_S)}$,\rule{0pt}{3ex}\\
$A_{\rm UT,I}^{\sin(\phi-\phi_S)\cos (0,1)\phi}$,\\
$A_{\rm UT,I}^{\cos(\phi-\phi_S)\sin\phi}$}&
0.03--0.35 & 1--10 & $<$0.7 &&
\parbox{6em}{\centering {\it 12}+12+12\\12+12\\{\it 12}} &
\parbox{6em}{\centering {\it 4}+4+4\\4+4\\{\it 4}} \\
CLAS & 2008 & \cite{Gavalian:2008aa} & $A_{\rm LU}(\phi)$ &
0.12--0.48 & 1.0--2.8 & 0.1--0.8 && 66 & 33\\
HERMES & 2009 & \cite{Airapetian:2009aa}  &
\parbox{10em}{$A_{\rm LU,I}^{\sin (1,2)\phi}$\rule{0pt}{4ex},
$A_{\rm LU,DVCS}^{\sin\phi}$,\\
$A_{\rm C}^{\cos(0,1,2,3)\phi}$}&
0.05--0.24 & 1.2--5.75 & $<$0.7 &&
\parbox{6em}{\centering 18+18+18\\18+18+{\it 18+18}} &
\parbox{6em}{\centering 6+6+6\\6+6+{\it 6+6}} \\
HERMES & 2010 & \cite{Airapetian:2010ab}  &
\parbox{10em}{$A_{\rm UL}^{\sin (1,2,3)\phi}$,\rule{0pt}{4ex}\\
$A_{\rm LL}^{\cos(0,1,2)\phi}$}&
0.03--0.35 & 1--10 & $<$0.7 &&
\parbox{6em}{\centering 12+12+{\it 12}\\12+{\it 12}+12} &
\parbox{6em}{\centering 4+4+{\it 4}\\4+{\it 4}+4} \\
HERMES & 2011 & \cite{Airapetian:2011uq} &
\parbox{10em}{%
$A_{\rm LT,I}^{\cos(\phi-\phi_S)\cos(0,1,2)\phi}$,\rule{0pt}{4ex}\\
$A_{\rm LT,I}^{\sin(\phi-\phi_S)\sin (1,2)\phi}$,\\
$A_{\rm LT,BH+DVCS}^{\cos(\phi-\phi_S)\cos(0,1)\phi}$,\\
$A_{\rm LT,BH+DVCS}^{\sin(\phi-\phi_S)\sin\phi}$}&
0.03--0.35 & 1--10 & $<$0.7 &&
\parbox{6em}{\centering {\it 12+12+12}\rule{0pt}{3ex}\\ {\it 12+12}\\
   {\it 12}+12\\ 12} &
\parbox{6em}{\centering {\it 4+4+4}\rule{0pt}{3ex}\\ {\it 4+4}\\
   {\it 4}+4 \\ 4}\\
HERMES & 2012 & \cite{Airapetian:2012mq} &
\parbox{10em}{$A_{\rm LU,I}^{\sin (1,2)\phi}\rule{0pt}{4ex}$,
$A_{\rm LU,DVCS}^{\sin\phi}$,\\
$A_{\rm C}^{\cos(0,1,2,3)\phi}$}&
 0.03--0.35 & 1--10 & $<$0.7 &&
\parbox{6em}{\centering 18+{\it 18+18}\\18+18+{\it 18+18}} &
\parbox{6em}{\centering 6+{\it 6+6}\\6+6+{\it 6+6}} \\
CLAS & 2015 & \cite{Pisano:2015iqa}  &
$A_{LU}(\phi)$, $A_{UL}(\phi)$, $A_{LL}(\phi)$ &
0.17--0.47 & 1.3--3.5 & 0.1--1.4 && 166+166+166 & 166+166+166 \\
CLAS & 2015 & \cite{Jo:2015ema}  &
 $\sigma(\phi)$, $\Delta\sigma(\phi)$ &
0.1--0.58 &1--4.6 & 0.09--0.52 && 2640+2640 & 2640+2640\\
Hall A & 2015 & \cite{Defurne:2015kxq} &
$\sigma(\phi)$, $\Delta\sigma(\phi)$ &
0.33--0.40 & 1.5--2.6 & 0.17--0.37 && 480+600 & 240+360 \\
\noalign{\smallskip}\hline
\textbf{}\end{tabular}
\end{center}
\end{table*}

\subsection{Fitting methods}
\label{sec:fitting-procedures}

In a perfect world, one would like to be in possession
of several well-motivated GPD models,
depending on a small number of parameters which have one-to-one
correspondence to physical properties of the nucleon, and one would perform
a simple global fit of this model(s) to all available data, with global
minimum of $\chi^2$ (or related goodness-of-fit estimator) revealing the
GPD-encoded nucleon structure. Having several models relying on qualitatively different dynamical descriptions allows one to make robust --- or discriminating --- predictions for future experiments, and to evaluate accurately the needed beam time. It also permits stringent tests of fitting techniques through fits to pseudo-data, generated by one model, and analysed with another one.

A similar approach, or {\it global fitting}, has been working out fine for PDFs,
where results of several groups, using different methods on
different data sets are generally in good agreement, or they can be compared to one another including their uncertainties. These results are extremely useful
for the whole high-energy physics community.
So naturally, one hopes that the extraction of GPDs could follow in these footsteps.
True, there are more GPD functions to be determined
then there are PDFs, but there are also more observables to
be measured, even considering only DVCS, thanks to the final state exclusivity
(neglecting, for the sake of argument, intrinsic difficulties
of exclusive measurements compared to inclusive ones).
This has been beautifully illustrated by HERMES measurements of the
almost complete set of DVCS observables.
Next, going from measurements to GPDs involves some sort of
deconvolution of factorization formula \refeq{eq:SIfactorization}, 
which is widely considered a
major obstacle in determination of GPDs.
But one should keep in mind that PDFs are also convoluted
almost exactly the same way!

A more serious hurdle is dealing with the dimensionality
of the domain space of the unknown functions: three
dimensions for GPD $H(x,\eta,t)$ in comparison to just one for
PDF $q(x)$, if we consider the dependence on the 
factorization scale, $Q^2$, to be known (therefore not listed among the arguments 
of the functions in this context).
This situation is well known in the
field of data analysis as the \emph{curse of
dimensionality}\footnote{``It is easy to find a coin lost
on a 100 meter line, but difficult to find it on a football
field.''
Here we could say that we deal with a haystack,
100 m per side.}.
As the dimensionality of the domain space increases,
any amount of available data becomes exponentially sparse.
So it is overly optimistic to expect any time soon in GPD
physics a situation like the present one with unpolarized PDFs. We have an excellent knowledge
of these functions in a wide range of their $x$ and $Q^2$ domains, from fits to experimental data carried out in the course of several decades, using a variety of functional forms and approaches including the neural networks, which have been accurately benchmarked so as to be readily comparable to one another.
Note in this context that all known constraints on the
GPDs, listed in \refsec{sec:gpd-definition-properties}, result in
the reduction of flexibility in choosing a GPD functional form, or a reduction in \emph{volume} of domain space,
without providing any help with the problem of
dimensionality. 
However, this situation will significantly improve in the near future, with the release of GPD-related data of unprecedented accuracy, in particular from Jefferson Lab. Even if not matching the accuracy and sophistication of PDF fits, highly precise data will bring the GPD field closer to the current mapping of PDFs. A careful look at the present situation, and a series of recommendations, will hopefully help bridging this gap faster.

Even more than in the case of PDFs,
the success of any attempt of global GPD fitting
depends on the choice of the fitting function,
or GPD model. This also means that the choice of the model
introduces a significant bias, which is difficult to
estimate quantitatively,
and care has to be exercised when stating
uncertainties of fitting results.
Taking the model parameter errors determined by
observing the variation of $\chi^2$ with their change, results in an error band that 
only renders a partial representation of the uncertainty
from the comparison of experiment with any particular fitting procedure, but it does not 
account for the entirety of the theoretical uncertainty, or what we could dub as
theory ``systematics''.

As will be reviewed in \refsec{sec:fits}, global fits existing
in the present literature are reasonably successful and
it looks like there are no major problems with the described
theoretical framework.
Still, many challenges lie ahead that will be important as forthcoming high-precision data from Jefferson Lab upgraded at 12~\GeV,
COMPASS~II, and, further down the road, from an EIC, become available. 
These include requiring that the various models and fits describe simultaneously
both the unpolarized and polarized target data, as well
as meson production data with terms beyond leading twist.
A subsequent step will also include a more detailed analysis of the dependence of the data on $Q^2$, which will include going 
beyond LO of the QCD perturbation series.
Data from the deuteron and other nuclear targets will also be available for global fits, and will allow for a precise flavor separation of the various GPDs.
Because of this, other fitting approaches have also
been tested, such as various versions of the so-called
\emph{local} fits, and fits using neural networks.

{\it Local fits} utilize the fact that several observables
can be measured at a single kinematical point.
Then one can search for \emph{values}
(as opposed to \emph{shapes}) of CFF functions
that can describe the data at this point or
in its close vicinity. In that respect, local fits correspond to CFF sampling.
Such a procedure can in principle be free of the
serious model biases of global fits, since so far it relies essentially on a leading-twist handbag formalism. 
As exemplified with \refeq{eq:CIunp}, there are well-established relations between the 
CFFs and the DVCS observables. These relations depend only on kinematical factors, or on the now well-known (or well enough for our purpose) nucleon form factors.
At a given ($E_{\textrm{beam}}, \xb, Q^2, t, \phi$) experimental point, these factors can be determined and the procedure thus consists, for each such experimental point, in fitting simultaneously the various observables available for this particular kinematics, taking the real and imaginary parts of CFFs as free parameters. In principle real and imaginary parts are related by dispersion relation such as \refeq{eq:DR-real}. However, the lack of data, and the smallness of the physical region which gives access only to a sub-interval of the integration domain, has been preventing so far the implementation of dispersion relations in local fits.

As already stressed, the main advantage of the local fits 
is that they are almost model-independent (in the limit that the leading-twist
assumption is correct), as the CFFs can vary freely. The main shortcoming is that
CFFs are fitted, and not GPDs themselves. At LO, the imaginary part of a CFF is equal to singlet and non-singlet GPD combinations (\ref{eq:singletGPD}) evaluated at $x = \eta$, but beyond LO this simple interpretation is lost. Therefore, in order to access GPDs, carrying out an additional model-dependent deconvolution (similar to a global fit) seems unavoidable. Nevertheless, in the light of the complicated
interplay of many observables and many GPDs that
can be difficult to disentangle in the global
fitting procedure, local
fits can provide quite direct information about
\nonpert structure functions in a given
kinematical-- region, and can serve as a good
consistency check of the whole framework. They can also be considered as a first (although not mandatory) step towards GPDs.

Another approach to the extraction of GPDs from data is to
harness some of the fast increasing number of machine
learning techniques, for example \emph{Artificial Neural Networks} (ANN).
The latter are designed to recognize structure in a given data set
and to quantify the statistical
properties of this structure.
They have already been successfully applied to the task
of fitting hadronic structure functions to the data,
for standard PDFs
\cite{Forte:2002fg,Ball:2008by}, or electromagnetic
form factors \cite{Graczyk:2014coa}.
It is a mathematical theorem that
neural networks are able to approximate any smooth
function \cite{cybenko89}, so they can be used as a GPD
model without danger of introducing bias from the model parameters.
Fitting neural networks to replicas of experimental
data gives a convenient method of propagating
(correlated and uncorrelated) experimental
uncertainties into GPDs.
Combined with the aforementioned lack of modelling bias, this
suggests that neural networks are a promising method for
obtaining GPDs with a realistic, or faithful uncertainty estimate which certainly deserves further studies.

ANN-based approaches have not yet been applied specifically
to quantitative fits of GPDs, but only of CFFs (see the preliminary study in \refcite{Kumericki:2011rz}). Many open questions remain at present, for instance how to
implement some of the GPD properties from
\refsec{sec:gpd-definition-properties}
in this framework, or how to handle the problem of the large dimensionality
of the space of unknown functions.

An intermediate and more affordable goal would be that, in a spirit similar to the local fits, an ANN-based approach could render a parametrization of the CFFs. ANN used for this intermediate step of the analysis would provide useful information in the form of a representation of experimental data which is closer to the sought-after GPDs than the actual observables. 
This information could then be used in searches for flexible-enough GPD
models to be used in traditional fitting procedures.

It is also important to notice that different forms of ANN-based algorithms are possible, and that subclasses of specific algorithms could work more appropriately for the complex multi-variable problem of GPD fitting. For instance, an alternative to the standard ANN approaches was developed in \refcites{Honkanen:2008mb,Askanazi:2014gxa} for PDF fits using Self-Organizing Maps (SOM).
A most important aspect of self-organizing algorithms is in their ability to project high dimensional input data onto lower dimensional representations while preserving the topological features present in the training data. This aspect makes the SOM algorithms particularly appealing for an application to future GPD fitting.


\subsection{Fits to the data}
\label{sec:fits}
The different types of fits described above have been applied to a variety of data sets, following different procedures. Only a few groups have attempted to extract information using directly DVCS data. In the global fit sector we list KM \cite{Kumericki:2007sa,Kumericki:2009uq,Kumericki:2013br}, GK \cite{Goloskokov:2007nt} (where fit was to DVMP data, but provides reasonable description of DVCS as well), and GGL \cite{Goldstein:2010gu}. Local fits were performed in \refcites{Guidal:2008ie,Moutarde:2009fg,Guidal:2009aa,Guidal:2010ig,Guidal:2010de,Kumericki:2013br,Boer:2014kya}. Finally a first application of an ANN based fit to GPDs was given in \refcite{Kumericki:2011rz}.  
As we discuss later on, the fitting procedures that were used so far are candidates for future more extensive data analyses, once subjected to an appropriate benchmarking.  In what follows we describe in more detail the various efforts to analyse data both globally and locally.


\subsubsection{Global fits}
\label{sec:fits-global}


\paragraph{Global fits restricted to low-$\boldsymbol\xb$ collider data}
\label{sec:fits-global-lowx}

First fits using Mellin-Barnes SO(3) partial wave expansion model
described in \refsec{sec:MB-SO3-PWE} were performed in
\cite{Kumericki:2007sa}, where only the leading partial wave
of \refeq{eq:lowx-KM-model} was used (so GPD doesn't depend on $\eta$).
For such rigid model the quark GPD skewness ratio
\begin{equation}
r \equiv \frac{H(x, \eta=x, 0)}{H(x,\eta=0,0)} \;,
\label{eq:skewness-ratio}
\end{equation}
(where denominator is calculated using the corresponding PDF, see \refeq{eq:FqFWD})
is fixed at its conformal (``Shuvaev''\cite{Shuvaev:1999ce}) value
\begin{equation}
r \approx 1.65\;,
\label{eq:Shuvaev-ratio}
\end{equation}
which is too large to correctly describe DVCS data at LO, where these
data point to $r\approx 1$. (By the way, it turns out that at NLO the conformal
value (\ref{eq:Shuvaev-ratio}) of skewness ratio is more realistic.)

Adding second partial wave, with  negative values
of skewness parameters $s_{2}^{\rm sea}$ and
$s_{2}^{G}$,  enabled successful
simultaneous description of HERA collider DIS and DVCS data,
at LO, NLO ($\overline{MS}$ and $\overline{CS}$ scheme) and
NNLO ($\overline{CS}$ scheme) \cite{Kumericki:2009uq}.
These fits to 85 DIS $F_2$ and 101 DVCS data points (where not
all were statistically independent) consistently had
$\chi^2/{\rm d.o.f.} < 1$.
Choice to directly fit also to DIS $F_2$ measurements and not
to use some standard published PDFs was motivated by wish
to work within a consistent framework for description of both processes,
including relatively simple prescription for treatment of
heavy flavors (\ie, ignoring them), where fixed number $N_f = 4$ of
light quarks was used.

Let us also mention that the first attempt of a global fit
that besides DVCS includes also exclusive
electroproduction of $\rho^{0}$ and $\phi$ mesons is described
in \refcite{Lautenschlager:2013uya}.
This complements the fits carried out through a decade, based on several variants of the hand-bag model by GK \cite{Goloskokov:2007nt}  for the description of deeply virtual meson production. 

\paragraph{Global fits restricted to fixed target data}
\label{sec:fits-global-largex}

Jefferson Lab 6~\GeV\ fixed target experiments have afforded us several high
precision data sets in a kinematic region characterized by larger $x$ values
and $Q^2$ in the multi-\GeV\ range, where valence
quark distributions are expected to dominate (see \reffig{fig:coverage}). 
Even so, DVCS data alone are not sufficient to reliably extract GPDs from data,
and complementary information from both exclusive measurements (nucleon form
factors) and DIS (PDFs) needs to be utilized. How
to practically use this information is open to question and several groups have
proposed different approaches. We refer in particular to GK which use the
hand-bag model within the double distribution ansatz \cite{Goloskokov:2007nt}, Diehl and Kroll 
who performed a quantitative zero-skewness GPD extraction based entirely on form factors data
and PDFs from DIS  \cite{Diehl:2013xca}, and GGL who introduced a {\em recursive
fit} based on the reggeized diquark model to organize information from the inherently
different types of data sets \cite{Goldstein:2010gu}. 
We give a more detailed description of the latter since it provides a 
flexible approach with tunable parameters affording a quantitative phenomenological extraction
of GPDs from data including the evaluation of the theoretical uncertainty.  The initial evaluations in ref. \cite{Goldstein:2010gu} are usable templates for future fits including a more extended set of data.
Results from this parametrization are summarized in \reftab{tab:GGL}. 
The recursive procedure works as follows: in a first phase one
fits the forward limit of the GPDs $H$ and $\tilde{H}$, given by PDFs from unpolarized and polarized DIS scattering, using only valence distribution functions. In this step the mass parameters $m$, $M_X$, $M_\Lambda$, the Regge parameter, $\alpha$, as well as the normalization
factors for $H$ and $\tilde{H}$, eqs. (\ref{GPDH},\ref{GPDHTILDE}), are determined. In \refcite{Goldstein:2010gu} it was chosen not to quote an error on these parameters because parametrizations of the valence components of PDFs, not the actual data,  were used in the fit. 
The parameters, $\alpha^\prime$, $p_q$, are subsequently obtained, as well the normalizations for $E$, and $\widetilde{E}$ by fitting the proton and neutron electromagnetic form factors, and the axial and pseudoscalar form factors, respectively (eqs. (\ref{eq:sumF1},\ref{eq:sumF2})). A very large number of data sets on the nucleon form factors is available. The sets that were used are listed in \refcites{Goldstein:2010gu,GonzalezHernandez:2012jv,Ahmad:2006gn,Ahmad:2007vw}.
Finally, the dependence on the skewness variable, $\eta$, is determined by fitting the parameters that determine the shape of the ERBL ($-\eta < x <\eta$) region with the set of data on $A_{LU}^{\sin \phi}$.  

\begin{table*}
\caption{%
\label{tab:GGL} 
Overview of parameters from GGL global fit. The fit returns values for the $u$ and $d$ valence quarks sector to be used along with the functional forms from eq. (\ref{fit_form}). The parameters fitted to PDFs are presented in the upper part of the table. They do not include the statistical error. The parameters fitted  to the nucleon electromagnetic, axial and pseudo-scalar form factors are quoted next, along with their uncertainty  and the number of data points used in this step of the recursive fit. An accurate flavor separation was possible using the information from \refcite{Cates:2011pz}. The number of DVCS data on the asymmetry, $A_{LU}^{\sin \phi}$, used in the last stage of the fit, is 12 \cite{Girod:2007aa} + 4 \cite{MunozCamacho:2006hx}. They constrain the ERBL region (\reffig{fig:spectator}). Numerical results on these parameters are not displayed in the table. A realistic statistical analysis can be carried out using the upcoming more abundant and precise data.}
\centering
\renewcommand{\arraystretch}{1.3}
\begin{tabular}{ccccc}
\hline

u quark            &  $H$                &  $E$                & $\widetilde{H}$       & $\widetilde{E} $  \\

\hline 
PDFs Fit \\
\hline
$m_u $ (GeV)                & 0.420                                 &  0.420                              &  2.624                &  2.624            \\
$M_X^u$ (GeV)            & 0.604                                 &  0.604                              &  0.474                &  0.474            \\
$M_\Lambda^u$ (GeV)    & 1.018                            &  1.018                              &  0.971                &  0.971            \\
$\alpha_u$                    & 0.210                                  &  0.210                              &  0.219                &  0.219            \\
\hline 
Nucleon Form Factors Fit,  177 points \cite{Ahmad:2006gn}  \\
\hline
$\alpha^\prime_u$      & 2.448  $\pm$ 0.0885       &  2.811 $\pm$ 0.765      &  1.543  $\pm$ 0.296     &  5.130  $\pm$ 0.101\\
$p_u$                            & 0.620  $\pm$ 0.0725       &  0.863 $\pm$ 0.482      &  0.346  $\pm$ 0.248     &  3.507 $\pm$ 0.054\\
${\cal N}_u$                  & 2.043                                 &  1.803                               &  0.0504                          &  1.074 \\ 
\\
$ \chi^2$/d.o.f. &  0.773   &  0.664 &   0.116    &     1.98  \\   
\\
\hline
d quark            &  $H$                &  $E$                & $\widetilde{H}$       & $\widetilde{E} $  \\ 
\hline 
PDFs Fit \\
\hline

$m_d $ (GeV)                      & 0.275                              & 0.275                                &  2.603                               &  2.603                 \\
$M_X^d$ (GeV)                  & 0.913                              & 0.913                                &  0.704                               &  0.704                 \\
$M_\Lambda^d$ (GeV)    & 0.860                               & 0.860                                &  0.878                               &  0.878                 \\
$\alpha_d$                          & 0.0317                            & 0.0317                              &  0.0348                            &  0.0348                \\
\hline 
Nucleon Form Factors Fit, 177 points \cite{Ahmad:2006gn} \\
\hline
$\alpha^\prime_d$             & 2.209 $\pm$ 0.156      & 1.362  $\pm$ 0.585       &  1.298  $\pm$ 0.245      &  3.385  $\pm$ 0.145    \\
$p_d$                                   & 0.658 $\pm$ 0.257      & 1.115  $\pm$ 1.150        &  0.974  $\pm$ 0.358      &  2.326  $\pm$ 0.137    \\
${\cal N}_d$                        & 1.570                              & -2.800                               & -0.0262                            &  -0.966  \\ 
\\
$ \chi^2$/d.o.f &   0.822  & 0.688   &  0.110     &    1.00   \\ \hline


\end{tabular}
\renewcommand{\arraystretch}{1.}
\end{table*}


 
\paragraph{Global world data fits}
\label{sec:fits-global-world}

To fit all available DVCS data (coming from both collider and
fixed-target experiments) in
\refcites{Kumericki:2009uq,Kumericki:2013br}
\emph{hybrid} modelling approach was used,
where the sea parton part is modelled by Mellin-Barnes
SO(3) partial wave expansion conformal space models
described in \refsec{sec:MB-SO3-PWE} (including
LO QCD evolution), while
the valence part is modelled using the dispersion relation
technique described in \refsec{sec:DR} (where PQCD evolution
is ignored).
In particular, the valence part of $H$ and $\tilde{H}$ GPDs
is modelled at the cross-over $\eta=x$ line using a diquark-inspired
model
\begin{multline}
  H_{v}(x,x,t) = \frac{n_{v}r_{v}}{1+x}
\left(\frac{2x}{1+x}\right)^{\alpha_{v}(t)}
\\ \times \left(\frac{1-x}{1+x}\right)^{b_{v}} \frac{1}{
1-\dfrac{1-x}{1+x}\dfrac{t}{M_{v}^2}} \;,
  \label{eq:KMvalence}
\end{multline}
(and similarly for $\tilde{H}_v$)
where normalization $n_{v}$ of the corresponding PDF
($n_{v} = 1.25$, $\tilde{n}_{v} = 0.6$) is factored out
so that the free parameter $r_v$ corresponds to the skewness ratio
\refeq{eq:skewness-ratio}.
Free parameters $b_v$ and $M_v$ control large-$x$ and residual
$t$ dependence, respectively.
For $\alpha_{v}(t)$, $\rho-\omega$
the Regge trajectory is used,
\begin{equation}
  \alpha_{v}(t) = 0.43 + 0.85\,t/\GeV^2 \;.
\label{eq:rho-omega-regge}
\end{equation}
This GPD gives the imaginary part of the CFF (see eq. (\ref{eq:DR-imag})),
and the dispersion relation (\ref{eq:DR-real})
gives the corresponding real part, apart from the subtraction
constant which is separately modelled like
\begin{equation}
\mathcal{C}_{\cal H}(t) = -\mathcal{C}_{\cal E}(t)
= \frac{C}{\left(1-\dfrac{t}{M_C^2}\right)^{\!\!2}} \;,
\end{equation}
giving two additional free parameters, $C$ and $M_{C}$.
In \refcites{Kumericki:2009uq,Kumericki:2013br}
GPD $E$ is modelled solely in terms of this subtraction
constant, \ie, $\mathcal{E} = \mathcal{C}_{\mathcal E}$,
while contribution of GPD $\tilde{E}$ is described using
pion-pole-inspired effective ansatz
\begin{align}\label{eq:pion-pole}
\re{\tilde{\cal E}}(\xi, t)& = \frac{r_{\pi}}{\xi}
\frac{2.164}{\left(0.0196-\dfrac{t}{\GeV^2}\right)
\left(1-\dfrac{t}{M_{\pi}^2}\right)^{\!\!2}}\,, \\
\im{\tilde{\cal E}}(\xi, t)& = 0 \;,
\end{align}
where $m_{\pi}^2 = 0.0196\,\GeV^2$, while $M_{\pi}$
and $r_{\pi}$ are free parameters.

In total, the hybrid models used in global fits have 11-18 free
parameters. Table ~\ref{tab:fits} gives an overview of
various instances of the published fits with their
$\chi^2/{\rm d.o.f.}$ characteristics and a list of
the measured observables that were used.
Fits are multi-step, as denoted by braces and parentheses in the table.
Thus, for example, for the model KM09a first a three-parameter
($N_{\rm sea}$, $\alpha_{\rm sea}$, $\alpha_{G}$) fit to 85
DIS $F_2$ data points is performed, fixing the leading SO(3)
partial wave at $t=0$ in (\ref{eq:qj}).
Second, the three-parameter ($M_{\rm sea}$, $s_{2}^{\rm sea}$,
$s_{2}^G$) fit to 45+56=101 $\sigma_{\rm DVCS}$ and $d\sigma_{
DVCS}/dt$ collider measurements is done fixing the shape
of sea-quark and gluon GPDs.
Finally, a five-parameter ($r_v$, $b_v$, $M_v$, $C_{\cal H}$,
$M_{C_{\cal H}}$) fit to 36 fixed-target beam spin and
charge asymmetries fixes the valence part of the model.
Preliminary fit to DIS $F_2$ (and for KM09a and KM09b also
second preliminary fit to collider DVCS data) is always with
$\chi^2/{\rm d.o.f.} \sim 1$ or better, and
only the $\chi^2/{\rm d.o.f}$ of the final fit is
displayed in the table.

\begin{table*}  
\caption{Overview of KM global fits with number of fitting
  parameters and number of experimental data points used.
  Fits are multi-step, as denoted by braces and parentheses,
  see explanation in the text.
}
\label{tab:fits}       
\centering
\renewcommand{\arraystretch}{1.3}
\setlength{\tabcolsep}{4pt}
\begin{tabular}{cccccccccc}
\hline\noalign{\smallskip}
Model & & KM09a & KM09b & KM10  & KM10a  & KM10b  & KMS11 & KMM12 & KM15\\
\noalign{\smallskip}\hline
Ref & & \cite{Kumericki:2009uq} & \cite{Kumericki:2009uq} &
\cite{Kumericki:2011zc} & \cite{Kumericki:2011zc} & \cite{Kumericki:2011zc} &
\cite{Kumericki:2011rz} & \cite{Kumericki:2013br} & \cite{Kumericki:2015lhb} \\
free params. & & \{3\}+(3)+5 & \{3\}+(3)+6 & \{3\}+15 &
\{3\}+10 & \{3\}+15 & NNet & \{3\}+15 & \{3\}+15  \\
$\chi^2/{\rm d.o.f.}$ & & 32.0/31 & 33.4/34 & 135.7/160 &
129.2/149 & 115.5/126 & 13.8/36 & 123.5/80 & 240./275  \\ \hline
$F_{2}$ & \cite{Aid:1996au} &
 \{85\} & \{85\} & \{85\} & \{85\} & \{85\} & &\{85\} &\{85\}\\
$\sigma_{\rm DVCS}$ &
\cite{Chekanov:2003ya,Aktas:2005ty,Aaron:2007ab,Chekanov:2008vy} &
    (45) & (45) & 51 & 51 & 45 & & 11 & 11 \\
$d\sigma_{\rm DVCS}/dt$ & \cite{Aktas:2005ty,Aaron:2007ab} &
    (56) & (56) & 56 & 56 & 56 & & 24 & 24 \\
$A_{LU}^{\sin\phi}$ & 
 \cite{Girod:2007aa,Ellinghaus:2007dw,Gavalian:2008aa,Pisano:2015iqa} &
   12+12 & 12+12 & 12 & 16 & 12+12 &  & 4 & 13 \\
$A_{LU,I}^{\sin\phi}$ & \cite{Airapetian:2009aa,Airapetian:2012mq} &
       &       & 18  & 18  &   & 18 & 6 & 6  \\
$A_{C}^{\cos 0\phi}$ & \cite{Airapetian:2009aa,Airapetian:2012mq} &
       &    & & & & & 6 & 6 \\
$A_{C}^{\cos \phi}$ &
 \cite{Airapetian:2008aa,Airapetian:2009aa,Airapetian:2012mq} &
    12 & 12 & 18 & 18 & 12 & 18 & 6 & 6 \\
$\Delta\sigma^{\sin \phi,w}$ & \cite{MunozCamacho:2006hx,Jo:2015ema,Defurne:2015kxq} &
       & & 12 &    &  & & 12 & 63 \\
$\sigma^{\cos 0\phi,w}$ & \cite{MunozCamacho:2006hx,Jo:2015ema,Defurne:2015kxq} &
       & & 4 &   & & & 4 & 58 \\
$\sigma^{\cos \phi,w}$ & \cite{MunozCamacho:2006hx,Jo:2015ema,Defurne:2015kxq} &
       & & 4 & &  & & 4 & 58 \\
$\sigma^{\cos \phi,w}/\sigma^{\cos 0\phi,w}$ & \cite{MunozCamacho:2006hx} &
       & 4 & & & 4 &  & \\
$A_{UL}^{\sin\phi}$ & \cite{Chen:2006na,Airapetian:2010ab,Pisano:2015iqa} &  
   &  &  &  &  &  & 10 & 17 \\
$A_{LL}^{\cos 0\phi}$ & \cite{Airapetian:2010ab,Pisano:2015iqa} &  
   &  &  &  &  &  & 4 & 14 \\
$A_{LL}^{\cos \phi}$ & \cite{Pisano:2015iqa} &  &  &  &  &  &  &  & 10 \\
$A_{UT,I}^{\sin(\phi-\phi_{S})\cos\phi}$ & \cite{Airapetian:2008aa} &
    &  &  &  &  &  & 4 & 4\\
\noalign{\smallskip}\hline
\end{tabular}
\renewcommand{\arraystretch}{1.}
\end{table*}

As can be seen from table~\ref{tab:fits}, most of the KM
global fits have used only data coming from
measurements on \emph{unpolarized} target, which was what the
just described hybrid model was designed for.
The principal difference between various instances of KM09
and KM10 models is in their treatment of
cross-section measurements by Hall A collaboration
\cite{MunozCamacho:2006hx}.
Models KM09a and KM10a don't use this data at all, whereas
KM09b and KM10b use the ratio of $n=1$ and $n=0$ weighted
cosine harmonics, see \refeq{eq:weighted}.
Model KM10 and newest published KMM12 model directly use all available
non-zero harmonics of Hall A data from
\refcite{MunozCamacho:2006hx}, and KMM12 experimentally adds
also \emph{polarized} target data.
KM10a,b models are an update of KM09a,b models, where for KM09x
sea-quark and gluon GPD components were pre-fitted to the
collider subset of data, while for KM10x and KMM12 true
simultaneous global DVCS fit was performed.

Special treatment of Hall A data was necessary because
of its extreme precision, posing quite a challenge for
models, especially with the size and strong $t$-dependence
of the unpolarized cross-section.
To come to terms with that, KM10 model has extremely large
$\tilde{\cal H}$ contribution which is considered an
effective parametrization of some large contribution to DVCS amplitude.
Such a large $\tilde{\cal H}$ leads to a conflict with
measurements on longitudinally polarized target, so
KMM12 model, which included also $A_{UL}$ data,
had to have some other way to deal with the Hall A data.
It was accomplished by simultaneous increase of
both $\mathcal{H}$ and pion-pole $\tilde{\cal E}$ contributions.

In the meantime, Hall A collaboration updated their
measurements of electroproduction cross-section
in \refcite{Defurne:2015kxq}. Change with respect
to 2006 data \cite{MunozCamacho:2006hx} is significant and
it looks like this updated data will be easier to
describe with existing models.
Indeed, the recent model KM15 from \refcite{Kumericki:2015lhb},
obtained by adding also the 2015 data coming from Hall A \cite{Defurne:2015kxq}
and CLAS \cite{Pisano:2015iqa,Jo:2015ema} collaborations to the fit, releases
some tensions present in older fits.

\begin{figure*}  
\centerline{\includegraphics[scale=1.0]{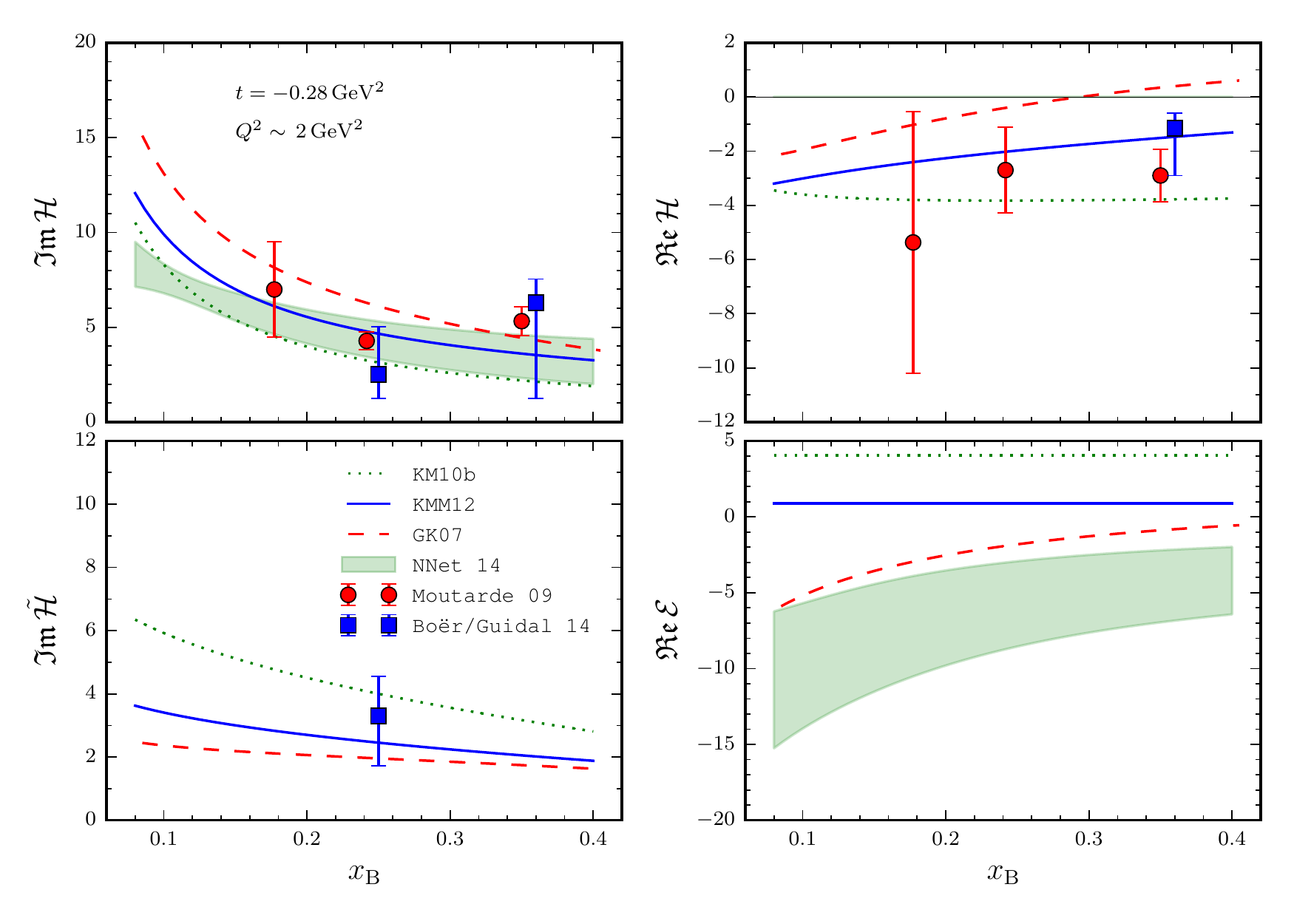}}
\caption{Selection of Compton form factors, for
moderate values of $x_B$, as extracted by
several different global \cite{Kumericki:2011zc,Kumericki:2013br,Goloskokov:2007nt},
neural network \cite{Kumericki:2015tqa},
and local \cite{Moutarde:2009fg,Boer:2014kya}
fitting procedures.
}
\label{fig:cffs}
\end{figure*}


\subsubsection{Local fits}
\label{sec:local-fits}

In the following we describe the efforts \cite{Guidal:2008ie,Moutarde:2009fg,Guidal:2009aa,Guidal:2010ig,Guidal:2010de,Moutarde:2009fg,Kumericki:2013br,Boer:2014kya} towards CFF fitting with different approaches that can all be seen as variations around the local fit strategy. 

\paragraph{Least squares minimization}

This method was pioneered in 2008 \cite{Guidal:2008ie}. In \refcites{Guidal:2008ie,Guidal:2009aa,Guidal:2010ig,Guidal:2010de}, only seven parameters were considered: $\re \mathcal{H}$, $\re \mathcal{E}$, $\re \tilde{\mathcal{H}}$, 
$\re \tilde{\mathcal{E}}$, $\im \mathcal{H}$, $\im \mathcal{E}$ and $\im \tilde{\mathcal{H}}$, and $\im \tilde{\mathcal{E}}$ was fixed to 0. This assumption has been recently removed \cite{Boer:2014kya}. Real and imaginary parts of CFFs were free to vary
within a 7- or 8-dimensional hypervolume, bounded by rather conservative
limits: $\pm$5 times the predictions of the VGG model. 

Local fits of CFFs are usually underconstrained problems. For example, let us consider the harmonic structure of the beam-spin asymmetry $A^-_{LU}$ following \refcite{Belitsky:2001ns}. Its exact structure may be different in the more refined formalisms of \refcites{Belitsky:2008bz,Belitsky:2010jw,Belitsky:2012ch} but that does not qualitatively change the argument. At leading twist, this asymmetry writes:
\begin{equation}
\label{eq:bsa-harmonic-structure-bmk02}
A^-_{LU}(\phi) = \frac{a \sin \phi}{1 + b \cos \phi + c \cos 2\phi + d \cos 3\phi} \;.
\end{equation}
Thus the information about the eight real quantities $\re \mathcal{H}$, $\re \mathcal{E}$, $\re \tilde{\mathcal{H}}$, 
$\re \tilde{\mathcal{E}}$, $\im \mathcal{H}$, $\im \mathcal{E}$, $\im \tilde{\mathcal{H}}$ and $\im \tilde{\mathcal{E}}$ is contained in just four coefficients $a$, $b$, $c$ and $d$, where some of them are kinematically suppressed. What is observed here in the case of the beam-spin asymmetry is general, and similar conclusions would be drawn for other observables. It is easy to obtain a good fit of experimental data, but many
combinations of the real and imaginary parts of CFFs can provide an equally good fit. Generically no information can reliably be extracted on any CFF unless several different observables measured at the \emph{same} kinematic configurations are studied simultaneously. Indeed, it was observed in \refcite{Guidal:2008ie}
that fitting both unpolarized
and beam-polarized Hall A cross sections resulted in a convergence of the fits for $\re \mathcal{H}$ and $\im \mathcal{H}$ only, while the other variables $\re \mathcal{E}$, $\re \tilde{\mathcal{H}}$, $\re \tilde{\mathcal{E}}$, $\im \mathcal{E}$, $\im \tilde{\mathcal{H}}$ and $\im \tilde{\mathcal{E}}$ are left undetermined. The fitting procedure also produces asymmetric error bars, which are evaluated by solving
the equation $\chi^2 = \chi^2_{\textrm{min}} +1$. These error bars actually reflect not only the statistical accuracy of the data
(which are precise at the few percent level) but the influence
of the subdominant fit parameters, left undetermined by the fit by lack of convergence. The error bars are thus a mixture of statistical and systematic uncertainties, related to the systematic uncertainties of the experimental data, but also to the systematic errors brought by the fitting procedure in itself. However this approach
may be considered as a conservative estimation of uncertainties and as one with minimal theory bias.

This fitting procedure has been successfully applied to Jefferson Lab and HERMES data, and some physical conclusions can be drawn:
\begin{itemize}
\item At fixed $t$, 
$\im \mathcal{H}$ increases as $\xb$ decreases (\ie going from Jefferson Lab to HERMES kinematics).
It is actually possible to extract $\im \mathcal{H}$ at the quasi-common value of 
$t\approx -0.28~\GeV^2$ from the Jefferson Lab Hall A, CLAS and HERMES data (with an interpolation between data points in some cases). We see in \reffig{fig:cffs} the $\xb$-dependence of $\im \mathcal{H}$.
\item The $t-$slope of $\im \mathcal{H}$ seems to increase 
with $\xb$ decreasing. The $t$-slope of the
GPD is related to the transverse spatial
densities of quarks in the nucleon.
This evolution with $\xb$ suggests that low-$x$ quarks (sea quarks) extend to the periphery of the nucleon while the high-$x$ quarks (valence quarks) tend to remain in the center of the nucleon. As discussed in \refcite{Guidal:2013rya}, this remark can be pushed further to produce some tentative transverse plane images of the nucleon structure. While an encouraging step with respect to what could be achieved with future data, the propagation of systematic uncertainties deserve a more detailed treatment.
\item $\im \tilde{\mathcal{H}}$ is in general
smaller than $\im \mathcal{H}$, see again \reffig{fig:cffs}, as expected for a polarized
quantity compared to an unpolarized one. Its $t$-dependence is also rather flat, suggesting
that the axial charge has a narrower 
distribution in the nucleon than the electromagnetic charge. \end{itemize}

\paragraph{Mapping and linearization}

In \refcite{Kumericki:2013br}, a complementary method has been described. It consists in deriving a set of relations associating DVCS observables to CFFs. This step is called \emph{mapping}. Under some reasonable approximations (leading-twist and LO description of DVCS, neglect of some $t/Q^2$ terms in analytical expressions, \ldots), these
relations can be made linear. Then, if a quasi-complete set of DVCS observables can be measured at a given ($\xb$, $Q^2$, $t$) point, one can build a system of eight linear equations
with eight unknowns, \ie the real and imaginary parts of the CFF $\mathcal{H}$, $\mathcal{E}$, $\tilde{\mathcal{H}}$ and $\tilde{\mathcal{E}}$. This system is 
then solved with standard matrix inversion and covariance error propagation techniques.

This approach has been applied to the HERMES data, which have the unique feature of having measured all beam-target single- 
and double-spin DVCS observables. However the absence of cross section measurement at HERMES implies that these observables are actually asymmetries, and that the mapping is not linear without further assumptions on the Bethe-Heitler and DVCS amplitudes. 

This mapping technique gives results that are in striking agreement with the least-square minimization technique discussed above.

\paragraph{Fitting with only $\boldsymbol H$}

One limitation of the two methods above is that every $(E_{\textrm{beam}}, \xb, Q^2, t)$ kinematic point is
considered individually and fitted independently of all others. In particular, nothing prevents the occurrence of oscillations when studying other data points, \ie when going from a local fit to a global sampling of the CFF functions. 

One attempt to enforce the smoothness of the CFFs, while introducing very little, controllable model dependence, was made in \refcite{Moutarde:2009fg}. This study used the CLAS beam spin asymmetries, the Jefferson Lab Hall A unpolarized and beam-polarized cross sections, and assumed the dominance of the GPD $H$. Smoothness is enforced by imposing a generic functional form on the GPD $H$. In \refcite{Moutarde:2009fg}, the singlet combination $H^+$ is described in the dual model framework (see \eg \refcite{Polyakov:2008aa} and refs therein). The $t$-dependence of the $B_{nl}$ coefficients of the partial wave expansion is parameterized as:
\begin{equation}
\label{eq-t-dependence-Bpl}
B_{nl}(t,Q^2_0) = \frac{a_{nl}}{1+b_{nl}(t-t_0)^2}
\end{equation}
with $t_0 =-0.28$ \GeV.

The fits to the Hall A and CLAS data were performed using both local and global procedures. The results for both kinds of fits are almost always compatible, which is a good consistency check. As expected, the results of the global fits are in general smoother, due to the implementation of the functional form for $H$. In contrast, this fitting strategy requires a large number of free parameters, which makes the fits rapidly unstable, and presumably forbids its extension to the treatment of the GPDs $E$, $\tilde{H}$, and $\tilde{E}$. This program has not been explored further due to this limitation.


\subsubsection{Neural network fits}
\label{sec:neural-network-fits}

The first study of neural network approach to
GPD fitting in \refcite{Kumericki:2011rz}
with limited set of HERMES beam spin and charge asymmetry DVCS
data \cite{Airapetian:2009aa}, is encouraging.
For these two observables it is expected that
GPD $H$ constitutes the most important contribution, so
in that study, CFFs $\im \mathcal{H}$ and $\re \mathcal{H}$ were
parametrized by neural networks as two independent functions of
kinematical variables $\xb$ and $t$. Simplifications of
parametrizing CFFs instead of GPDs and ignoring the
evolution, led to the good convergence of neural
network back-propagation learning algorithm.
Obtained CFF functions are in good agreement with those
extracted by traditional least-squares model fits in
\refcite{Kumericki:2009uq}.
More recently, preliminary results are obtained \cite{Kumericki:2015tqa}, with 
more neural-network represented CFFs fitted to global set of fixed-target data
and displayed on fig.~\ref{fig:cffs}.


\section{Proposals for future directions}
\label{sec:preparing-future}

\subsection{Treatment of uncertainties}
\label{sec:uncertainties}

With precision cross-section data coming from Hall A
collaboration, we already caught a glimpse of what future
experiments could offer.
We look forward to many precise
measurements at range of kinematical points and
researchers wishing to extract GPD-related information
from this data should better be prepared.
GPD models presently on the market will most likely not
be flexible enough for the task.
Model building in the double-distribution representation could
step up from the Radyushkin's DD ansatz,
while conformal-space models could be improved by
completing the program of consistent SO(3) partial
wave expansion using proper Wigner functions and going
away from small-$\eta$ approximation.
Pursuing models in different representations is
essential because it gives us at least some idea of
systematic bias introduced by the choice of the model.
Some quantitative understanding of this bias would
be very welcome before any error bands on plots
of GPDs could be interpreted as total uncertainty
of given GPD, and not only as propagated experimental
uncertainty.
For this, neural networks can prove helpful.

Coming to the subject of experimental uncertainties, one
should be aware that features
of GPD phenomenology bring along some specifics not
present in PDF fitting.
For example, theory of leptoproduction shows that
Fourier harmonics of asymmetries or cross sections
are useful intermediary objects and fitting to those
is better than fitting directly to observables
depending on azimuthal angle $\phi$.
Let us elaborate this important point some more.
Mathematically speaking, of course, Fourier harmonics
contain in principle the same information
as $\phi$-space functions.
Still, present state
of the field is such that higher harmonics are
neither used in models, nor are they
visible in the data.
This will stay true for the foreseeable future.
In combination with the fact that most observables are 
dominated by just the first cosine or sine harmonics,
this means that all models are trivially agreeing
with the data concerning frequency and phase of
the $\phi$-oscillations, which leads to incorrect
assessment of compatibility of various models
with the data.
Essentially, only the amplitude
of oscillations is relevant for GPD extraction.
On large arrays of vertically squeezed plots with $\phi$-abscissae
it is easy for models to look fine.
More importantly, in fitting procedures, models
can ``build-up'' good value of $\chi^2$ by good
description of ``trivial'' points, like for
example zeros of beam spin asymmetry for
$\phi=0$, $\pi$, and  $2\pi$.
Additionally, in global fits, there would be a
mismatch of statistical weight of measurements
available only as harmonics (like those from
HERMES) if they are combined with much larger
number of measurements of $\phi$-dependent
quantities, see last columns of table \ref{tab:exp-fixed}.
So, even if we disregard the fact that harmonics
have more direct connection with GPDs, they
are preferred purely from the point of
statistical model appraisal.

Consequently, harmonics should not be treated as a
less important by-product of the measurements
of $\phi$-de\-pen\-dent observables,
but additional attention should be paid to things
such as, \eg, estimates of systematic error of
harmonics.
Namely, given the $\phi$-dependent data, Fourier transform
itself can always be performed afterwards, when needed,
even by theorists. This even has the advantage
that the choice where to truncate the Fourier series can
be postponed and one can work with various scenarios.
Statistical uncertainties pose also no problem
and can be propagated from $\phi$-space to harmonics
using some standard procedure.

However, systematic error is different. To propagate
it correctly to Fourier harmonics it is first necessary
to know if the uncertainty for different values of
azimuthal angle $\phi$ is correlated or not.
Uncorrelated systematic uncertainties are for
fitting purposes usually added in quadrature to
the statistical ones, before Fourier transform is
performed.
Correlated systematic uncertainties, on the other hand,
should in principle be added to systematic ones
\emph{after} the Fourier transform.
In many analyses
such correlated uncertainties are lumped under the name ``normalization
uncertainty'' and are stated simply as global percentage of
measured values.
However, it is important to take into consideration
possible variation of this uncertainty with $\phi$.
Namely, although the true normalization uncertainty (\eg, due
to the luminosity uncertainty) will simply proportionally
influence all harmonics, behavior of
$\phi$-dependent correlated systematics can be
more complicated.
For example, one percent ($\Delta=0.01$) systematic error
of 2015 Hall A measurements stemming from the
parametrization choice
is dominantly $\cos\phi$ modulated
(see fig. 20 in \cite{Defurne:2015kxq}).
If we approximate it by pure $\cos\phi$ modulation,
then, due to this uncertainty, total cross section,
decomposed into dominant $\cos\phi$ harmonics,
$\sigma^{\cos n\phi} \equiv c_n$, will
variate like
\begin{equation}
 (c_0 + c_1 \cos\phi + \dots) \cdot  (1 + \Delta \cos\phi) \,.
\label{eq:cos-syst}
\end{equation}
This results in relative variations of leading
cross section harmonics like
\begin{align}
\frac{\delta c_0}{c_0}&
 \sim \left(\frac{c_1}{2 c_0}\right) \Delta \;, \\
\frac{\delta c_1}{c_1}&
 \sim \left(\frac{c_0}{c_1}\right) \Delta  \;.
\label{eq:cn-var}
\end{align}
Since $c_0 > c_1$, (for Hall A their ratio
is about 2),  this means that
such modulated uncertainty will
average itself out of the constant harmonic
while uncertainty of the first $\sigma^{\cos\phi}$
harmonic will get enhanced
(from one to about two percent in the discussed case of Hall A).

This becomes even more important when one works with
weighted harmonics, \refeqs{eq:w}{eq:weighted}, because
faster convergence of the series means that the enhancement
ratio $c_0/c_1$ becomes significantly larger.
Note that this first $\cos\phi$ harmonic of cross section
measured by Hall A collaboration is one of the observables
most difficult to describe in fits, so all
this may be relevant even today, when statistical
errors dominate experiments.
With increased amount and precision of data awaiting in the
future, increased attention to treatment of systematic error
is called for.

\subsection{Dissemination of experimental and theory results}
\label{sec:uniformization}

As can be seen from tables~\ref{tab:exp-collider} and
\ref{tab:exp-fixed}, experimental efforts of last
fifteen years yielded thousands of measurements
of DVCS and DVCS-related leptoproduction cross sections
and asymmetries.
Performing any kind of global analysis
on this data implies some sort of organization and
standardization.
Thinking about the amount of data expected from future
measurements, time seems ripe for deliberations
about the format this data should take in order to facilitate easy
communication between researchers themselves, as well as
between researchers and their computers.

Obviously, this short review cannot assume the authority of a
community agreement, but as a first step towards
such standardization we give in the
appendix~\ref{app:data-file-format} description and
an example of such a file format that has proven to be useful
to some of us over past years.
It is presently used for database of experimental measurements,
but it could be used in the same form also for easy
dissemination of model predictions.

Whether it is possible also in this area to completely
follow the lead of PDF fitting groups and have standardized
formats for numerical grids giving complete description
of GPDs in the relevant kinematic region, like
Les Houches PDF accord \cite{Buckley:2014ana},
is an open question.
Larger dimensionality of GPD support space in
comparison to PDFs
($x$, $\eta$, $t$, and sometimes $\Q^2$ versus
$x$, and sometimes $Q^2$) brings along
some problems. To get required precision, grids have
to be dense, and can become forbiddingly large,
so maybe analytic description of models with only
parameters in numerical form may prove to be the best
way to go.

In further contrast to situation with DIS and PDFs, formulas
connecting GPDs with observables are quite complex and there
are several sets of them available in the literature, using
different approximations. When combined with models of
GPDs in different representations and evolution code in
different schemes, the whole framework becomes
quite elaborate. A flexible computing platform for GPD phenomenology has recently been described in ref. \cite{Berthou:2015oaw}. It is not easy for a researcher to
numerically reproduce the results of others, which is a
necessary prerequisite for a healthy phenomenology
To help with this, establishment of a benchmark toy-GPD
models with published numerical characteristics all the
way to observables for a several benchmark kinematic
situations would be a great tool to have.
Here again Les Houches PDF accord \cite{Buckley:2014ana}
can serve as a role model.


\section{Conclusions}
\label{sec:conclusions}

Motivated by firm foundations in the theory of QCD
and the lure of possible access to 3D proton structure
and resolution of proton spin puzzle, phenomenology
of GPDs developed over last two decades into a mature field.
Many experiments were performed, resulting in decent
amount of DVCS data, and many attempts were made to
describe this data using different GPD models, with
varying success, as reviewed here.
Admittedly, knowledge of GPDs that would enable confident
application of proton spin sum rule or give us
reliable 3D parton probability density $q(x, \vec{b})$,
is not exactly around the corner.
Still, clear progress is visible, and we expect that
the next generation of experiments will bring along
data that will seriously constrain models and
lead to GPD shapes with reliability that we have
learned to expect from the PDF fitting.
For this to happen, more progress in theory and model
building is needed, to bridge over the kinematical
regions not directly accessible in DVCS (or other
GPD-related) experiments, and to give reliable
assessment of uncertainties entailed in model-dependent
GPD extractions.
Less model-dependent approaches, such as local fits and
neural network parametrizations, can also be pursued, to
give us Compton form factors as intermediate step
towards GPD extraction.
To mature further, field would certainly benefit from
some amount of standardization in data dissemination
and from some benchmark toy-model cases to facilitate
comparison of results coming from different
phenomenological approaches.
We hope the improved set of tools will be in place
before new generation of experimental data starts
flowing in.


\begin{acknowledgement}
\textbf{Acknowledgements.}
The authors thank Harut Avakian, Aurore Courtoy, Maxime Defurne, Nicole D'Hose, Michel Gar\c{c}on, Francois Xavier Girod, Gary Goldstein, Osvaldo Gonzalez Hernandez, Michel Guidal, Peter Kroll, Fabienne Kunne, C\'edric Mezrag, Dieter M\"{u}ller, Franck Sabati\'e, Silvia Pisano and Jakub Wagner
for numerous useful discussions.

This work was partly supported by the Department of Energy Grant DE-FG02-01ER41200, DE-AC05-06OR23177, by the Croatian Science Foundation under the project no. 8799,
by the QuantiXLie Center of Excellence, and by the
Commissariat à l'Energie Atomique et aux Energies
Alternatives, the ANR-12-MONU-0008-01 "PARTONS".
\end{acknowledgement}

\appendix

\section{Data file format}
\label{app:data-file-format}

Here we describe and give an example of a
specific data file format, originally developed
for fits in \refcite{Kumericki:2009uq},
which can be used to consistently represent
all present and future numerical data relevant
for GPD phenomenology, including
experimental measurements, theory predictions
and model GPD or CFF values.

One of the principal features of this format is
that it is both human- and computer-readable
which is of great practical convenience.
Syntactic rules are simple:

\begin{enumerate}
  \item Empty lines and lines starting with
    hash sign (\texttt{\#}) are ignored by computer
    parsers and can be used for comments meant
    for human readers.
  \item First part of the file is \emph{preamble},
    consisting of lines with structure
    \begin{center}
        \texttt{key = value}
    \end{center}
    where \texttt{key} should be regular computer
    variable identifier, \ie, should consist only
    of letters and numbers and should not start
    with number. (Special signs should be avoided
    to make file easy to parse by different programming
    languages and computing environments.)
  \item second and final part of the file is a
    \emph{grid} of numbers.
\end{enumerate}

\noindent Semantic rules are:
\begin{enumerate}
  \item There is world-unique ID number of the file,
    given by key \texttt{id}, and contact data of
    person who created the file, given by key
    \texttt{editor}. If there are further edits
    by other people keys such as \texttt{editor2} are used.
  \item Other information describing origin of the
    data can be given using keys such as
    \texttt{collaboration}, \texttt{year}, \texttt{reference},
    etc. These keys can be used for automatic plots generation.
  \item Coordinate frame used is given by
     key \texttt{frame}, equal to either \texttt{Trento}
     or \texttt{BMK}.
  \item Scattering process is described using keys
    \texttt{in1particle}, \texttt{in2particle}, \ldots
    \texttt{out1particle}, \ldots, set equal to
    usual symbols for HEP particle names.
  \item Kinematical and polarization properties of
    a particle \texttt{in1} are then given using keywords
    \texttt{in1energy},\\
    \texttt{in1polarizationvector} (\texttt{L}
    for longitudinal, \texttt{T} for trans\-ver\-sal,
    \texttt{U} or unspecified for unpolarized) etc.
  \item Key \texttt{in1polarization} describes the amount
    of polarization and is set to 1 if
    polarization is $100\%$ or if measurements are
    already renormalized to take into account
    smaller polarization (which they mostly are).
  \item Sign of \texttt{in1polarization} describes how the
    asymmetries are formed, by giving polarization of the
    first term in the asymmetry numerator (and similarly for \texttt{in1charge}).
  \item For convenience, type of the process is summarized
    by keys \texttt{process} (equal to \texttt{ep2epgamma}
    for leptoproduction of photon,
    \texttt{gammastarp2gammap} for DVCS,\\
    \texttt{gammastarp2rho0p} for DV$\rho^0$P, etc.)
    and \texttt{exptype} (equal to \texttt{fixed target}
    or \texttt{collider}).
  \item Finally, columns of numbers grid are described
    using keys such as \texttt{x1name} giving the column
    variable and \texttt{x1value = columnK},
    where \texttt{K} is the corresponding grid column number 
    counting from 1.
    Here \texttt{x1}, \texttt{x2} \ldots are used for 
    kinematics (``x-axes'',
    such as $x_{\rm B}$, $\Q^2$, $t$, $\phi$),
    while \texttt{y1} is for the measured observable.
  \item Units should be specified by keys such as \texttt{in1unit},
    and in particular for angles it should be stated whether
    their unit is \texttt{deg} or \texttt{rad}.
  \item Uncertainties are given by keys such as
    \texttt{y1error} etc., as displayed in
    example below.
  \item For Fourier harmonics, special column names are used:
    \texttt{FTn} for harmonic of azimuthal angle $\phi$ between lepton
    and reaction plane and \texttt{varFTn} for harmonic
    of azimuthal angle $\phi_S$ of target polarization vector. Then
    in the grid, positive numbers $0, 1, 2, \cdots$ denote
    $\cos 0\phi$, $\cos\phi$, $\cos 2\phi, \cdots$ harmonics,
    while negative numbers $-1, -2, \cdots$ denote
    $\sin\phi$, $\sin 2\phi, \cdots$ harmonics.
  \item If some kinematical value is common to the whole data
  set then instead of \texttt{x1value = columnK} we can
  specify, \eg, \texttt{x1value = 0.36}.
\item It is important that names for observables be
  standardized. We use names formed as given in examples in
  table~\ref{tab:identifiers}.
\end{enumerate}

\begin{table}
\begin{center}
\begin{tabular}{ccccc}
\hline
  $\sigma$  & \texttt{X}  &\hspace*{3ex}& $A_{C}$ & \texttt{AC} \\
$\Delta\sigma$ & \texttt{XLU} && $A_{UL}$ & \texttt{AUL} \\
$\sigma^{w}$ & \texttt{Xw} && $A_{\rm LL}$ & \texttt{ALL} \\
$A_{\rm LU}$ & \texttt{ALU} && $A_{\rm UT,I}$ & \texttt{AUTI} \\
$A_{\rm LU,I}$ & \texttt{ALUI} && $A_{\rm UT,DVCS}$ & \texttt{AUTDVCS} \\
$A_{\rm LU,DVCS}$ & \texttt{ALUDVCS} 
&& $A_{\rm LT,I}$ & \texttt{ALTI} \\
\noalign{\smallskip}\hline
\end{tabular}
\caption{Identifiers for DVCS observables}
\label{tab:identifiers}
\end{center}
\end{table}

As an example, we now give example of data
file corresponding to the HERMES collaboration
measurement of several Fourier harmonics
of $A_{UT, I}$ \cite{Airapetian:2008aa}, 
where numbers grid is abridged to save space.

\begin{verbatim}
## BEGIN file AUTI-HERMES-08.dat
id = 66
editor = John Doe (john@lab.org)

### Experiment

collaboration = HERMES
process = ep2epgamma
exptype = fixed target
year = 2008
reference = arXiv:0802.2499
texkey = Airapetian:2008aa
reference2 = Table 1b

### Scattering Process

frame = Trento

in1particle = e+
in1energy = 27.6
in1energyunit = GeV

in2particle = p
in2polarizationvector = T
in2polarization = +1

out1particle = e+
out2particle = p
out3particle = gamma


### Observable

# A^{sin,cos(varphi)}_{UT,I}
y1name = AUTI
y1unit = 1
y1value = column6
y1errorstatistic = column7
y1errorsystematic = column8

# y1errorsystematicplus = column8
# y1errorsystematicminus = column9

### x-axes

x1name = tm
x1unit = GeV^2
x1value = column1

x2name = xB
x2unit = 1
x2value = column2

x3name = Q2
x3unit = GeV^2
x3value = column3

x4name = varFTn
x4unit = 1
x4value = column4

x5name = FTn
x5unit = 1
x5value = column5

### Data

#-t   x_B  Q2 varFTn FTn  A    stat.  syst.
############################################

#  A_{UT,I}^{sin(phi-phi_S)}

0.03  0.08  1.9  -1  0  -0.030  0.031  0.008
0.10  0.10  2.5  -1  0   0.022  0.044  0.021
0.20  0.11  2.9  -1  0   0.133  0.050  0.025
0.42  0.12  3.5  -1  0   0.085  0.082  0.028

0.10  0.05  1.5  -1  0   0.083  0.051  0.021
0.10  0.08  2.2  -1  0   0.037  0.048  0.021

    [...]

#  A_{UT,I}^{sin(phi-phi_S) cos(phi)}

0.03  0.08  1.9  -1  1  -0.152  0.068  0.026
0.10  0.10  2.5  -1  1  -0.073  0.068  0.008
0.20  0.11  2.9  -1  1  -0.244  0.078  0.028

    [...]

#  A_{UT,I}^{cos(phi-phi_S) sin(phi)}

0.03  0.08  1.9   1 -1  -0.100  0.069  0.044
0.10  0.10  2.5   1 -1   0.054  0.076  0.030

    [...]

## END file AUTI-HERMES-08.dat
\end{verbatim}



\end{document}